\documentclass[12pt]{article}
\usepackage{amsmath}
\usepackage{graphicx}
\usepackage{enumerate}
\usepackage{inputenc}
\usepackage{amssymb}
\usepackage{longtable}
\usepackage{natbib}
\usepackage{url} 
 \usepackage{color}
 \usepackage{algorithm,algorithmic}
\usepackage{amsthm}
\newtheorem{remark}{Remark}

\usepackage{float,appendix,lmodern}
\usepackage[colorlinks,citecolor=blue,urlcolor=blue]{hyperref}
\usepackage{multirow}
\newcommand{\blind}{1}

\newcommand{\cM}{{\cal M}}

\newcommand{\cD}{\mbox{$\mathcal{D}$}}

\newtheorem{definition}{Definition}\newtheorem{theorem}{Theorem}
\newtheorem{lemma}{Lemma}

\renewcommand{\hat}{\widehat}
\def\singlespace{\def\baselinestretch{1}\@normalsize}

\newcommand{\argmin}{{\rm argmin}}
\newcommand{\argmax}{{\rm argmax}}

\newcommand{\tr}{\mbox{tr}}

\newcommand{\bA}{{\mathbf A}}
\newcommand{\bB}{{\mathbf B}}

\newcommand{\bE}{{\mathbf E}}
\newcommand{\bG}{{\mathbf G}}
\newcommand{\bH}{{\mathbf H}}
\newcommand{\bI}{{\mathbf I}}

\newcommand{\bL}{{\mathbf L}}
\newcommand{\bM}{{\mathbf M}}

\newcommand{\bQ}{{\mathbf Q}}
\newcommand{\bP}{{\mathbf P}}

\newcommand{\bS}{{\mathbf S}}
\newcommand{\bU}{{\mathbf U}}
\newcommand{\bV}{{\mathbf V}}

\newcommand{\bX}{{\mathbf X}}

\newcommand{\bZ}{{\mathbf Z}}
\newcommand{\ba}{{\mathbf a}}

\newcommand{\bc}{{\mathbf c}}

\newcommand{\bff}{{\mathbf f}}

\newcommand{\bq}{{\mathbf q}}

\newcommand{\bs}{{\mathbf s}}
\newcommand{\bu}{{\mathbf u}}
\newcommand{\bv}{{\mathbf v}}
\newcommand{\bw}{{\mathbf w}}
\newcommand{\bx}{{\mathbf x}}
\newcommand{\by}{{\mathbf y}}
\newcommand{\bz}{{\mathbf z}}

\newcommand{\bSigma}{\boldsymbol{\Sigma}}

\newcommand{\bve}{\mbox{\boldmath$\varepsilon$}}

\newcommand{\bC}{{\mathbf C}}

\newcommand{\bzero}{{\mathbf 0}}

\newcommand{\calD}{{\mathcal D}}

\newcommand{\calF}{{\mathcal F}}

\newcommand{\calM}{{\mathcal M}}

\def\6bullets{\bullet\bullet\bullet\bullet\bullet\bullet}

\begin{document}

\def\spacingset#1{\renewcommand{\baselinestretch}%
{#1}\small\normalsize} \spacingset{1}


\if1\blind
{
  \title{\bf Regularized Estimation of Loading Matrix in Factor Models for High-Dimensional Time Series}
  \author{Xialu Liu\thanks{
    The authors contribute equally and are listed in alphabetical order. {The authors thank Rongmao Zhang, Xiaoran Wu, and Baojun Dou for sharing the R code.}}\hspace{.2cm}\\
    Department of Management Information Systems, San Diego State University\\
    and \\
    Xin Wang \\
    Department of Mathematics and Statistics, San Diego State University}
      \date{}
  \maketitle
} \fi

\if0\blind
{
  \bigskip
  \bigskip
  \bigskip
  \begin{center}
    {\LARGE\bf Regularized Estimation of Loading Matrix in Factor Models for High-Dimensional Time Series}
\end{center}
  \medskip
} \fi

\bigskip
\begin{abstract}
High-dimensional data analysis using traditional models suffers from overparameterization. Two types of techniques are commonly used to reduce the number of parameters -- regularization and dimension reduction. In this project, we combine them by imposing a sparse factor structure and propose a regularized estimator to further reduce the number of parameters in factor models. One issue of factor analysis is that factors are hard to interpret, as both factors and the loading matrix are unobserved. To address this, we introduce a penalty term when estimating the loading matrix for a sparse estimate. As a result, each factor only drives a smaller subset of time series that exhibit the strongest correlation, improving the factor interpretability  and predictive performance. Compared with existing work, the proposed method yields an estimator which enjoys the oracle property, and works in more general settings. The proposed method does not require the loading matrix to be orthogonal or factors to be independent. The theoretical properties of the proposed estimator are investigated. The simulation results are presented to confirm that our algorithm performs well. We apply our method to Hawaii tourism data. 
\end{abstract}

\noindent%
\vfill

\newpage
\spacingset{1.5} 
\label{sec:intro}
\section{Introduction}
High-dimensional time series data are widely observed in various disciplines, including finance, economics, business, and medical science. However, when analyzing high-dimensional data, traditional models, such as vector autoregressive models and regression models, suffer from overparameterization as noted in \cite{lutkepohl2005} and \cite{chang2015}. Consequently, traditional methods designed for these models are often not satisfactory in terms of prediction accuracy and model interpretability \citep{tibshirani1996}. To address this problem, two types of methods are commonly used to reduce the number of parameters --- regularization and dimension reduction \citep{liu2022}. Regularization methods resort to a sparsity assumption on the coefficient matrices and incorporate a penalty in the optimization function to obtain meaningful estimators; see examples \cite{Basu:Michailidis:2015} and \cite{medeiros2016}, while dimension reduction assumes that the high-dimensional data can be represented by a low-dimensional process. Among various dimension reduction tools, factor analysis is a popular choice and has been extensively investigated in the literature \citep{pena1987,li1996,bai2002,stock2002a,diebold2006,lam2011,ando2017clustering}. In this paper, we combine both methods by imposing a sparse factor structure on the observed data and propose a regularized estimator for the loading matrix to further reduce the number of parameters. Our proposed method will not only enhance the model interpretation but also build a more parsimonious model to improve the predictive performance.

Factor analysis has been an active research area in statistics and econometrics with a comprehensive theoretical framework \citep{bai2002,bai2003,lam2011,lam2012} and numerous extensions, including factor models with structural breaks \citep{chen2015,baltagi2017, barigozzi2018,baltagi2021}, threshold mechanism \citep{massacci2017,liu2020}, regime switching \citep{liu2016}, and constraints \citep{chen2019constrained}, and factor models for matrix data \citep{wang2019} and tensor data \citep{chen2022,han2022,han2023,han2024}. 
One issue of factor analysis is that factors are difficult to interpret as both factors and the loading matrix are latent. Ideally, observed variables can be used to proxy unobserved common factors. \cite{bai2006evaluating} proposed statistical tests to evaluate the latent and observed factors. However, identifying suitable proxy variables is often challenging, especially when dealing with new data or data that are not well understood yet. 
Another solution is to interpret factors via loadings. In this paper, we incorporate a penalty function into the estimation of the loading matrix to shrink certain loadings to zero. This encourages each factor to load on a smaller subset of time series that exhibit the strongest correlation and thus enhances interpretability, thereby providing guidance on how to find meaningful proxy variables.

Two widely used statistical methods aimed at improving the interpretability of factor models are particularly worth mentioning --- varimax rotation and sparse principal component analysis (SPCA). It is well known that the loading matrix is not uniquely defined and can rotate in the loading space, which is obtained by principal component analysis (PCA) or other matrix decomposition methods. Varimax seeks the rotation that maximizes the variance of the squared elements in the loading matrix. Hence, loadings yielded by varimax are either very large in magnitude or close to 0. Compared to varimax, the method we propose has two key advantages. First, while varimax produces estimated loadings that are near-zero, our method ensures an estimate with zero elements, enhancing factor interpretability. Second, our method searches for the solution in a more general setting, yielding a more sparse loading matrix estimate. Unlike varimax, which keeps the loading space invariant, we allow the estimator to deviate slightly from the loading space obtained from PCA or its variants. SPCA is an extension of PCA \citep{zou2006} and requires the principal components to be orthogonal. In contrast to SPCA, our method relaxes this assumption, leading to a more sparse loading matrix and improving the interpretability of the factor model. Although PCA is commonly used as a tool to conduct
factor analysis, the objective of factor analysis is distinct: it seeks to identify latent factors that
explain the relationships among observed variables.

In regression models, regularization methods, where some constraints on the parameters were added to solve an inference problem, are widely used to enhance model interpretability, reduce model variation, and thus improve prediction accuracy. LASSO proposed by \cite{tibshirani1996}, uses an $L_1$ penalized likelihood for linear regression models to conduct variable selection. Subsequently, other sparsity-inducing penalties, such as smoothly clipped absolute deviation (SCAD) \citep{fan2001variable} and minimax concave penalty (MCP) \citep{zhang2010}, were also considered. Estimators obtained by SCAD and MCP both enjoy the oracle properties, meaning they perform as well as if the true underlying model were known in advance \citep{zou2006b}, which is theoretically appealing \citep{bertsimas2020}.
However, regularization methods have not been fully studied in the context of factor models. \cite{kristensen2017diffusion}, \cite{pelger2022interpretable}, \cite{uematsu2022estimation}, \cite{uematsu2022inference}, and \cite{wu2026sparse} studied the factor models with sparse loadings and most of them formulated the estimation as regularized optimization problems with an $L_1$-norm penalty term. Compared with the existing work, our proposed method has the following advantages. First, we use MCP penalty and the proposed estimator enjoys the oracle property, while their estimators do not. Our method with MCP penalty addresses two limitations of methods with $L_1$-norm penalty: estimation bias and model selection inconsistency. Secondly, our method works in more general settings, and yields a more sparse estimate, which builds a more parsimonious model to improve model interpretation and predictive performance. We relax the orthogonal assumption on the loading matrix and the assumption that factors are independent. As mentioned earlier, the loading matrix can rotate in the loading space. The existing work first selects a specific rotation, which makes columns of the loading matrix orthogonal and/or the factor covariance matrix an identity matrix, and then imposes the sparsity assumption on that particular loading matrix. However, these restrictions that factors are independent and the loading matrix
is orthogonal are often too strong for empirical analysis. For example, the Fama-French three-factor models include an overall market factor and factors related to firm size (SMB) and book-to-market equity (HML). Empirical evidence shows that the correlation between the overall market factor and HML is significantly different from 0, which is -0.38 in \cite{fama1993common}, -0.622 in \cite{durand2011fear}, -0.30 in \cite{fama2015five}. Another example is the geographic location (i.e. Europe, Asia, Africa, etc) and the economic condition including the level of economic development (i.e. advanced vs emerging), which are considered as two important determinants for international business activities \citep{kose2003international,sanyal2005determinants,fernhaber2008international}. Since Europe has the most advanced economies and Africa has the most emerging economies, it is not appropriate to assume that the loadings of these two factors are orthogonal. In this paper, we relax these restrictions and explore the entire loading space to identify the most sparse loading matrix, which is more general and leads to a more sparse estimate. {Numerical experiments and real data analysis in Section \ref{sec:sim} and Section \ref{sec:example} show that our method outperforms the existing methods.}

The rest of the paper is organized as follows. Section \ref{sec:model} introduces factor models with a sparse loading matrix. Section \ref{sec:Est} presents our proposed estimation algorithm. The theoretical properties of the proposed estimators are discussed in Section \ref{sec:them}. Numerical experiments and real data analysis are provided in Section \ref{sec:sim} and Section \ref{sec:example}, respectively. Section \ref{sec:conclusion} concludes. The technical proofs and results are included in the Supplementary Material.

\section{Model}

We introduce some notations first. For a vector $\bz$, we use $z_i$ to denote its $i$-th element. For a $p_1\times p_2$ matrix $\bZ$, its ($i,j$)-th element is denoted by $z_{ij}$ and its $i$-th column is denoted by $\bz_i$. Furthermore, we use ${\cal M}(\bZ)$ to denote the space spanned by the columns of $\bZ$. Let $\Vert \bZ\Vert_F$ be the Frobenius norm of $\bZ$, where $\Vert \bZ\Vert_F = \sqrt{\sum_{i=1}^{p_1}\sum_{j=1}^{p_2}z_{ij}^2}$, $\Vert \bZ\Vert_2$ be the $L$-2 norm of $\bZ$, where $\Vert \bZ\Vert_2 = \left[\lambda_{\max}(\bZ^\top \bZ)\right]^{1/2} $ and $\lambda_{\max}(\cdot)$ is the maximum eigenvalue of a square matrix, and $\|\bZ\|_{\min}$ is the nonzero minimum singular value of $\bZ$. We also define $L_1$, $L_{\infty}$, and max norm of a matrix as follows: $\Vert{\bf Z}\Vert_{1}  =\max_{1\leq j\leq p_2}\sum_{i=1}^{p_1}\vert z_{ij}\vert$, $\Vert{\bf Z}\Vert_{\infty}  =\max_{1\leq i\leq p_1}\sum_{j=1}^{p_2}\vert z_{ij}\vert$, and $\Vert{\bf Z}\Vert_{\max} =\max_{ij}\vert z_{ij}\vert$. We follow \cite{cape2019two} and define two-to-infinity norm as $\Vert{\bf z}\Vert_{\infty}=\max_{i}\vert z_{i}\vert$ for a vector $\bz$ and as $\Vert{\bf Z}\Vert_{2\rightarrow\infty}=\sup_{\Vert\bx\Vert_{2}=1}\Vert{\bf Z}{\bf x}\Vert_{\infty}$ for a matrix $\bZ$. For a scalar $z$, $|z|$ denotes its absolute value; for a set $\mathcal{Z}$, $|\mathcal{Z}|$ is the number of elements in the set; for a vector $\bz$ or a matrix $\bZ$, $|\bz|$ or $|\bZ|$ is the number of its nonzero elements. For any $\{a_n\}$ and $\{b_n\}$, ``$a_n\asymp b_n$'' means $\lim_{n \to \infty} a_n/b_n = c$, where $c$ is a positive constant, and ``$a_n \gtrsim b_n$'' means $a_n^{-1}b_n = o(1)$.

\label{sec:model}
Let $\bx_t$ be an observed $p \times 1$ time series
$t=1,\ldots, n$. The general form of a factor model for a $p$-dimensional time series is
\begin{equation}
\bx_t=\bA \bff_t +\bve_t, \label{eq::model}    
\end{equation}
where $\bx_t$ is the $p$-dimensional time series, $\bff_t=(f_{t1}, f_{t2}, \ldots, f_{tr})^\top$ is a set of unobserved
(latent) factor time series with dimension $r$ that is much smaller than $p$,
the matrix $\bA$ is the loading matrix of the common factors, the
term $\bA \bff_t$
can be viewed as the signal component of $\bx_t$ and called the common component,
and $\bve_t$ is a noise process. The loading matrix $\bA$ represents the impact of common
factors $\bff_t$ on the observed process $\bx_t$.

A key characteristic of factor models is that both the factors $\bff_t$ and the loading matrix $\bA$ are unobserved, leading to two challenges. First, the interpretation of latent factors is inherently difficult. Second, there is an ambiguity issue. Specifically, $(\bA, \bff_t)$ in (\ref{eq::model}) can be replaced by $(\bA\bV, \bV^{-1}\bff_t)$, where $\bV$ is an invertible $r\times r$ matrix. Hence, $\bA$ and $\bff_t$ are not uniquely defined. However, the column space spanned by $\bA$, denoted by ${\cal M}(\bA)$ and known as the loading space, remains unique. 

In this paper, to improve the model interpretability, we re-write the factor model in (\ref{eq::model}) as
\begin{equation}
 \bx_t=\bA^s \bff_t^s +\bve_t, \label{eq::model112}   
\end{equation}
where $\bA^s$ is one of the matrices with most zero elements in the loading space ${\cal M}(\bA)$, and satisfies the following conditions
\begin{enumerate}
\item $\|\ba_i^s\|_2=\|\ba_i\|_2$;
\item Let $m_{i}$ be the number of nonzero elements in $\ba^s_i$, and $0 \leq m_1 \leq m_2  \leq \ldots \leq m_r$.
\end{enumerate}
Condition 1 assumes that the norm of $\ba_i^s$ is equal to that of $\ba_i$, serving two purposes: it ensures that factor strength remains the same after the rotation, and it makes $\ba_i^s$ identifiable. Condition 2 indicates that the sparsity level of $r$ columns in $\bA^s$ is in a descending order.

We can also re-write the model with a standardized loading matrix as follows,
\begin{equation}
   \bx_t=\bQ \bz_t +\bve_t, \label{eq::model2} 
\end{equation}
where $\bq_1=\frac{\ba_1^s}{\|\ba_1^s\|_2}$ and $\bq_i=\frac{\ba_i^s}{\|\bP_i \ba_i^s\|_2}$, for $i=2,\ldots, r$, $\bP_i =\bI- \bQ_{(i)} (\bQ_{(i)}^\top \bQ_{(i)})^{-1} \bQ_{(i)}^\top$ and $\bQ_{(i)}=(\bq_1, \ldots, \bq_{i-1})$. We control the norm of $\bq_i$ by adding a constraint that the remainder has a norm of 1 if we project $\bq_i$ on the space spanned by $\{\bq_1, \ldots, \bq_{i-1}\}$, for $i=2,\ldots, r$. Note that the sparsity of $\bq_i$ and $\ba_i^s$ is the same. Compared to model (\ref{eq::model}), model (\ref{eq::model2}) not only provides a clearer interpretation of factors but also alleviates the ambiguity issue.

It is worth noting that ${\cal M}(\bA)={\cal M}(\bA^s)={\cal M}(\bQ)$. Our goal is to estimate the loading space and find one of the sparsest representatives in the loading space ${\cM}(\bA)$ ---factor matrix $\bQ$, and to recover the factor process. If $m_1 < m_2  < \ldots < m_r$, $\bQ$ can be uniquely identified, and we expect that our proposed algorithm can estimate it accurately. Otherwise, the proposed method may not be able to identify $\bQ$. However, it will find a loading matrix in $\cM(\bQ)$ with the same level of sparsity as $\bQ$. Fortunately, with that matrix, we can easily enumerate all possible choices for the sparse loading matrix when $r$ is fixed.

\noindent{\bf Example 1.} This example illustrates the difference between $\bA$, $\bA^s$, and $\bQ$. 
\begin{eqnarray}\label{eq::AQQ}
\bA= \left(
\begin{array}{ccc}
1   &1  &1\\
1   &1  &1\\
0   &1 &1\\
0   &0  &1\\
1   &2  &2
\end{array}
\right), 
\quad 
\bA^s= \left(
\begin{array}{ccc}
0   &0  &\sqrt{\frac{8}{3}}\\
0   &0  &\sqrt{\frac{8}{3}}\\
0   &\sqrt{\frac{7}{2}} &0\\
\sqrt{3}  &0  &0\\
0   &\sqrt{\frac{7}{2}}  &\sqrt{\frac{8}{3}}\\
\end{array}
\right), 
\quad
\bQ= \left(
\begin{array}{ccc}
0   &0  &1\\
0   &0  &1\\
0   &\sqrt{2}/2  &0\\
1   &0  &0\\
0   &\sqrt{2}/2  &1 \\
\end{array}
\right).
\end{eqnarray}
Note that $\bA=\bQ\bV$, where
\[
\bV= \left(
\begin{array}{ccc}
0   &0  &1\\
0   &\sqrt{2} &0\\
1  &1 &1 
\end{array}
\right).
\]
Hence, $\bz_t$ in (\ref{eq::model2}) satisfies $\bz_t=\bV^{-1}\bff_t$.
${\mathcal M}(\bA)={\mathcal M}(\bA^s)={\mathcal M}(\bQ)$ but $\bA^s$ and $\bQ$ are more sparse, making the interpretation of the model much clearer. From the loading matrix $\bQ$, we can see that only the 4-th time series loads on factor 1, the 3-rd and the 5-th time series loads on factor 2, and the 1-st, the 2-nd, and the 5-th time series load on factor 3. Note that, the columns of $\bQ$ are non-orthogonal in this example. 

If the setting in \cite{wu2026sparse} is adopted, where the columns in the loading matrix are eigenvectors of $\bM$ in (\ref{eq::M_def}) and are orthonormal, and for simplicity we further assume that $h_0=1$, $\{z_{t1}\}$, $\{z_{t2}\}$, and $\{z_{t3}\}$ are independent processes, and their auto-covariance matrix at lag 1 is identity, then the loading matrix will be written as $[\bv_1, \bv_2, \bv_3]$, where
\[
\bv_1=\frac{1}{\sqrt{30-10\sqrt{5}}}\left(
\begin{array}{ccc}
\sqrt{5}-1\\
\sqrt{5}-1\\
3-\sqrt{5}\\
0\\
2\\
\end{array}\right),\quad
\bv_2=\left(
\begin{array}{ccc}
0\\
0\\
0\\
1\\
0\\
\end{array}\right),\quad
\bv_3=\frac{1}{\sqrt{30+10\sqrt{5}}}\left(
\begin{array}{ccc}
-\sqrt{5}-1\\
-\sqrt{5}-1\\
3+\sqrt{5}\\
0\\
2\\
\end{array}\right).
\]
This loading matrix is less sparse than ours, $\bQ$ in (\ref{eq::AQQ}). From this example, we can see that our proposed method will reduce more parameters in factor models and build a more parsimonious factor model, which improves model interpretation and prediction accuracy.

Due to the latent nature of factors, various methods exist in the literature for separating factors from the noise process. Two assumptions are commonly used \citep{wang2019}. The first one assumes that the factors have impacts on most of the time series, and thus the noise process can be weakly serially dependent and weakly cross-sectionally dependent; see \cite{chamberlain1983, bai2002,bai2003,stock2002b,bai2006confidence,hallin2007,bai2008forecasting,stock2011}, among others. The second one assumes that factors capture all dynamics of the data and the noise process has no serial dependence but can accommodate strong cross-sectional dependence, see \cite{pan2008,lam2011,lam2012, chang2015,liu2016,wang2019,liu::elynn2022,chen2022}. Among the existing studies on factor models with sparse loadings, \cite{kristensen2017diffusion}, \cite{pelger2022interpretable}, \cite{uematsu2022estimation}, and \cite{uematsu2022inference} follow the first, while \cite{wu2026sparse} uses the second.
In this paper, we adopt the second assumption, assuming no serial dependence in the noise process but allowing the strong cross-sectional dependence in the noise process. However, our method can be extended to cases where the first assumption holds. 

Without loss of generality, we set the mean of the factor process to $\mathbf{0}$.  

\section{Estimation}
\label{sec:Est}
In this section, we first briefly review a standard estimation method for the loading matrix proposed by \cite{lam2011} in Section \ref{subsec:reviewlam} and then present our algorithm for obtaining a sparse estimate in Section \ref{subsec:est_algorithm}.

\subsection{The standard estimation method}
\label{subsec:reviewlam}
Define
\begin{align}
\bSigma_x(h)=\frac{1}{n-h} \sum_{t=1}^{n-h} {\rm E} (\bx_t \bx_{t+h}^\top),\quad
\bM= \sum_{h=1}^{h_0} \bSigma_x(h) \bSigma_x(h)^\top,\label{eq::M_def}
\end{align}
where $h_0$ is a pre-specified positive integer. Since $\{\bve_t\}$ has no serial dependence, we have
\begin{align}
\bM= \bA^s \left(\sum_{h=1}^{h_0} \bSigma_f^s(h) \bA^{s\top} \bA^s \bSigma_f^s(h)^\top \right) \bA^{s\top}, \label{eq::M}
\end{align}
where $\bSigma_f^s(h)=\sum_{t=1}^{n-h} {\rm E}(\bff_t^s \bff_{t+h}^{s\top})/(n-h)$. If the matrix in parentheses of (\ref{eq::M}) is of full rank, the space spanned by the eigenvectors of $\bM$ corresponding to non-zero eigenvalues is ${\calM}(\bA^s)$.

\cite{lam2011} defines the sample version of these matrices as follows
\[
\hat{\bSigma}_x(h)= \frac{1}{n-h} \sum_{t=1}^{n-h}\bx_t \bx_{t+h}^\top, \quad \hat{\bM}= \sum_{h=1}^{h_0} \hat{\bSigma}_x(h) \hat{\bSigma}_x(h)^\top.
\]

Thus, the loading space ${\cal M}(\bA^s)$ or ${\cal M}(\bQ)$ is estimated by ${{\cal M}(\hat{\bS})}$, where $\hat{\bS}=\{\hat{\bs}_1,\ldots, \hat{\bs}_r\}$ and $\hat{\bs}_i$ is the eigenvector of $\hat{\bM}$ corresponding to the $i$-th largest eigenvalue. In other words,
\begin{equation}
 \hat{\bS}= \argmax_{\bS^\top\bS=\bI_r} \text{tr}(\bS \hat{\bM} \bS^\top). \label{eq::obj}   
\end{equation}

\begin{remark} 
In practice, $r$ is unknown and needs to be estimated. There are plenty of studies on the estimation of the number of factors; see \cite{bai2002,onatski2009,kapetanios2010,lam2012}. However, the presence of sparsity in the loading matrix may pose a new challenge. \cite{trapani2018randomized} proposed a randomized sequential procedure to determine the number of factors, which is very general and does not require any assumptions on factors, loadings or errors. Under our settings, the randomized sequential test can be used to select the number of factors.
\end{remark}


\subsection{Estimation with Regularization}
\label{subsec:est_algorithm}

In this section, we first introduce the optimization problem for estimating the loading matrix and then present the proposed algorithm. 

\subsubsection{The optimization problem}

To obtain the desired estimate, the column space of this estimate should be close to ${\cal M}(\hat{\bS})$ and the number of zero elements in this estimate is sufficiently large. Therefore, we formulate an optimization problem that minimizes the distance between our estimate and $\cM(\hat{\bS})$, while incorporating a penalty term that promotes sparsity in the estimate.

The distance of two linear spaces $\calM (\bU_1)$ and $\calM (\bU_2)$ with dimension of $r$ is defined as
\begin{align}
\label{eq:distance}
\calD (\calM(\bU_1), \calM(\bU_2))= \left(1- \frac{\tr(\bH_1 \bH_1^\top \bH_2 \bH_2^\top)}{r} \right)^{1/2},
\end{align}
where the columns of $\bH_i$ are an orthonormal basis of $\calM(\bU_i)$ for $i=1,2$ \citep{chang2015}. It is a quantity between 0 and 1. It is 1 if the two spaces are orthogonal and 0 if $\calM (\bU_1)=\calM (\bU_2)$.




As for the penalty term, we use MCP with the following form: $\mathcal{P}_{\gamma}(x;\lambda) = \lambda\vert x\vert-\frac{x^{2}}{2\gamma}$ if $\vert x\vert\leq\gamma\lambda$, and $\mathcal{P}_{\gamma}(x;\lambda) = \frac{1}{2}\gamma\lambda^{2}$ if $\vert x\vert>\gamma\lambda$,
where $\gamma$ is fixed at 3 as in different literature \citep{breheny2011coordinate, breheny2015group} and $\lambda$ is a tuning parameter, which will be selected based on data-driven criteria. Compared with $L_1$-norm penalty term used in \cite{wu2026sparse}, MCP satisfies the oracle property. MCP provides asymptotically unbiased estimation while $L_1$-norm penalty term cannot.

 From Lemma D.2 in the Supplementary Material, minimizing the distance of $\calM (\bU_1)$ and $\calM (\bU_2)$ in (\ref{eq:distance}) is equivalent to minimizing $\sum_{i=1}^r \Vert \bH_1\bH_1^\top - \mathbf{h}_{2i}\mathbf{h}_{2i}^\top\Vert_F^2$. Hence, we use the following steps to estimate the columns of $\bQ$ in a sequential way, where $q_{ij}$ is the $(i,j)$-th element in $\bQ = (\bq_1,\bq_2,\dots, \bq_r)$:
\begin{enumerate}
\item We estimate $\bq_1$ by
\begin{equation}
\label{eq:step1}
   \hat{\bq}_1=\arg \min_{\bq_1} \frac{1}{2}\Vert \hat{\bS} \hat{\bS}^\top -  \bq_1 \bq_1^\top\Vert_F^2 + \sum_{j=1}^p\mathcal{P}_{\gamma}(\vert q_{1j}\vert ;\lambda),  \text{ subject to } \Vert \bq_1 \Vert_2 =1.
\end{equation}

\item Let $\tilde{\bs}_1=\hat{\bq}_1$. For $i=2, \ldots, r$, we do the following
\begin{enumerate}
    \item Let $\tilde{\bS}_i= (\tilde{\bs}_1, \ldots, \tilde{\bs}_{i-1})$.
    \item Estimate $\hat{\bq}_i$ by
    \begin{align}
    \label{eq:step2}
        \hat{\bq}_i= \arg \min_{\bq_i} \frac{1}{2}\Vert \hat{\bS} \hat{\bS}^\top -\bs_i \bs_i^\top \Vert_F^2 + \sum_{j=1}^p\mathcal{P}_{\gamma}(\vert q_{ij}\vert ;\lambda) \\
        \text{subject to } \bs_i= (\bI- \tilde{\bS}_i \tilde{\bS}_i^\top)\bq_i \text{ and } \Vert \bs_i\Vert_2 = 1.  \nonumber
    \end{align}
    \item $\tilde{\bs}_i = (\bI- \tilde{\bS}_i \tilde{\bS}_i^\top)\hat{\bq}_i$.

\item $\hat{\bQ}= (\hat{\bq}_1, \ldots, \hat{\bq}_r)$.
\end{enumerate}




    
\end{enumerate}

In \eqref{eq:step1} and \eqref{eq:step2}, the penalty function $\mathcal{P}_{\gamma}(\cdot)$ is applied to the elements in $\bq_i$, such that as $\lambda$ increases, some of the elements will be shrunk to zero to obtain a sparse estimator $\hat{\bq}_i$.

\begin{remark} Varimax rotation is a widely used technique for improving the interpretability of factor models. It seeks a rotation within ${\cal M}(\hat{\bS})$ that maximizes the variance of the squared elements of the loading matrix, i.e.
\[
\max_{{\cal M}(\bS) = {\cal M}(\hat{\bS})} \left\{\frac{1}{p} \sum_{j=1}^r \left[  \sum_{i=1}^p s_{ij}^4-\frac{1}{p} \left(\sum_{i=1}^p s_{ij}^2 \right)^2 \right]\right\},
\]
where $s_{ij}$ is the ($i,j$)-th element in $\bS$. Consequently, many elements in the loading matrix obtained via varimax rotation are close to zero but not exactly zero. Our method offers two advantages: (1) our estimate has zero elements due to the inclusion of a penalty term in the objective function and (2) we allow the column space of the estimate to deviate slightly from $\cM(\hat{\bS})$ in exchange for a more sparse structure.
\end{remark}

\begin{remark} 
In the literature, there are several studies on factor models with sparse loadings \citep{kristensen2017diffusion,pelger2022interpretable,uematsu2022estimation,uematsu2022inference,wu2026sparse}. As mentioned earlier, the loading matrix is not uniquely defined and can rotate in the loading space. All existing work first chooses a specific rotation where the covariance matrix of factors is an identity matrix and/or the columns in the loading matrix are orthogonal, and then impose the sparsity assumption on the particular loading matrix. In this study, we relax these  assumptions. Our method searches for the most sparse loading matrix in the whole loading space, which yields a more sparse estimate for a more parsimonious model to improve model interpretation and predictive performance. We compare these two approaches in Section \ref{sec:sim} and Section \ref{sec:example}. The results show that our method outperforms others.
\end{remark}


\subsubsection{The algorithm}
The optimization problems in \eqref{eq:step1} and \eqref{eq:step2} can be formatted as the following general minimization problem,
\begin{equation}
    \label{eq:optimizaiton0}
    \min_{\bq}\frac{1}{2}\Vert \bG -  \bB\bq \bq^\top \bB \Vert _F^2 + \sum_{j=1}^p \mathcal{P}_{\gamma}(\vert q_j\vert;\lambda),  \text{ subject to }  \bq^\top \bB\bq = 1.
\end{equation}
In particular, $\bG = \hat{\bS} \hat{\bS}^\top$, $\bB = \bI$ for the problem in \eqref{eq:step1}, and $\bB = \bI- \tilde{\bS}_i \tilde{\bS}_i^\top$ for the problem in \eqref{eq:step2}, which satisfies $\bB\bB = \bB$. To use the ADMM algorithm \citep{boyd2011distributed}, we rewrite the optimization problem in \eqref{eq:optimizaiton0} as 
\begin{align}
    \label{eq:optimizaiton}
    \min_{\bq,\bs}\frac{1}{2}\Vert \bG -  \bB\bq  \bs^\top\Vert _F^2 + \sum_{j=1}^p \mathcal{P}_{\gamma}(\vert q_j\vert;\lambda),\quad
    \text{subject to }  \bs = \bB \bq, \text{ and } \bs^\top \bs = 1. 
\end{align} 
Note that the original optimization problem in \eqref{eq:optimizaiton0} is with respect to $\bf{q}$, but the equivalent optimization problem in \eqref{eq:optimizaiton} is with respect to $\bf{q}$ and $\bf{s}$. We introduce $\bs$ because the ADMM algorithm decomposes the original optimization into several sub-optimization problems, and introducing $\bs$ simplifies solving these sub-optimization problems. In the ADMM algorithm, the augmented Lagrangian for \eqref{eq:optimizaiton} has the following form,
\begin{align}
    L\left(\bs, \bq, \bv\right)  =& \frac{1}{2}\Vert \bG -  \bB\bq  \bs^\top\Vert _F^2 + \sum_{j=1}^p \mathcal{P}_{\gamma}(\vert q_j\vert ;\lambda) + \left\langle\bv,\bs - \bB \bq \right\rangle + \frac{\rho}{2}\Vert \bs - \bB \bq\Vert^2\\
        &\text{subject to } \bs^\top \bs = 1, \nonumber
\end{align}
where $\bv$ is a $p$-dimensional vector containing all the Lagrange multipliers and $\rho$ is a fixed penalty parameter. Here, we set it at 1 as in \citet{ma2017concave} and \citet{wang2023spatial}.
Then, we can update $\bs, \bq, \bv$ iteratively. At the $(l+1)$-th iteration, given the current values of $\bs^{(l)}, \bq^{(l)}$, and $\bv^{(l)}$, the updates of $\bs, \bq, \bv$ are
\begin{align}
\bs^{(l+1)} & =\argmin_{\bs^\top \bs = 1}  L\left(\bs, \bq^{(l)}, \bv^{(l)}\right),\label{eq:updates0}\\
\bq^{(l+1)} & =\argmin_{\bq}  L(\left(\bs^{(l+1)}, \bq, \bv^{(l)}\right), \label{eq:updateq0}\\
\bv^{(l+1)} & =\bv^{(l)}+\rho \left(\bs^{(l+1)} -\bB\bq^{(l+1)}\right). \label{eq_update}
\end{align}

To update $\bs$, minimizing \eqref{eq:updates0} is equivalent to minimizing the following objective function with respect to $\bs$,
\[
-\bs^\top \bG \bB \bq^{(l)} - \bs^\top \bB \bq^{(l)} \rho + \bs^\top \bv^{(l)},
\]
with $\bs^\top \bs = 1$. Let $\bc_1 =  \bG \bB \bq^{(l)} + \rho \bB \bq^{(l)}  - \bv^{(l)}$. By Cauchy-Schwarz inequality, the update of $\bs^{(l+1)}$ is
\begin{equation}
    \label{eq:update_s}
    \bs^{(l+1)} = \frac{\bc_1}{\Vert \bc_1\Vert_2}.
\end{equation}

Note that $\bB \bB = \bB$, and $\bB = \bB^\top$. To update $\bq$ in \eqref{eq:updateq0}, it is equivalent to minimizing 
\begin{equation}
    \frac{\rho}{2} \bq^\top \bB \bq - \bq^\top (\bB \bv^{(l+1)} + \rho \bB \bs^{(l+1)} + \bB \bG \bs^{(l+1)}) + \sum_{j=1}^p\mathcal{P}_{\gamma}(\vert q_j\vert ;\lambda).
\end{equation}
We can re-write the objective function in the following format,
\begin{equation}
\label{eq:update_q}
 \frac{1}{2}\Vert  \frac{1}{\sqrt{\rho}}( \bv^{(l+1)} + \rho  \bs^{(l+1)} +  \bG \bs^{(l+1)})  - \sqrt{\rho} \bB \bq\Vert_2^2 + \sum_{j=1}^p\mathcal{P}_{\gamma}(\vert q_j\vert ;\lambda).
\end{equation}
\eqref{eq:update_q} can be minimized using the gradient algorithm with the MCP penalty. We use the R package {\it ncvreg} \citep{breheny2011coordinate} to obtain the solution for a fixed value of $\lambda$. 

 In summary, the computational algorithm can be described as follows in Algorithm \ref{alg}.
\begin{algorithm}[H]
	\caption{The optimization algorithm}
    \label{alg}
	\begin{algorithmic}[1]
     	\REQUIRE: Initialize $\bq^{(0)}$ and $\bv^{(0)} = \bf{0}$. 
	    \FOR {$i=1$}
         \STATE Set $\bB = \bI$
         \FOR{$l=1,2,\dots,...$}
          \STATE Update $\bs_1$ by \eqref{eq:update_s},  $\bq_1$ by minimizing \eqref{eq:update_q} and  $\bv$ by \eqref{eq_update}.
         \STATE Stop and get $\tilde{\bs}_1$ and $\hat{\bq}_1$ if convergence criterion is met.
        \ENDFOR
        \ENDFOR
        \FOR {$i=2,\dots ,r$}
        \STATE{Compute $\tilde{\bS}_i = (\tilde{\bs}_1,\dots, \tilde{\bs}_{i-1})$ and $\bB = \bI - \tilde{\bS}_i\tilde{\bS}_i^\top.$}
           \FOR{$l=1,2,\dots,...$}
        \STATE Update $\bs_i$ by \eqref{eq:update_s}, $\bq_i$ by minimizing \eqref{eq:update_q} and $\bv$ by \eqref{eq_update}.
        \STATE Stop and get $\tilde{\bs}_i$ and $\hat{\bq}_i$ if convergence criterion is met.
          \ENDFOR
		\ENDFOR
  \STATE Obtain $\hat{\bQ} = (\hat{\bq}_1,\dots, \hat{\bq}_r)$.
	\end{algorithmic}
\end{algorithm}

\begin{remark}
    The initial values we use are from the results of varimax rotation. We order the estimated basis from varimax based on the $L_1$ norm for each column, denoted as $(\bu_{v1}, \bu_{v2},\dots ,\bu_{vr})$. Then, set $\bq_1^{(0)} = \bu_{v1}$. 
    We use $\bu_{vi}$ as an initial for $\bs_i$ and $\bq_i^{(0)} = \bB^{-}\bu_{vi}$ for $i=2,\ldots, r$ to satisfy all the constraints, where $\bB^-$ is the Moore-Penrose generalized inverse of $\bB$. These initial values work well in both the simulation study and the real data analysis; See Section \ref{sec:sim} and Section \ref{sec:example}.
\end{remark}

\begin{remark}
  The stopping criterion is $\Vert \bs - \bB \bq\Vert_2\leq \epsilon$ as in the literature \citep{ma2017concave,wang2023spatial}, where $\epsilon$ is a small positive value. Here we use $\epsilon =  10^{-5}$.
\end{remark}

\begin{remark}
  We use BIC to select tuning parameters, which is also used in \cite{uematsu2022estimation}. BIC is defined as 
  \begin{equation}
      BIC(\lambda) = \log(\frac{1}{np} \sum_{t=1}^n\Vert \bx_t - \hat{\bx}_t  \Vert^2 ) + \frac{\log(np)}{np}
    \vert \hat{\bQ} (\lambda) \vert,
  \end{equation}
  where $\hat{\bx}_t = \hat{\bQ} (\hat{\bQ}^\top \hat{\bQ})^{-1} \hat{\bQ}^\top\bx_t$ and $\vert \hat{\bQ} (\lambda) \vert$ is the number of nonzero elements in $\hat{\bQ} (\lambda)$. In particular, a sequence of $\lambda$ values will be evaluated. The $\lambda$ value minimizing the BIC will be used to obtain the final estimate.
\end{remark}

\section{Theoretical properties}
\label{sec:them}

In this section, we will study the asymptotic properties of our proposed estimators. 



The regularity conditions we need are listed below.
\begin{enumerate}
\renewcommand{\labelenumi}{\textbf{(C\arabic{enumi})}}

\item  Let $\calF_i^j$ be the $\sigma$-field generated by $\{\bff_t^s: i \leq t \leq j\}$.
The joint process $\{\bff_t^s\}$ is
$\alpha$-mixing with mixing coefficients satisfying
$
\sum_{t=1}^{\infty} \alpha(t)^{1-2/\gamma}<\infty,
$
for some $\gamma>2$, where
$\alpha(t)=\sup_{i} \sup_{A \in \calF_{-\infty}^i, B \in \calF_{i+t}^{\infty}} |P(A \cap B) -P(A)P(B)|$.  \label{cond_alphamix}

\item For any $i=1,\ldots, r$, $t=1, \ldots, n$, $E(|f_{t,i}^s|^{2\gamma})< \sigma_f^{2\gamma}$, where $f_{t,i}^s$ is the $i$-th element of $\bff_t^s$,  $\sigma_f>0$ is a constant, and $\gamma$ is given in Condition (C\ref{cond_alphamix}).  \label{cond_fbound}

\item  $\bve_{t}$ and $\bff_t^s$ are uncorrelated given ${\cal F}_{-\infty}^{t-1}$. Let $\bSigma_{e,t}$ be the covariance of $\bve_t$. $|\sigma_{e,t,ij}|<\bSigma_{\epsilon}^2 <\infty$ for $i,j=1,\ldots,p$, and $t=1,\ldots,n$. In other words, the absolute value of each element of $\bSigma_{e,t}$ remains bounded by a constant $\sigma_{\epsilon}^2$ as $p$ increases to infinity, for $t=1,\ldots, n$. \label{cond_cov}


\item  There exists a constant $\delta \in [0,1]$ such that $\|\bA^s\|_2^2  \asymp \|\bA^s\|^2_{\min}\asymp m^{1-\delta}$, as $p$ goes to infinity, where $m=\sum_{i=1}^r m_i$ is the number of nonzero elements in $\bA^s$. Furthermore, $\Vert \bA^s\Vert_{\max} \leq C_1$, where $C_1$ is a positive constant. In addition, $m_1 \asymp m_2 \asymp \ldots \asymp m_r \asymp m$. \label{cond_strength}

\item  $\bM$ has $r$ distinct nonzero eigenvalues. \label{cond_eigenM}

\item $\bve_t$'s are independent sub-Gaussian random vectors. Each random vector in the sequence $\{\bff_t^s\}$ follows a sub-Gaussian distribution. \label{cond_subgaussian}
\end{enumerate}

As mentioned earlier, there are two ways to separate the noise process and the factor. The first one assumes that the idiosyncratic error has weak serial dependence and weak cross-sectional dependence with $\sum_{i=1}^p\sum_{j=1}^p |\sigma_{e,t,ij}|\leq Cp$ for any $t=1,\ldots,n$ and $C$ is a positive constant; see \cite{bai2002}, \cite{bai2003}, \cite{bai2006evaluating}, \cite{bai2008forecasting}, \cite{uematsu2022estimation},\cite{uematsu2022inference} and among others. The second one assumes that the noise process has no serial dependence but the strong cross-sectional dependence is allowed with $|\sigma_{e,t,ij}|<C$ for any $i,j=1,\ldots,p$ and $t=1,\ldots,n$; see \cite{lam2011,lam2012,chang2015,wang2019,chen2022}. We follow the second assumption in this study and believe that our approach is also adaptable to the first one; exploring this extension is left for future work. For the second assumption, Conditions (C\ref{cond_alphamix})--(C\ref{cond_cov}) and Condition (C\ref{cond_eigenM}) are quite standard \citep{lam2011, lam2012, chang2015, liu2016,wang2019,liu2022} and used to ensure that the estimated autocovariance matrices converge. 


To measure the strength of factors,  \cite{lam2012} introduced a strength factor index $\delta$ and assumed $\|\ba_i^s\|^2\propto p^{1-\delta}$ for $i=1,\dots, r$, where $\ba^s_i$ is the $i$-th column of $\bA^s$ and $\delta \in [0,1]$. When $\delta=0$, the factors are strong; when $\delta>0$, the factors are weak.  \cite{chang2015} proposed a similar measure that is $\|\bA^s\|^2_2 \propto \|\bA\|_{\min}^2 \propto p^{1-\delta}$. Since the strength of factors is defined through the norm of loadings, it is reasonable to consider only nonzero elements and replace $p$ with $m$ when imposing the sparsity assumption on the loading matrix in (C\ref{cond_strength}). In fact, $\delta$ reflects the scale of the elements in the loading matrix, and this can be confirmed by the settings of numerical experiments in \cite{lam2011}, where each element in the loading matrix is generated from a uniform random variable on the interval $[-1, 1]$ divided by $p^{\delta/2}$. In simulation section, we will generate random variables from a truncated standard normal distribution and then divide them by $(m/r)^{\delta/2}$ to obtain elements in $\bA^s$.  In addition, we impose an elementwise bound on $\bA^s$ with $\Vert \bA^s\Vert_{\max} \leq C_1$ and assume that the sparsity level of loading vectors remains the same in Condition (C\ref{cond_strength}). The relationship between $m$ and $p$ definitely plays an important role in the convergence rate of our estimators. We will discuss it in Section \ref{sec:them}.

Condition (C\ref{cond_subgaussian}) is a commonly used assumption in models for high-dimensional data analysis, such as regression models in \cite{ma2017concave} and \cite{wang2023spatial}, and factor models for functional time series, as in \cite{guo2021factor} and \cite{fang2022finite}. Since the tails of a sub-Gaussian random variable are dominated by the tails of a normal random variable, Condition (C\ref{cond_subgaussian}) helps bound the tails of the noise and factor processes. The definition is provided in the Supplemental Materials Section E.

As explained earlier, $\bQ$ is not necessarily an orthogonal matrix. Hence, we impose an assumption to ensure that column vectors in $\bQ$ are well separated as the dimension grows. To achieve this, we first obtain the orthogonal basis of $\cM(\bQ)$ using Gram-Schmidt orthonormalization. Specifically, let $\bS=(\bs_1, \bs_2,\ldots, \bs_r)$, where $\bs_1=\bq_1$, and
$
\bs_i=(\bI-\bS_i\bS_i^\top) \bq_i, 
$
where $\bS_i=(\bs_1,\ldots, \bs_{i-1})$ for $i=2,\ldots, r$. Let $\mathcal{V}_{{i}}$
denote the nonzero indices of ${\bf q}_{i}$ and $\mathcal{V}_{s_{i}}$
denote the nonzero indices of ${\bf s}_{i}$.
We define $\mathcal{V}_{i}^*=\mathcal{V}_{s_{1}}\cup\mathcal{V}_{s_{2}}\dots\cup\mathcal{V}_{s_{i-1}}\cup\mathcal{V}_{i}$, and $\mathcal{N}_{i}^*=\mathcal{V}_{i}^*\backslash\mathcal{V}_{i}$. 
$\mathcal{N}_i^*$ contains indices where the corresponding elements in $\bq_i$ are zero while the corresponding elements in at least one of $\{\bs_i\mid i=1,\ldots, i-1\}$ are nonzero. Note that $\mathcal{N}_i^*$ cannot be an empty set. Otherwise, $(\mathcal{V}_{s_{1}}\cup\mathcal{V}_{s_{2}}\dots\cup\mathcal{V}_{s_{i-1}}) \subset \mathcal{V}_i$, which means that there exists a vector $\bv \in \mathbb{R}^{(i-1)}$ such that $(\bq_i-\bS_i\bv)$ is more sparse than $\bq_i$ and thus $(\bq_1, \ldots, \bq_i-\bS_i\bv)$ is more sparse than $(\bq_1,\ldots,\bq_i)$. If that is true, $\bQ$ would not be one of the loading matrices with most zero elements in $\cM(\bA)$.

Let ${\bf S}_{i,1}={\bf S}_{i[\mathcal{N}_{i}^{*}]}$, we also have the following two assumptions. 

\begin{enumerate}
\renewcommand{\labelenumi}{\textbf{(C\arabic{enumi})}}
\setcounter{enumi}{6}
\item $\Vert\mathbf{S}_{i,1}\Vert_{\min}
\asymp 1$. \label{cond_si1_eigen} 

\item  There exists a positive constant $C_{\mu} > 1$ such that $\Vert{\bf S}\Vert_{2\rightarrow\infty}\leq C_{\mu}\sqrt{\frac{r}{m}}$. \label{cond_coherence} 

\end{enumerate}

Condition (C\ref{cond_si1_eigen}) indicates that the column vectors in $\bQ$ are far apart and each column vector provides enough information about zero elements as the dimension grows.

The bounded coherence assumption in Condition (C\ref{cond_coherence}) is widely used in matrix theory; see examples in \cite{fan2018pertur} and \cite{cape2019two}. \cite{cape2019two} assumes $\Vert\bS\Vert_{2\rightarrow} \leq C_{\mu}\sqrt{\frac{r}{p}}$ for a $p \times r$ orthonormal matrix $\bS$. Since $\bQ$ in our setting is sparse with $m$ nonzero elements, we replace $p$ with $m$ and assume that the sparsity level of $\bS$ is $O(m)$. Condition (C\ref{cond_coherence}) implies that each element in $\bS$ is bounded by $C_{\mu}\sqrt{\frac{r}{m}}$.



First, we have Theorem \ref{thm_s_nosparse} for the asymptotic property for $\hat{\bS}$ obtained in \eqref{eq::obj}, where is the estimator proposed in \cite{lam2011} without considering the sparsity.

\begin{theorem}
    \label{thm_s_nosparse}
     Under Conditions (C\ref{cond_alphamix})-(C\ref{cond_eigenM}) and $m^{\delta-1}p n^{-1/2}=o(1)$, it holds that
\[
\| \cM(\hat{\bS})-\cM(\bS)\Vert_2 =O_p(m^{\delta-1}p n^{-1/2}).
\]
\end{theorem}

Theorem \ref{thm_s_nosparse} shows the impact of the sparsity level on the convergence rate of the estimated loading space by \cite{lam2011}. If the number of nonzero elements in $\bQ$ grows as fast as the dimension, i.e., $m=O(p)$, $\cM(\hat{\bS})$ converges to $\cM(\bS)$ at the rate of $p^{\delta} n^{-1/2}$, which is the same as the results in \cite{lam2011}. If $\delta<1$, the more sparse $\bQ$ is, the more bias $\hat{\bS}$ includes, and thus, the slower $\cM(\hat{\bS})$ converges to $\cM(\bS)$. 

Next, we will study the asymptotic properties of our proposed estimator $\hat{\bQ}$. The following theorem shows that our estimator converges faster than the one proposed by \cite{lam2011} when the loading matrix is sparse.



Let $b = \min_i\min_{j\in \mathcal{V}_i} \vert q_{ij}\vert$, which is the minimal signal of $\bQ$,  and $\phi_{n,p,m} = \max\left(m^{2\delta-2}p^{2}n^{-1/2},m^{\delta}\right)$. We define $\tau_{n,p,m}=\phi_{n,p,m}\sqrt{\frac{\log p}{n}}$ if $m=o(p)$, and $\tau_{n,p,m}=p^{\delta}n^{-1/2}$ if $m=O(p)$. We have the following result for the proposed estimator.


\begin{theorem}
    \label{thm_general}
Assume that $m_1 < m_2 < \ldots < m_r$ and $b > \gamma\lambda$. If $\lambda \gtrsim \tau_{n,p,m}$ and $\tau_{n,p,m}= o(1)$ as $n\rightarrow \infty$ and $p\rightarrow\infty$. Under Conditions (C\ref{cond_alphamix})-(C\ref{cond_coherence}), then 
\begin{align*}
    &\Vert\hat{{\bf Q}}-{\bf Q}\Vert_2= O(\tau_{n,p,m})= \begin{cases}
      O_{p}\left(\phi_{n,p,m}\sqrt{\frac{\log p}{n}}\right) & \text{if } m=o(p), \\  
      O_{p}\left(m^{\delta - 1} p n^{-1/2}\right) = O_{p}\left(p^\delta n^{-1/2}\right) & \text{if } m = O(p),\\
    \end{cases} \\
  & P (\hat{\mathcal{V}}_i = \mathcal{V}_i) = 1, \mbox{ for } i=1,\ldots, r,
\end{align*}
as $n$ and $p$ go to infinity, where and $\hat{\mathcal{V}}_i$ contains the indexes of nonzero elements in $\hat{\bq}_i$.
\end{theorem}


\begin{remark}
   From the first step of our proof for Theorem \ref{thm_general}, we show that the proposed estimator has the same convergence rate as the oracle estimator and enjoys the oracle property. 
\end{remark}



Theorem \ref{thm_general} shows that the proposed estimators for the loading matrix and the nonzero indexes of the loading matrix are both consistent under some mild conditions. It also demonstrates the impact of the sparsity level of the loading matrix on its estimation error. If $m=O(p)$, $\hat{\bQ}$ converges at the same rate as the estimator proposed in \cite{lam2011}. If $m=o(p)$, the convergence rate of $\hat{\bQ}$ is determined by two terms: the first term, $O(m^{2\delta-2}p^2n^{-1/2}\sqrt{\frac{\log p}{n}})$, is related to the estimation bias and the second term, $O(m^\delta \sqrt{\frac{\log p}{n}})$, is related to the estimation variance. If $m^{\delta-2}p^2 n^{-1/2}\geq O(1)$, i.e., the loading matrix is quite sparse, $\hat{\bS}$ has a larger bias as shown in Theorem \ref{thm_s_nosparse}, so the estimation error of $\bQ$ is dominated by the first term. In that case, the more sparse the matrix, the larger the error is. If $m^{\delta-2}p^2 n^{-1/2}=o(1)$, i.e., the loading matrix is relatively dense, the estimation error is dominated by the second term (variance). In that case, the more sparse the matrix, the smaller the error is.


\begin{remark}
Theorem \ref{thm_general} shows the estimators are consistent when the sparsity levels of $\{\bq_i \mid i=1,\ldots, r\}$ are all distinct. However, even if it is not true and $\bQ$ is not uniquely defined, our method can estimate the loading space well and capture one of the most sparse loading matrices; see the example in Section \ref{subsec:r3}.
\end{remark}

\begin{remark}
From Theorem \ref{thm_general}, we have $\Vert \tilde{\bS}_i - \bS_i\Vert_2 = O_p(\tau_{n,p,m})$, $\Vert  \tilde{\bS}_i\tilde{\bS}_i^\top  - \bS_i\bS^\top\Vert_2  = O_p(\tau_{n,p,m})$ and $\mathcal{D}\left(\mathcal{M}\left(\tilde{{\bf S}}_i\right),\mathcal{M}\left({\bf S}_i\right)\right)=O_p(\tau_{n,p,m})$.
\end{remark}


\begin{theorem}
    \label{thm_estimate}
    If all eigenvalues of ${\bf \Sigma}_{e,t}$ are uniformly bounded from infinity as $p\rightarrow \infty$, it holds that 
    \begin{equation}
        p^{-1/2}\Vert\hat{{\bf Q}}\hat{{\bf z}}_{t}-{\bf Q}{\bf z}_{t}\Vert_{2}=O_{p}\left(p^{-1/2}m^{1/2-\delta/2}\Vert\hat{{\bf Q}}-{\bf Q}\Vert_{2}+p^{-1/2}\right), \label{eq::common}
    \end{equation}
    as $n$ and $p$ go to infinity, where $\hat{{\bf z}}_{t} = (\hat{\bQ}^\top \hat{\bQ})^{-1} \hat{\bQ}^\top\bx_t$.
\end{theorem}

Theorem \ref{thm_estimate} specifies the convergence rate for the estimated common component. 
If $m=O(p)$, the rate is the same as that shown in \cite{lam2011}, and the right-hand side of (\ref{eq::common}) is $O_p(p^{-\delta/2}\Vert\hat{\bQ}-\bQ\Vert_2+p^{-1/2})$. 
If $m=o(p)$, the right-hand side of (\ref{eq::common}) can be written as $O_p(\sqrt{\frac{m}{p}}m^{-\delta/2}\Vert \hat{\bQ}-\bQ\Vert_2+p^{-1/2})$, which is smaller than the one in \cite{lam2011}, and our method produces a better estimator than \cite{lam2011}. 

\section{Simulation Study}
\label{sec:sim}

In this section, we use four examples to illustrate the performance of our proposed method. The first three examples are used to compare our proposed approach with the method proposed by \cite{lam2011}, varimax rotation, and the method proposed by \cite{wu2026sparse}. Since these methods adopt different ways to define the loading matrix, we cannot directly evaluate the estimation error of the loading matrix. Here we report the estimation error of the loading space for a fair comparison, $\cD({\cM}(\bQ), {\cM}(\hat{\bQ}))$, which is defined in (\ref{eq:distance}).  In all examples columns of $\bQ$ are non-orthogonal.
In Section \ref{subsec:r3}, we set different values for $\delta$, $p$, $n$, and $m$ grows to infinity as fast as $p$. 
In Section \ref{subsec:m}, we allow $m$ to grow at different rates of $p$ to demonstrate the impact of $m$ on the estimation results. 
In Section \ref{subsec:sparse}, the sparsity level in each column of $\bQ$ is different and we present the estimation of $\bQ$ with our method. ``eigen'' represents the method by \cite{lam2011}, 
``SFM'' represents the method in \cite{wu2026sparse} when the sparsity level is fixed at the true, and ``sparse'' represents our proposed method with tuning parameter selected based on BIC.  For varimax rotation, to obtain sparse estimates, we set loadings with absolute values below a threshold to zero, and we consider two thresholds: 0.01 (``varmax1'') and 0.05 (``varimax2''). 

Datasets are simulated from model \eqref{eq::model112}. The nonzero elements in $\bA^s$ are simulated from a truncated standard normal distribution with absolute values bounded above by 0.1 divided by $(m/r)^{\delta/2}$. We set $r=3$, and $\mathbf{f}_t^s$ is generated from three independent AR(1) processes with AR coefficient of $0.9$ and innovation variance of 1. The diagonal elements of $\bSigma_{e,t}$ are all 1, and its off-diagonal elements are 0.5.  The number of factors is assumed to be known. For each setting, we generate 300 samples. 

\subsection{Study on $\delta$, $p$ and $n$}
\label{subsec:r3}

\begin{figure}[H]
    \centering
    \includegraphics[scale = 0.9]{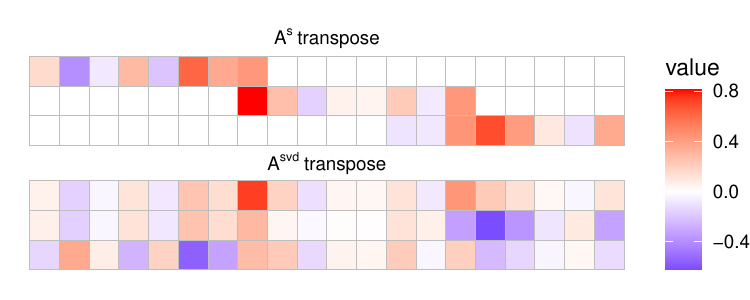}
    \caption{An example of structures of loading matrices when $p=20$.}
        \label{fig::sim_example}
\end{figure}

In this example,  we consider different combinations of $p$ and $n$, where $p = 20, 50, 100, 200, 500$ and $n = 100, 200, 500$. In $\bA^s$, the first $0.4p$ elements in the first column, and the last $0.4p$ elements in the third column are nonzero,  while all other elements in these two columns are zero.  In the second column, the first $0.3p$ elements are zero, the following $0.4p$ elements are nonzero, and the remaining $0.3p$ elements are zero. Note that the second column of $\bA^s$ contains nonzero elements that overlap with those in both the first and third columns, making the whole matrix non-orthogonal.  Figure \ref{fig::sim_example} gives an example of the simulated loading matrix when $p=20$. To illustrate the difference between our settings and the ones in \cite{wu2026sparse},  we apply singular value decomposition to $\bM$ to obtain $\bA^{\text{svd}}$, the loading matrix assumed in \cite{wu2026sparse}, which are orthogonal. We also standardize the columns of $\bA^s$ so that each column has unit $L_2$ norm. Figure \ref{fig::sim_example} shows that $\bA^s$,  the loading matrix in our settings, is very sparse, while $\bA^{\text{svd}}$ is not at all. Although many elements in $\bA^{\text{svd}}$ have small magnitude, thy are not exactly zeros.


Table \ref{tab_dist_r3} shows the average estimation error of the loading space for different methods with $\delta=0$ and $\delta=0.25$. Our proposed estimator outperforms others, especially when the sample size is small. It is interesting that when the threshold for the varimax is appropriately selected, varimax performs better than the method of \cite{lam2011}, which does not account for the sparsity of the loading matrix. However, the optimal threshold value depends on $p$, $n$, and $\delta$, so selecting the threshold value is particularly challenging when analyzing real data.



\begin{table}[H]
\centering
\caption{Mean and standard deviation (in parentheses) of the distance between the estimated loading space and the true loading space for the example in Section \ref{subsec:r3}.}
\label{tab_dist_r3}
{\scriptsize
\begin{tabular}{llllllll}
  \hline
$\delta$ & $n$ & $p$ & eigen & varimax1 & varimax2 & SFM & sparse \\ 
  \hline
  \multirow{15}{*}{0}
&100 & 20 & 0.115(0.056) & 0.115(0.056) & 0.107(0.057) & 0.346(0.213) & 0.095(0.057) \\ 
 & 100 & 50 & 0.110(0.038) & 0.109(0.038) & 0.103(0.033) & 0.242(0.180) & 0.088(0.036) \\ 
 & 100 & 100 & 0.106(0.031) & 0.103(0.032) & 0.120(0.019) & 0.182(0.125) & 0.084(0.028) \\ 
 & 100 & 200 & 0.103(0.028) & 0.093(0.030) & 0.174(0.012) & 0.152(0.084) & 0.082(0.024) \\ 
 & 100 & 500 & 0.104(0.028) & 0.085(0.028) & 0.310(0.011) & 0.144(0.071) & 0.084(0.024) \\ 
  \cline{2-8}
  &200 & 20 & 0.066(0.022) & 0.066(0.023) & 0.066(0.021) & 0.274(0.224) & 0.052(0.023) \\ 
  &200 & 50 & 0.062(0.016) & 0.060(0.017) & 0.076(0.013) & 0.127(0.114) & 0.047(0.014) \\ 
  &200 & 100 & 0.061(0.015) & 0.056(0.016) & 0.105(0.010) & 0.121(0.077) & 0.046(0.013) \\ 
  &200 & 200 & 0.061(0.015) & 0.050(0.015) & 0.164(0.010) & 0.118(0.064) & 0.046(0.012) \\ 
  &200 & 500 & 0.061(0.014) & 0.049(0.010) & 0.303(0.009) & 0.114(0.064) & 0.046(0.012) \\ 
  \cline{2-8}
   &500 & 20 & 0.035(0.010) & 0.035(0.010) & 0.047(0.013) & 0.208(0.215) & 0.026(0.010) \\ 
  &500 & 50 & 0.035(0.008) & 0.033(0.008) & 0.066(0.012) & 0.091(0.091) & 0.024(0.007) \\ 
  &500 & 100 & 0.035(0.008) & 0.032(0.008) & 0.099(0.010) & 0.081(0.056) & 0.024(0.006) \\ 
  &500 & 200 & 0.035(0.007) & 0.029(0.006) & 0.159(0.009) & 0.086(0.061) & 0.024(0.006) \\ 
  &500 & 500 & 0.035(0.007) & 0.035(0.004) & 0.301(0.009) & 0.090(0.063) & 0.024(0.006) \\
  \hline
   \multirow{15}{*}{0.25}
&100 & 20 & 0.152(0.074) & 0.152(0.074) & 0.143(0.076) & 0.362(0.203) & 0.130(0.077) \\ 
&  100 & 50 & 0.164(0.063) & 0.163(0.063) & 0.147(0.065) & 0.276(0.172) & 0.139(0.065) \\ 
&  100 & 100 & 0.170(0.053) & 0.168(0.054) & 0.159(0.046) & 0.232(0.126) & 0.145(0.053) \\ 
&  100 & 200 & 0.178(0.051) & 0.173(0.053) & 0.204(0.033) & 0.211(0.094) & 0.154(0.049) \\ 
&  100 & 500 & 0.203(0.058) & 0.191(0.062) & 0.334(0.023) & 0.225(0.086) & 0.181(0.055) \\ 
  \cline{2-8}
&200 & 20 & 0.087(0.030) & 0.087(0.030) & 0.082(0.030) & 0.282(0.218) & 0.072(0.031) \\ 
&  200 & 50 & 0.091(0.023) & 0.089(0.024) & 0.089(0.018) & 0.149(0.115) & 0.073(0.022) \\ 
&  200 & 100 & 0.097(0.024) & 0.093(0.025) & 0.117(0.014) & 0.147(0.080) & 0.078(0.021) \\ 
&  200 & 200 & 0.106(0.025) & 0.096(0.027) & 0.174(0.012) & 0.147(0.056) & 0.087(0.021) \\ 
&  200 & 500 & 0.118(0.028) & 0.098(0.029) & 0.312(0.011) & 0.153(0.055) & 0.100(0.023) \\ 
  \cline{2-8}
 & 500 & 20 & 0.046(0.013) & 0.046(0.013) & 0.053(0.014) & 0.211(0.212) & 0.036(0.014) \\ 
 & 500 & 50 & 0.051(0.012) & 0.049(0.012) & 0.071(0.012) & 0.100(0.089) & 0.038(0.011) \\ 
 & 500 & 100 & 0.056(0.013) & 0.050(0.013) & 0.104(0.011) & 0.095(0.055) & 0.042(0.011) \\ 
 & 500 & 200 & 0.060(0.013) & 0.049(0.013) & 0.163(0.009) & 0.102(0.055) & 0.046(0.011) \\ 
 & 500 & 500 & 0.067(0.014) & 0.053(0.010) & 0.304(0.009) & 0.110(0.056) & 0.052(0.011) \\ 
   \hline
\end{tabular}
}
\end{table}

Table \ref{tab_nonzero_r3} shows the average value of $|m-\hat{m}|$, with the standard deviation given in parentheses, where $\hat{m}$ is the number of nonzero elements in the estimated loading matrix. 
Overall, our proposed method effectively recovers the sparsity of the loading matrix. The performance of ``sparse'' is much better than other methods. Although varimax rotation may be able to estimate the loading space accurately when the threshold is appropriately selected, it cannot recover the sparsity as effectively as our proposed method.  

Table \ref{tab_dist_r3} and Table \ref{tab_nonzero_r3} show that as $n$ grows, the loading space and zero elements are estimated more accurately. As $p$ increases, both estimation errors also increase unless the factors are strong with $\delta=0$. These observations align with our conclusions in Theorem \ref{thm_general}.

\begin{table}[H]
\centering
\caption{Mean and standard deviation (in parentheses) of $|m-\hat{m}|$ for the example in Section \ref{subsec:r3}}
\label{tab_nonzero_r3}
{\scriptsize
\begin{tabular}{ll|llll|llll}
  \hline
& & \multicolumn{4}{c|}{$\delta=0$} & \multicolumn{4}{c}{$\delta=0.25$}\\
\hline
$n$ & $p$ & varimax1 & varimax2 & SFM & sparse  & varimax1 & varimax2 & SFM & sparse\\
  \hline
100 & 20 & 26.1(3.5) & 4.4(3.2) & 27.9(7.0) & 2.6(3.1) & 27.7(3.2) & 6.1(3.9) & 27.7(6.9) & 3.4(3.8) \\ 
  100 & 50 & 49.6(9.1) & 5.3(2.9) & 70.5(17.5) & 5.0(6.9) & 58.8(8.1) & 5.2(4.9) & 69.6(17.6) & 7.0(8.3) \\ 
  100 & 100 & 65.8(20.1) & 22.2(4.1) & 143.3(37.0) & 7.8(12.3) & 96.0(18.9) & 20.5(5.3) & 141.7(37.3) & 12.9(19.3) \\ 
  100 & 200 & 63.3(34.9) & 72.8(6.1) & 282.8(73.0) & 10.7(14.8) & 140.2(39.4) & 72.4(7.4) & 279.1(72.4) & 21.5(28.6) \\ 
  100 & 500 & 31.0(40.7) & 300.1(9.3) & 731.1(183.6) & 19.3(29.9) & 195.3(103.0) & 302.6(9.3) & 718.6(184.9) & 64.2(70.9) \\ 
  \hline
  200 & 20 & 22.2(4.5) & 2.6(2.5) & 29.1(7.1) & 2.3(3.5) & 24.2(4.1) & 3.2(2.8) & 29.1(7.0) & 3.0(3.9) \\ 
  200 & 50 & 33.9(10.6) & 6.2(2.6) & 75.7(15.5) & 2.2(3.3) & 44.5(9.9) & 5.8(2.6) & 75.4(15.3) & 4.0(6.0) \\ 
  200 & 100 & 34.9(17.3) & 22.6(3.8) & 144.8(35.2) & 3.2(5.4) & 62.1(19.7) & 22.6(3.9) & 143.9(35.2) & 5.9(10.0) \\ 
  200 & 200 & 18.0(16.8) & 72.4(6.4) & 284.7(73.0) & 5.6(9.2) & 68.2(34.9) & 72.6(6.3) & 282.8(72.6) & 11.4(20.2) \\ 
  200 & 500 & 23.9(6.0) & 298.6(9.7) & 723.1(185.4) & 9.6(13.8) & 41.5(49.8) & 299.6(9.7) & 716.8(184.5) & 26.8(31.0) \\ 
  \hline
    500 & 20 & 18.2(5.3) & 2.0(2.0) & 30.5(6.3) & 1.4(2.9) & 20.2(4.9) & 2.2(2.1) & 30.4(6.3) & 2.1(3.7) \\ 
  500 & 50 & 22.0(9.4) & 6.1(2.5) & 77.4(15.1) & 1.2(3.6) & 29.5(9.5) & 6.0(2.6) & 77.3(15.1) & 2.5(5.3) \\ 
  500 & 100 & 17.1(12.4) & 23.0(4.0) & 152.3(33.7) & 1.4(3.8) & 29.5(14.8) & 23.0(4.1) & 152.0(33.7) & 3.3(7.9) \\ 
  500 & 200 & 6.1(8.3) & 72.1(6.3) & 301.8(67.9) & 2.1(5.2) & 16.6(15.7) & 72.3(6.5) & 300.7(67.9) & 5.1(13.4) \\ 
  500 & 500 & 24.6(4.5) & 299.0(9.4) & 745.5(175.8) & 5.2(13.0) & 23.6(6.1) & 299.5(9.5) & 743.1(174.7) & 7.9(18.4) \\ 
   \hline
\end{tabular}
}
\end{table}

\subsection{Study on $m$}
\label{subsec:m}

In this section, we evaluate the impact of $m$ (the total number of nonzero loadings in $\bA^s$) on the estimation errors and let $m$ grow to infinity at different rates relative to $p$. We consider four different growth rates with $m= 3p^{1/2}, 3p^{2/3}, 3p^{3/4}, 1.8p$ and set $p=50,100,200,500$,  $\delta=0, 0.25$ and $n=1000$.  In $\bA^s$, each column has $m/3$ nonzero elements. Specifically, the first $m/3$ elements in its first column, the last $m/3$ elements in the third column are nonzero, while all other elements are zero. In the second column, the first $p/3$ elements are zero, the following $m/3$ elements are nonzero, and the left elements are also zero.
Table \ref{tab_dist_m} and Table \ref{tab_nonzero_m} show the average and standard deviation of the estimation errors for the loading space and the sparsity level, respectively. 
When $p$ is fixed, as $m$ increases,  the estimation error first decreases and then increases, which aligns with our conclusion in Theorem \ref{thm_general}. When the loading matrix is sparse, the estimation error is dominated by the bias; when the loading matrix is dense, it is dominated by the variance. 

\begin{table}[H]
\centering
\caption{Mean and standard deviation (in parentheses) of the distance between the estimated loading space and the true loading space for the example in Section \ref{subsec:m}.}
\label{tab_dist_m}
\begin{tabular}{llll}
  \hline
$m$ & $p$ & $\delta = 0$ & $\delta = 0.25$ \\ 
  \hline
$3p^{1/2}$ & 50 & 0.019(0.010) & 0.029(0.017) \\ 
  $3p^{2/3}$ & 50 & 0.016(0.006) & 0.025(0.009) \\ 
  $3p^{3/4}$ & 50 & 0.016(0.005) & 0.025(0.008) \\ 
  $1.8p$ & 50 & 0.017(0.004) & 0.027(0.006) \\ 
  \hline
  $3p^{1/2}$ & 100 & 0.018(0.013) & 0.028(0.023) \\ 
  $3p^{2/3}$ & 100 & 0.016(0.004) & 0.025(0.007) \\ 
  $3p^{3/4}$ & 100 & 0.016(0.004) & 0.027(0.008) \\ 
  $1.8p$ & 100 & 0.016(0.004) & 0.029(0.007) \\ 
  \hline
  $3p^{1/2}$ & 200 & 0.018(0.013) & 0.032(0.027) \\ 
  $3p^{2/3}$ & 200 & 0.015(0.003) & 0.026(0.007) \\ 
  $3p^{3/4}$ & 200 & 0.016(0.004) & 0.029(0.008) \\ 
  $1.8p$ & 200 & 0.016(0.003) & 0.031(0.006) \\ 
  \hline
  $3p^{1/2}$ & 500 & 0.018(0.012) & 0.042(0.044) \\ 
  $3p^{2/3}$ & 500 & 0.016(0.003) & 0.031(0.011) \\ 
  $3p^{3/4}$ & 500 & 0.016(0.003) & 0.032(0.008) \\ 
  $1.8p$ & 500 & 0.016(0.003) & 0.035(0.007) \\ 
  \hline
\end{tabular}
\end{table}

\begin{table}[H]
\centering
\caption{Mean and standard deviation (in parentheses) of $\vert m - \hat{m}\vert $ for the example in Section \ref{subsec:m}.}
\label{tab_nonzero_m}
\begin{tabular}{llrll}
  \hline
$m$ & $p$ & $m/3$ & $\delta = 0$ & $\delta = 0.25$ \\ 
  \hline
$3p^{1/2}$ & 50 & 7 & 3.3(13.1) & 5.8(16.7) \\ 
  $3p^{2/3}$ & 50 & 14 & 1.5(7.5) & 3.6(11.5) \\ 
  $3p^{3/4}$ & 50 & 19 & 0.8(4.6) & 1.8(6.1) \\ 
  $1.8p$ & 50 & 30 & 2.2(3.4) & 2.6(4.1) \\ 
  \hline
  $3p^{1/2}$ & 100 & 10 & 2.9(18.7) & 10.1(34.1) \\ 
  $3p^{2/3}$ & 100 & 22 & 0.9(8.4) & 2.1(11.1) \\ 
  $3p^{3/4}$ & 100 & 32 & 0.8(6.2) & 3.9(15.4) \\ 
  $1.8p$ & 100 & 60 & 5.0(4.6) & 4.2(5.1) \\ 
  \hline
  $3p^{1/2}$ & 200 & 14 & 5.0(34.7) & 25.2(82.4) \\ 
  $3p^{2/3}$ & 200 & 34 & 0.2(0.6) & 1.7(15.3) \\ 
  $3p^{3/4}$ & 200 & 53 & 0.4(0.9) & 3.4(19.0) \\ 
  $1.8p$ & 200 & 120 & 8.3(7.2) & 6.7(8.5) \\ 
  \hline
  $3p^{1/2}$ & 500 & 22 & 11.3(85.4) & 94.5(245.2) \\ 
  $3p^{2/3}$ & 500 & 63 & 0.5(0.9) & 11.5(76.2) \\ 
  $3p^{3/4}$ & 500 & 106 & 0.7(1.4) & 6.9(41.4) \\ 
  $1.8p$ & 500 & 300 & 13.6(18.0) & 12.9(22.7) \\ 
   \hline
\end{tabular}
\end{table}

\subsection{Study on sparsity}
\label{subsec:sparse}

In this section, we use an example to illustrate the performance of our method on sparsity estimation of the loading matrix.  In $\bA^s$, the first $0.4p$ elements in the first column, the middle $0.5p$ elements in the second column, and the last $0.6p$ elements in the third column are nonzero,  while all other elements are zero. Note that the second column of $\bA^s$ contains nonzero elements that overlap with those in both the first and third columns. We consider $\delta=0$, $n=200, 500$ and $p = 20, 100$ and $500$. Let $m_i$ represent the true number of nonzero elements in ${\bf q}_i$ and $\hat{m}_i$ represent the estimated number of nonzero elements in $\hat{\bq}_i$. In order to evaluate the accuracy of sparsity estimation, we also report false negative value (FN, the number of elements falsely identified as zero), false positive (FP, the number of elements falsely identify as nonzero) and F1 score, a number between 0 and 1, which measures the classification accuracy (nonzero or zero). The higher the value of F1 score is, the better the identification is. Table \ref{tab_col} shows the summary results of different measures about identifying non/zero elements for each loading vector.  Overall, the proposed approach can identify non/zero elements for each loading vector well. In particular, as sample size increases, the accuracy gets better. 

\begin{table}[ht]
\centering
\caption{Mean and standard deviation (parentheses) of measures of identifying non/zero elements for each loading vector}
\label{tab_col}
{\scriptsize
\begin{tabular}{rllllll}
  \hline
loadings & $n$ & $p$ & $\vert \hat{m}_i - m_i\vert$ & FN & FP & F1 \\ 
  \hline
 \multirow{5}{*}{1} & 200 & 20 & 0.993(1.567) & 0.130(0.356) & 1.010(1.663) & 0.940(0.086) \\ 
     & 200 & 100 & 2.943(5.572) & 0.317(0.657) & 3.053(5.775) & 0.963(0.056) \\ 
     & 200 & 500 & 10.498(14.599) & 1.813(2.099) & 12.057(15.458) & 0.967(0.035) \\ 
      \cline{2-7}
     & 500 & 20 & 0.657(1.479) & 0.020(0.140) & 0.663(1.518) & 0.965(0.074) \\ 
     & 500 & 100 & 0.987(2.377) & 0.040(0.228) & 0.993(2.418) & 0.988(0.027) \\ 
     & 500 & 500 & 4.447(6.292) & 0.097(0.478) & 4.537(6.562) & 0.989(0.016) \\ 
    \hline
     \multirow{5}{*}{2} & 200 & 20 & 0.943(1.424) & 0.410(0.724) & 1.100(1.787) & 0.932(0.096) \\ 
     & 200 & 100 & 1.627(3.353) & 0.617(0.966) & 1.590(3.669) & 0.979(0.034) \\ 
     & 200 & 500 & 5.391(10.350) & 3.174(2.972) & 6.311(12.222) & 0.982(0.025) \\ 
      \cline{2-7}
     & 500 & 20 & 0.693(1.368) & 0.183(0.459) & 0.823(1.668) & 0.956(0.085) \\ 
     & 500 & 100 & 0.457(1.067) & 0.050(0.233) & 0.460(1.140) & 0.995(0.012) \\ 
     & 500 & 500 & 1.540(2.415) & 0.153(0.480) & 1.593(2.607) & 0.997(0.006) \\ 
    \hline
     \multirow{5}{*}{3} & 200 & 20 & 0.763(0.965) & 0.787(0.851) & 0.923(1.292) & 0.930(0.072) \\ 
     & 200 & 100 & 1.340(1.900) & 1.017(1.158) & 0.937(2.266) & 0.984(0.022) \\ 
     & 200 & 500 & 3.926(3.520) & 4.739(3.564) & 2.753(4.971) & 0.988(0.012) \\ 
      \cline{2-7}
     & 500 & 20 & 0.540(0.823) & 0.477(0.738) & 0.830(1.254) & 0.948(0.073) \\ 
     & 500 & 100 & 0.527(1.747) & 0.190(0.650) & 0.557(2.167) & 0.994(0.020) \\ 
     & 500 & 500 & 0.837(1.875) & 0.260(0.638) & 0.877(2.116) & 0.998(0.004) \\ 
   \hline
\end{tabular}
}
\end{table}

\section{Real data analysis}
\label{sec:example}

We apply the proposed algorithm to the Hawaii tourism data and present the results to demonstrate how our proposed method enhances the model interpretability.
Tourism is the largest single source of Hawaii’s GDP. Therefore, a deep understanding of its dynamics is crucial for the local industry and economy \citep{liu2015}.

We downloaded the data from the official website of Hawaii's government (\url{https://dbedt.hawaii.gov/visitor/}). The dataset contains the number of domestic visitors to Hawaii from Washington, D.C., and all 49 U.S. states with $p=50$. The data are recorded monthly from January 2009 to December 2019 with $n=132$.  To stabilize the variance and remove the increasing trend, we first took the logarithm transformation and then took a difference to pre-process the original data. Set $h_0=1$. We use the sequential test proposed in \cite{trapani2018randomized} to estimate the number of factors and it yields $\hat{r}=1$. To better compare our methods and others, we present the results with two factors and $\hat{r}=2$.

First, we use one-step-ahead rolling forecasting to compare different models, including the standard method \citep{lam2011}, the method in \cite{wu2026sparse} with the tuning parameter selected based on cross-validation (``SFM (CV)") and based on BIC (``SFM (BIC)"), and our proposed method. In particular, we use VAR(1) to predict the factor processes. 
Table \ref{tbl::hawaii} reports the mean squared error (MSE) of one-step-ahead rolling forecasting of Hawaii data based on {the predictions of the last 30 observations} and the number of zeros in the estimated loading matrix based on the full data set. It can be seen that our proposed method produces the most sparse loading matrices and the lowest prediction errors.

\begin{table}[H]
\caption{Hawaii tourism: Comparison of different methods}\label{tbl::hawaii}
\begin{center}
\begin{tabular}{l|c c}
&  MSE      &\#of 0\\ \hline
Standard method &1.020  &0\\
SFM (CV) & 1.109 & 0\\
SFM (BIC) & 1.135 & 16\\
Our method              &1.016  &41\\
\hline
  \end{tabular}
\end{center}
\end{table}

\begin{figure}[H]
    \centering
    \includegraphics[width=0.75\linewidth]{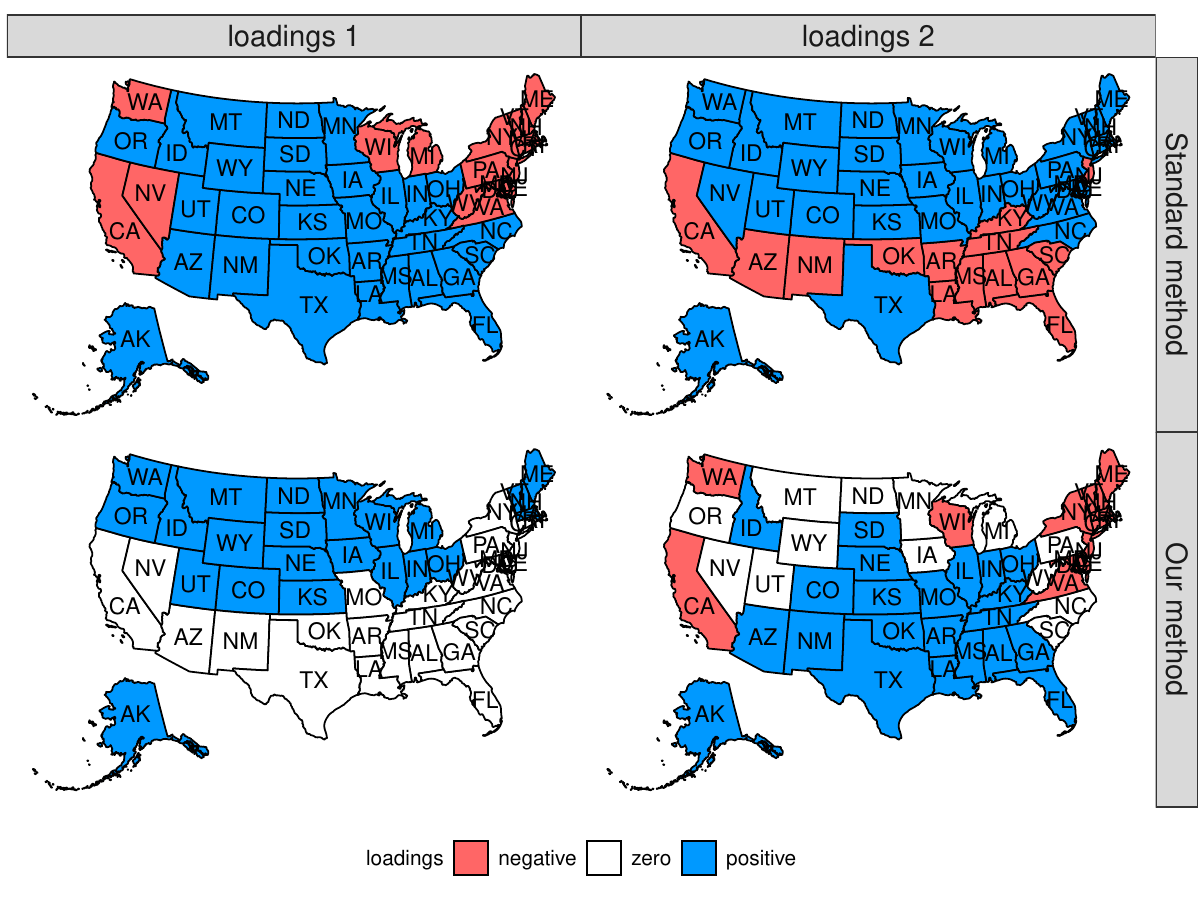}
    \caption{Estimated loadings for Hawaii tourism data. }
    \label{fig:sign}
\end{figure}

Figure \ref{fig:sign} plots the estimated loadings for two factors obtained from our method, with loadings for factor 1 in the left panels and loadings for factor 2 in the right panels. In the plot, positive loadings are in blue, negative loadings in red, and zero loadings in white. The top panels show the estimated loadings by a standard method with no sparsity assumption in \cite{lam2011}. None of loadings is zero, and factors are linear combination of all series, making the factors difficult to interpret. 
The bottom panels show the estimated loadings of our method with 41 zero loadings. 
We can clearly see the patterns of the factors from the sparse loading matrix. The states that load positively on Factor 1 are all in high latitude. For Factor 2, most of the states that load positively locate inland or are in relatively low latitude, while the states that load negatively have ocean or Great Lakes coastlines. It implies that the tourism of Hawaii may be driven by two groups of visitors --- people who would like to escape the cold and who would like to enjoy the beach and water activities. 



\begin{figure}[H]
    \centering
    \includegraphics[width=0.55\linewidth]{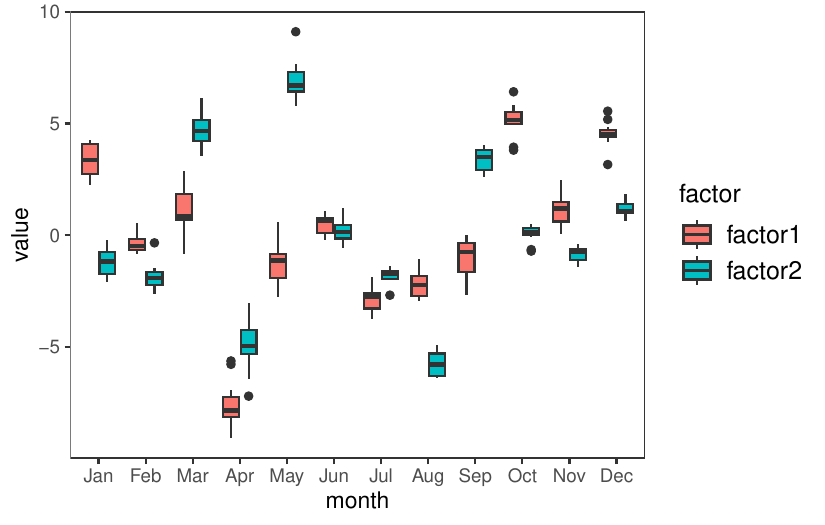}
    \caption{Boxplots of estimated factors for different months for Hawaii tourism data.}
    \label{fig:box}
\end{figure}

To confirm our interpretation of factors, we make boxplots of the estimated factors for different months, as shown in Figure \ref{fig:box}. The scales of the two factors are very close, making them comparable. Their means are 0.0253 and -0.0252, with standard deviations of 3.422 and 3.603, respectively. Factor 1 is much larger from October to February and in August, corresponding to fall and winter months when the temperature is low. Factor 2 is notably larger in March-May, July, and September, suggesting that these visitors prefer to avoid the rainy season in Hawaii (October-April). These findings support our interpretation of the factors.





The results show that our proposed method captures the dynamics of Hawaii tourism data and improves the prediction accuracy. Moreover, our method yields the most sparse estimate and thus enhances the model interpretability.

\section{Conclusion}
\label{sec:conclusion}

In this paper, we introduce a new approach to redefining the factor models for high-dimensional time series with sparse loadings and develop an algorithm to obtain a regularized estimator for the loading matrix. We study the theoretical properties of our proposed estimators and validate the algorithm's performance using both simulated and real data. The regularized estimator with zero loadings improves the model interpretability and enhances the predictive performance of the model. Compared to the existing methods, our proposed approach offers the following advantages: (1) The proposed enjoys the oracle property while existing methods with $L_1$-norm penalty suffers from estimation bias and model inconsistency; (2) unlike the existing methods which select a specific rotation of the loading matrix first and then impose the sparsity assumption, we do not impose an orthogonality constraint on the loading matrix or assume that factors are independent. Relaxing these assumptions results in a more sparse estimate, leads to a more parsimonious model, and thus further improves model interpretation and predictive accuracy.


\begin{center}
{\large\bf SUPPLEMENTARY MATERIAL}
\end{center}

This supplementary material includes detailed proofs for Theorems 1, 2 and 3 (Section \ref{appendix_thm}), lemmas and their proofs (Section \ref{appendix_lemma}, \ref{appendix_lemma_tech}) and some existing definitions and results (Section \ref{appendix_existing}).

\renewcommand{\thesection}{\Alph{section}}

\addtocounter{section}{-7}
\section{Proof of Theorems}
\label{appendix_thm}

We also include our conditions here. The regularity conditions we need are listed below.
\begin{enumerate}
\renewcommand{\labelenumi}{\textbf{(C\arabic{enumi})}}

\item  Let $\calF_i^j$ be the $\sigma$-field generated by $\{\bff_t^s: i \leq t \leq j\}$.
The joint process $\{\bff_t^s\}$ is
$\alpha$-mixing with mixing coefficients satisfying
\[
\sum_{t=1}^{\infty} \alpha(t)^{1-2/\gamma}<\infty,
\]
for some $\gamma>2$, where
$\alpha(t)=\sup_{i} \sup_{A \in \calF_{-\infty}^i, B \in \calF_{i+t}^{\infty}} |P(A \cap B) -P(A)P(B)|$.  \label{cond_alphamix}

\item For any $i=1,\ldots, r$, $t=1, \ldots, n$, $E(|f_{t,j}^s|^{2\gamma})< \sigma_f^{2\gamma}$, where $f_{t,i}^s$ is the $i$-th element of $\bff_t^s$,  $\sigma_f>0$ is a constant, and $\gamma$ is given in Condition (C\ref{cond_alphamix}).  \label{cond_fbound}

\item  $\bve_{t}$ and $\bff_t^s$ are uncorrelated given ${\cal F}_{-\infty}^{t-1}$. Each element of $\bSigma_{e,t}$ remains bounded by a positive constant $\sigma_{\epsilon}^2$ as $p$ increases to infinity, for $i=1,2$, and $t=1,\ldots, n$, where $\bSigma_{e,t}$ is the covariance of $\bve_t$.\label{cond_cov}


\item  There exists a constant $\delta \in [0,1]$ such that $\|\bA^s\|_2^2  \asymp \|\bA^s\|^2_{\min}\asymp m^{1-\delta}$, as $p$ goes to infinity, where $m$ is the number of nonzero elements in $\bA^s$. Furthermore, $\Vert \bA^s\Vert_{\max} \leq C_1$, where $C_1$ is a positive constant.  \label{cond_strength}

\item  $\bM$ has $r$ distinct nonzero eigenvalues. \label{cond_eigenM}

\item $\bve_t$'s are independent sub-Gaussian random vectors. Each random vector in the sequences $\bff_t^s$ follows a sub-Gaussian distribution. \label{cond_subgaussian}

\item $\Vert\mathbf{S}_{i,1}\Vert_{\min}
\asymp 1$. \label{cond_si1_eigen} 

\item  There exists a positive constant $C_{\mu} > 1$ such that $\Vert{\bf S}\Vert_{2\rightarrow\infty}\leq C_{\mu}\sqrt{\frac{r}{m}}$. \label{cond_coherence} 

\end{enumerate}

\subsection{Proof of Theorem 1}

\begin{proof}
Following the proof of Theorem 1 in \cite{lam2011} with the results in Lemma \ref{lemma::M}, we have
\[
\Vert \hat{\bS} -\bS\Vert_2 = O_p(\| \hat{\bM}- \bM\|_2/\|\bM\|_{\min})  = O_p(m^{\delta-1}p n^{-1/2}).
\]
With Theorem 2.3 in \cite{chang2015}, we reach the conclusion. 
\end{proof}

\subsection{Proof of Theorem 2}

\begin{proof}

 Let $\tau_{n,p,m} = \max\left(m^{2\delta-2}p^{2}n^{-1/2},m^{\delta}\right)\sqrt{\frac{\log p}{n}} $ when $m = o(p)$ and $\tau_{n,p,m} = p^\delta n^{-1/2}$ when $m=O(p)$. We will prove the results in two parts. In part 1, we will prove the results for $\bq_1$, and prove the results for $\bq_i$ for $i=2,\dots, r$ in part 2. 

\paragraph{Part 1} We prove the results for $\bq_1$ in two steps. In step 1, we study the property of the oracle estimator. Then, we show that the oracle estimator is the local minimizer.    

{\bf Step 1: } When the sparsity, $\mathcal{V}_1$, is known, the oracle estimator is defined as 
\begin{align*}
\hat{{\bf q}}_{1}^{or} & =\arg\min_{{\bf q}_{1}}\Vert\hat{{\bf S}}\hat{{\bf S}}^{\top}-{\bf q}_{1}{\bf q}_{1}^{\top}\Vert_{F}^{2}\\
 & \text{subject to }{\bf q}_{1[\mathcal{N}_{1}]}={\bf 0} \text{ and } \Vert {\bq_1} \Vert_2 =1 .
\end{align*}

For simplicity, denote ${\bf q}_{1}^{*}={\bf q}_{1[\mathcal{V}_{1}]}$, ${\bf S}^{*}={\bf S}_{[\mathcal{V}_{1}]}=\left({\bf s}_{1[\mathcal{V}_{1}]},\dots,{\bf s}_{r[\mathcal{V}_{1}]}\right)$
and $\hat{{\bf S}}^{*}=\hat{{\bf S}}_{[\mathcal{V}_{1}]}=\left(\hat{{\bf s}}_{1[\mathcal{V}_{1}]},\dots,\hat{{\bf s}}_{r[\mathcal{V}_{1}]}\right)$.
Note that ${\bf q}_{1[\mathcal{V}_{1}]}={\bf s}_{1[\mathcal{V}_{1}]}$.
The problem above is equivalent to 
\[
\hat{{\bf q}}_{1}^{*}=\arg\min_{{\bf q}_{1}^{*}}\Vert\hat{{\bf S}}^{*}\hat{{\bf S}}^{*\top}-{\bf q}_{1}^{*}{\bf q}_{1}^{*\top}\Vert_{F}^{2} \text{  subject to } \Vert \bq_1^* \Vert_2 =1. 
\]

From Lemma \ref{lemma_obj}, we know that this is equivalent to maximizing $\bq_1^{*\top}\hat{{\bf S}}^{*}\hat{{\bf S}}^{*\top} \bq_1^*$ with respect to $\Vert \bq_1^*\Vert_2 = 1$.  This corresponds to the maximal eigenvector of $\hat{{\bf S}}^{*}\hat{{\bf S}}^{*\top}$.

Let $\hat{{\bf L}}=\hat{{\bf S}}^{*}\hat{{\bf S}}^{*\top}$ and ${\bf L}={\bf S}^{*}{\bf S}^{*\top}$. It is known that $\bL = {\bf S}^{*}{\bf S}^{*\top}{\bf q}_{1}^{*}={\bf q}_{1}^{*}$ based on the definition of ${\bf S}^{*}$. Thus, this allows us to apply Lemma \ref{lemma_eigenvalue} and Lemma \ref{lemma_Sstar}, which concludes that $\Vert\hat{{\bf q}}_{1}^{*}-{\bf q}_{1}^{*}\Vert_2=O_p\left( \tau_{n,p,m}\right)$.  This implies that 
$\Vert\hat{{\bf q}}_{1}^{or}-{\bf q}_{1}\Vert_2=O_p\left(\tau_{n,p,m}\right)$.

{\bf Step 2: } The next step is to show that $\hat{{\bf q}}_{1}^{or}$ is a local minimizer
of $G\left({\bf q}_{1}\right)=\frac{1}{2}\Vert\hat{{\bf S}}\hat{{\bf S}}^{\top}-{\bf q}_{1}{\bf q}_{1}^{\top}\Vert_{F}^{2}+\sum_{j=1}^{p}\mathcal{P}_{\gamma}\left(\vert q_{1j}\vert;\lambda\right)$
subject to $\Vert{\bf q}_{1}\Vert_2=1$.

Consider a neighbor of ${\bf q}_{1}$ such that $\Vert{\bf u}-{\bf q}_{1}\Vert_2=O_{p}\left(\tau_{n,p,m}\right)$ and $\Vert{\bf u}\Vert_2=1$.  Define ${\bf u}^{*}[\mathcal{V}_{1}]={\bf u}_{[\mathcal{V}_{1}]}$
and ${\bf u}^{*}[-\mathcal{V}_{1}]=\bzero$, and $\alpha=\Vert{\bf u}^{*}\Vert_2$.
Let $\tilde{{\bf u}}={\bf u}^{*}/\alpha$ , which indicates that $\tilde{{\bf u}}[-\mathcal{V}_{1}]={\bf 0}$ and $\Vert \tilde{\bu}\Vert_2 =1$ based on the definition of $\tilde{\bu}$.

First we will compare $G\left(\hat{{\bf q}}_{1}^{or}\right)$ and
$G\left(\tilde{{\bf u}}\right)$.

We have $\Vert{\bf u}^{*}-{\bf q}_{1}\Vert_2=O_{p}\left(\tau_{n,p,m}\right)$,
$\Vert{\bf u}_{[-\mathcal{V}_{1}]}\Vert_2=\Vert{\bf u}-{\bf u}^{*}\Vert_2=O_{p}\left(\tau_{n,p,m}\right)$,
and $\alpha=\Vert{\bf u}^{*}\Vert_2\geq\Vert{\bf q}_{1}\Vert_2 - \Vert{\bf u}^{*}-{\bf q}_{1}\Vert_2=1-\Vert{\bf u}^{*}-{\bf q}_{1}\Vert_2$ with $\alpha <1$. We can have 
\begin{align*}
\tilde{{\bf u}}-{\bf q}_{1} & =\frac{{\bf u}^{*}}{\alpha}-{\bf q}_{1}=\frac{{\bf u}-\left({\bf u}-{\bf u}^{*}\right)}{\alpha}-{\bf q}_{1} =\frac{{\bf u-{\bf q}_{1}-\left({\bf u}-{\bf u}^{*}\right)}}{\alpha}+\left(\frac{1}{\alpha}-1\right){\bf q}_{1}.
\end{align*}
Thus 
\begin{align*}
\Vert\tilde{{\bf u}}-{\bf q}_{1}\Vert_2 & \leq\frac{1}{\alpha}\Vert{\bf u}-{\bf q}_{1}\Vert_2+\frac{1}{\alpha}\Vert{\bf u}-{\bf u}^{*}\Vert_2+\frac{1}{\alpha}-1\\
&= \frac{\Vert{\bf u}-{\bf q}_{1}\Vert_2+\Vert{\bf u}-{\bf u}^{*}\Vert_2+\Vert{\bf u}-{\bf q}_{1}\Vert_2}{1-\Vert \bu^* - \bq_1 \Vert_2 } = O_p(\tau_{n,p,m}).
\end{align*}

Based on the assumption about the minimal signal and the assumption about $\lambda$, for $j\in\mathcal{V}_{1}$, we have
$\vert\hat{q}_{1j}^{or}\vert\geq\vert q_{1j}\vert-\vert q_{1j}-\hat{q}_{1j}^{or}\vert>\gamma\lambda$
since $\vert q_{1j}-\hat{q}_{1j}^{or}\vert=O_{p}\left(\tau_{n,p,m}\right)$
from the result in step 1. Similarly we have $\vert\tilde{u}_{j}\vert>\gamma\lambda$. Then, $\mathcal{P}_{\gamma}\left(\vert\hat{q}_{1j}^{or}\vert;\lambda\right)=\mathcal{P}_{\gamma}\left(\vert\tilde{u}_{j}\vert;\lambda\right)=\frac{1}{2}\gamma\lambda^{2}$ based on the definition MCP.
So we have $\sum_{j=1}^{p}\mathcal{P}_{\gamma}\left(\vert\hat{q}_{1j}^{or}\vert;\lambda\right)=\sum_{j=1}^{p}\mathcal{P}_{\gamma}\left(\vert\tilde{u}_{j}\vert;\lambda\right)$.

Based on the definition of $\hat{{\bf q}}_{1}^{or}$, we have $\Vert\hat{{\bf S}}\hat{{\bf S}}^{\top}-\hat{{\bf q}}_{1}^{or}\left(\hat{{\bf q}}_{1}^{or}\right)^{\top}\Vert_{F}^{2}<\Vert\hat{{\bf S}}\hat{{\bf S}}^{\top}-\tilde{{\bf u}}\tilde{{\bf u}}^{\top}\Vert_{F}^{2}$
for $\tilde{{\bf u}}\neq\hat{{\bf q}}_{1}^{or}$. This implies that
$G\left(\hat{{\bf q}}_{1}^{or}\right)<G\left(\tilde{{\bf u}}\right)$.

Next, we will compare $G\left(\tilde{{\bf u}}\right)$ and $G\left({\bf u}\right)$.
We have that 
\begin{align}
\label{eq_gdiff}
G\left(\tilde{{\bf u}}\right)-G\left({\bf u}\right) & =-\tilde{{\bf u}}^{\top}\hat{{\bf S}}\hat{{\bf S}}^{\top}\tilde{{\bf u}}+{\bf u}^{\top}\hat{{\bf S}}\hat{{\bf S}}^{\top}{\bf u}+\sum_{j=1}^{p}\mathcal{P}_{\gamma}\left(\vert\tilde{u}_{j}\vert;\lambda\right)-\sum_{j=1}^{p}\mathcal{P}_{\gamma}\left(\vert u_{j}\vert;\lambda\right).
\end{align}

Let's consider the first two components in \eqref{eq_gdiff}.  Since $-\tilde{{\bf u}}^{\top}\hat{{\bf S}}\hat{{\bf S}}^{\top}\tilde{{\bf u}}=-\frac{1}{\alpha^{2}}{\bf u}^{*^{\top}}\hat{{\bf S}}\hat{{\bf S}}^{\top}{\bf u}^{*}\leq-{\bf u}^{*^{\top}}\hat{{\bf S}}\hat{{\bf S}}^{\top}{\bf u}^{*}$. Thus, 
\begin{align}
\label{eq_part1}
-\tilde{{\bf u}}^{\top}\hat{{\bf S}}\hat{{\bf S}}^{\top}\tilde{{\bf u}}+{\bf u}^{\top}\hat{{\bf S}}\hat{{\bf S}}^{\top}{\bf u} & \leq-{\bf u}^{*^{\top}}\hat{{\bf S}}\hat{{\bf S}}^{\top}{\bf u}^{*}+{\bf u}^{\top}\hat{{\bf S}}\hat{{\bf S}}^{\top}{\bf u} \nonumber \\
 & = {\bf u}^{*\top}\hat{{\bf S}}\hat{{\bf S}}^{\top}\left({\bf u}-{\bf u}^{*}\right)+\left({\bf u}-{\bf u}^{*}\right)^{\top}\hat{{\bf S}}\hat{{\bf S}}^{\top}{\bf {\bf u}} \nonumber\\
 & \leq\Vert{\bf u}^{*\top}\hat{{\bf S}}\hat{{\bf S}}^{\top}\left({\bf u}-{\bf u}^{*}\right)\Vert_2+\Vert\left({\bf u}-{\bf u}^{*}\right)^{\top}\hat{{\bf S}}\hat{{\bf S}}^{\top}{\bf u}\Vert_2.
\end{align}
Denote ${\bf E}=\hat{{\bf S}}\hat{{\bf S}}^{\top}-{\bf S}{\bf S}^{\top}$, we have
\begin{align*}
\Vert{\bf u}^{*\top}\hat{{\bf S}}\hat{{\bf S}}^{\top}\left({\bf u}-{\bf u}^{*}\right)\Vert_2 & =\Vert{\bf u}^{*\top}\left({\bf S}{\bf S}^{\top}+{\bf E}\right)\left({\bf u}-{\bf u}^{*}\right)\Vert_2\\
 & \leq\Vert{\bf u}^{*\top}{\bf S}{\bf S}^{\top}\left({\bf u}-{\bf u}^{*}\right)\Vert_2+\Vert{\bf u}^{*\top}{\bf E}\left({\bf u}-{\bf u}^{*}\right)\Vert_2.
\end{align*}

Let ${\bf u}^{*}={\bf u}^*- {\bf q}_1 + {\bf q}_1={\bf q}_{1}+{\bf e}^*$, where  ${\bf e}^*={\bf u}-{\bf q}_{1}$, thus, the first part can be bounded by 
\begin{align*}
\Vert{\bf u}^{*\top}{\bf S}{\bf S}^{\top}\left({\bf u}-{\bf u}^{*}\right)\Vert_2 & \leq\Vert{\bf q}_{1}^{\top}{\bf S}{\bf S}^{\top}\left({\bf u}-{\bf u}^{*}\right)\Vert_2+\Vert{\bf e}^{*\top}{\bf S}{\bf S}^{\top}\left({\bf u}-{\bf u}^{*}\right)\Vert_2\\
 & \leq0+O_{p}\left(\tau_{n,p,m} \right)\Vert{\bf u}-{\bf u}^{*}\Vert_2. 
\end{align*}

Furthermore $\Vert{\bf u}^{*\top}{\bf E}\left({\bf u}-{\bf u}^{*}\right)\Vert_2= \Vert{\bf u}^{*\top}{\bf E}\Vert_{\max} \sum_{j\notin \mathcal{V}_1}\vert u_j\vert$. From Lemma \ref{lemma_uE}, $\Vert{\bf u}^{*\top}{\bf E}\left({\bf u}-{\bf u}^{*}\right)\Vert_2= O_p(\tau_{n,p,m}) \sum_{j\notin \mathcal{V}_1}\vert u_j\vert$.
 Thus, the first part in \eqref{eq_part1} can be bounded as follows
$\Vert{\bf u}^{*\top}\hat{{\bf S}}\hat{{\bf S}}^{\top}\left({\bf u}-{\bf u}^{*}\right)\Vert_2=O_p(\tau_{n,p,m}) \sum_{j\notin \mathcal{V}_1}\vert u_j\vert$.

Similarly, ${\bf u}={\bf u}-{\bf q}_{1}+{\bf q}_{1}={\bf e}+{\bf q}_{1}$
, the second part in \eqref{eq_part1} can be bounded by 
\begin{align*}
\Vert\left({\bf u}-{\bf u}^{*}\right)^{\top}\hat{{\bf S}}\hat{{\bf S}}^{\top}{\bf u}\Vert_2 & =\Vert{\bf u}^{\top}\left({\bf S}{\bf S}^{\top}+{\bf E}\right)\left({\bf u}-{\bf u}^{*}\right)\Vert_2\\
 & \leq\Vert{\bf u}^{\top}{\bf S}{\bf S}^{\top}\left({\bf u}-{\bf u}^{*}\right)\Vert_2+\Vert{\bf u}^{\top}{\bf E}\left({\bf u}-{\bf u}^{*}\right)\Vert_2\\
 & \leq\Vert{\bf e}^{\top}{\bf S}{\bf S}^{\top}\left({\bf u}-{\bf u}^{*}\right)\Vert_2+\Vert{\bf q}_{1}^{\top}{\bf S}{\bf S}^{\top}\left({\bf u}-{\bf u}^{*}\right)\Vert_2\\
 & +\Vert{\bf e}^{\top}{\bf E}\left({\bf u}-{\bf u}^{*}\right)\Vert_2+\Vert{\bf q}_{1}^{\top}{\bf E}\left({\bf u}-{\bf u}^{*}\right)\Vert_2 \leq O_{p}\left(\tau_{n,p,m}\right)\sum_{j\notin\mathcal{V}_{1}}\vert u_{j}\vert.
\end{align*}

Thus, we have the following result for  \eqref{eq_part1},
\begin{equation}
\label{eq_bound1}
-\tilde{{\bf u}}^{\top}\hat{{\bf S}}\hat{{\bf S}}^{\top}\tilde{{\bf u}}+{\bf u}^{\top}\hat{{\bf S}}\hat{{\bf S}}^{\top}{\bf u}\leq O_{p}\left(\tau_{n,p,m}\right)\sum_{j\notin\mathcal{V}_{1}}\vert u_{j}\vert.
\end{equation}

Next, we will consider the bound for the last two components in \eqref{eq_gdiff}. 
Since $\Vert{\bf u}-{\bf q}_{1}\Vert_2=O_{p}\left(\tau_{n,p,m}\right)$
and $\Vert\tilde{{\bf u}}-{\bf q}_{1}\Vert_2=O_{p}\left(\tau_{n,p,m}\right)$
and $\lambda/\tau_{n,p,m} \to \infty$ as assumed, thus, $\vert\tilde{u}_{j}\vert>\gamma\lambda$
and $\vert u_{j}\vert>\gamma\lambda$ for $j\in\mathcal{V}_{1}$. Thus we have, 
\[
\sum_{j=1}^{p}\mathcal{P}_{\gamma}\left(\vert\tilde{u}_{j}\vert;\lambda\right)-\sum_{j=1}^{p}\mathcal{P}_{\gamma}\left(\vert u_{j}\vert;\lambda\right)=-\sum_{j\notin\mathcal{V}_{1}}\mathcal{P}_{\gamma}\left(\vert u_{j}\vert;\lambda\right)=-\sum_{j\notin\mathcal{V}_{1}}\left(\lambda\vert u_{j}\vert-\frac{\vert u_{j}\vert^{2}}{2\gamma}\right).
\]

Thus,
\begin{align*}
G\left(\tilde{{\bf u}}\right)-G\left({\bf u}\right) & \leq O_{p}\left(\tau_{n,p,m}\right)\sum_{j\notin\mathcal{V}_{1}}\vert u_{j}\vert-\sum_{j\notin\mathcal{V}_{1}}\left(\lambda\vert u_{j}\vert-\frac{\vert u_{j}\vert^{2}}{2\gamma}\right)\\
 & =-\sum_{j\notin\mathcal{V}_{1}}\left(\lambda-\frac{\vert u_{j}\vert}{2\gamma}-O_{p}\left(\tau_{n,p,m}\right)\right)\vert u_{j}\vert.
\end{align*}

Since $\vert u_{j}\vert=O_{p}\left(\tau_{n,p,m}\right)$ for
$j\notin\mathcal{\mathcal{V}}_{1}$, thus $\lambda \gtrsim \vert u_{j}\vert$. 
This implies that $G\left(\tilde{{\bf u}}\right)-G\left({\bf u}\right)<0$ for $\tilde{\bf u} \neq {\bf u}$.
Thus, we have shown that $G\left(\hat{{\bf q}}_{1}^{or}\right)<G\left(\tilde{{\bf u}}\right)<G\left({\bf u}\right)$ for $\bu \neq \hat{\bq}_1^{or}$,
which implies that $\hat{{\bf q}}_{1}^{or}$ is a local minimizer
of the objective function. This completes the proof. 

\paragraph{Part 2} In this part, we prove the results for $\bq_i$, $i=2,\dots,r$, which also depends on the results in Part 1. 

The objective function is 
\begin{align*}
\hat{\bq}_i = \argmin_{\bq_i}\frac{1}{2}\Vert\widehat{{\bf S}}\widehat{{\bf S}}^{\top}-{\bf s}_{i}{\bf s}_{i}^{\top}\Vert_{F}^{2}+\sum_{j=1}^{p}\mathcal{P}_{\gamma}\left(\vert q_{ij}\vert;\lambda\right)\\
\text{subject to }{\bf s}_{i}=\left({\bf I}-\tilde{{\bf S}}_{i}\tilde{{\bf S}}_{i}^{\top}\right){\bf q}_{i}\text{ and }\Vert{\bf s}_{i}\Vert=1,
\end{align*}
where $\tilde{\bf S}_i = (\tilde{\bf s}_1,\dots \tilde{\bf s}_{i-1})$, the estimated space from the previous step. 

We have proved that $\Vert \hat{{\bf q}}_1 - {\bf q}_1\Vert_2= O_p(\tau_{n,p,m})$ in Part 1, which implies that  $\Vert \tilde{\bf S}_i - {\bf S}_i\Vert_2 = O_p(\tau_{n,p,m})$  for $i=2$. Next, we will show $\Vert \hat{{\bf q}}_i - {\bf q}_i\Vert_2= O_p(\tau_{n,p,m})$
if $\Vert \tilde{\bf S}_i - {\bf S}_i\Vert_2 = O_p(\tau_{n,p,m})$ for $i=2,\dots,r$. The result can be proved using the recursive procedure.

Let ${\bf H}_i={\bf I}-{\bf S}_{i}{\bf S}_{i}^{\top}$,
\textbf{$\hat{{\bf H}}_i={\bf I}-\tilde{{\bf S}}_{i}\tilde{{\bf S}}_{i}^{\top}$}. Also $\bs_i = (\bI - {\bS}_i {\bS}_i^\top)\bq_i = \bH \bq_i$ and $\hat{\bs}_i = (\bI - \tilde{\bS}_i \tilde{\bS}_i^\top)\hat{\bq}_i = \hat{\bH}_i\hat{\bq}_i$.

The optimization algorithm is equivalent to the following based on Lemma \ref{lemma_obj},
\begin{align}
\label{eq_ob_qi}
\hat{{\bf q}}_{i}=\arg\min_{{\bf q}_{i}}-{\bf q}_{i}^{\top}\hat{{\bf H}}_i\hat{{\bf S}}\hat{{\bf S}}_i^{\top}{\bf \hat{H}}_i{\bf q}_{i}+\sum_{j=1}^{p}\mathcal{P}_{\gamma}\left(\vert q_{ij}\vert;\lambda\right)\\
\text{subject to }{\bf q}_{i}^{\top}\hat{{\bf H}}_i{\bf \hat{H}}_i{\bf q}_{i}=1. \nonumber
\end{align}

{\bf Step 1}: We consider the oracle property if the sparsity of ${\bf q}_{i}$ is given.
Let $\hat{{\bf q}}_{i}^{or}$ be the oracle estimator when $\mathcal{V}_{i}$ is known. Then, the objective function in \eqref{eq_ob_qi} becomes 
\begin{align}
\label{eq_ob_qi_or}
\hat{{\bf q}}_{i}^{or}=\arg\min_{{\bf q}_{i}}-{\bf q}_{i}^{\top}\hat{{\bf H}}_i\hat{{\bf S}}\hat{{\bf S}}_i^{\top}{\bf \hat{H}}_i{\bf q}_{i}\\
\text{subject to }{\bf q}_{i}^{\top}\hat{{\bf H}}_i{\bf \hat{H}}_i{\bf q}_{i}=1 \text{ and } \bq_{i[-\mathcal{V}_i]} ={\bf 0}. \nonumber
\end{align}

When $\mathcal{V}_i$ is known, the sparsity of ${\bf s}_{i}$ is a subset of $\mathcal{V}_i^* = \mathcal{V}_{s_{1}}\cup\cdots\cup\mathcal{V}_{s_{i-1}}\cup\mathcal{V}_{i}$,
which is an order of $m$. We know that ${\bf s}_{i}=\hat{{\bf H}}_i {\bf q}_i = \left({\bf I}-\tilde{{\bf S}}_{i}\tilde{{\bf S}}_{i}^{\top}\right){\bf q}_{i}\text{ and }\Vert{\bf s}_{i}\Vert_2=1$. Based on the discussion above Lemma \ref{lemma:inv_s2} , we know that ${\bf s}_i^* = {\bf s}_{i[\mathcal{V}_i^*]}$ satisfies,
\[
{\bf s}_{i}^{*}=\left(\begin{array}{cc}
\tilde{{\bf S}}_{i,1}\left(\tilde{{\bf S}}_{i,1}^{\top}\tilde{{\bf S}}_{i,1}\right)^{-1}\tilde{{\bf S}}_{i,1}^{\top} & {\bf 0}\\
{\bf 0} & \mathbf{I}
\end{array}\right)\mathbf{s}_{i}^{*}=\tilde{{\bf A}}{\bf s}_{i}^{*}
\]

Since $\widehat{{\bf H}}_i{\bf \widehat{H}}_i=\widehat{{\bf H}}_i$, when $\mathcal{V}_{i}^*$ is known,  \eqref{eq_ob_qi} is equivalent to the following optimization problem
\begin{align}
\label{eq_obj_si}
\hat{{\bf s}}_{i}=\arg\max_{ {\bf s}_{i}}  {\bf s}_{i}^{\top}\widehat{{\bf H}}_i\widehat{{\bf S}}\widehat{{\bf S}}^{\top}{\bf \widehat{H}}_i{\bf s}_{i}\\
\text{subject to } \Vert{\bf s}_{i}\Vert =1 ,\, {\bf s}_{i[-\mathcal{V}_i^*]}={\bf 0} \text{ and } {\bf s}_i^* = \tilde{{\bf A}}{\bf s}_{i}^{*}. \nonumber
\end{align}

Let ${{\bf S}}^* ={{\bf S}}_{[\mathcal{V}_i^*]}$ and $\hat{{\bf S}}^* =\hat{{\bf S}}_{[\mathcal{V}_i^*]}$ be the subrows of ${\bf S}$ and $\hat{{\bf S}}$, respectively. And let 
${\bf H}^{*}_i={\bf H}_{[\mathcal{V}_i^*]}$ and  $\widehat{{\bf H}}^{*}_i=\widehat{{\bf H}}_{[\mathcal{V}_i^*]}$ be the  $\mathcal{V}_i^*$ subrows and subcolumns of ${\bH}_i$ and $\hat{\bH}_i$, respectively. The optimization problem in \eqref{eq_obj_si} is equivalent to following
\begin{align*}
\hat{{\bf s}}_{i}^{*}=\arg\min_{{\bf s}_{i}^{*}}-{\bf s}_{i}^{*\top}\tilde{{\bf A}}\hat{{\bf H}}_{i}^{*}\hat{{\bf S}}^{*}\hat{{\bf S}}^{*\top}{\bf \hat{H}^{*}}_{i}\tilde{{\bf A}}{\bf s}_{i}^{*}\\
\text{subject to }\Vert{\bf s}_{i}^{*}\Vert_2=1.
\end{align*}

This is finding the leading eigenvector of $\tilde{{\bf A}}\hat{{\bf H}}_{i}^{*}\hat{{\bf S}}^{*}\hat{{\bf S}}^{*\top}{\bf \hat{H}^{*}}_{i}\tilde{{\bf A}}$. Based on the definition, we know that 
\begin{align*}
\tilde{{\bf A}}\hat{{\bf H}}_{i}^{*}\hat{{\bf S}}^{*} & =\left(\begin{array}{cc}
\tilde{{\bf S}}_{i,1}\left(\tilde{{\bf S}}_{i,1}^{\top}\tilde{{\bf S}}_{i,1}\right)^{-1}\tilde{{\bf S}}_{i,1}^{\top}-\tilde{{\bf S}}_{i,1}\tilde{{\bf S}}_{i,1}^{\top} & -\tilde{{\bf S}}_{i,1}\tilde{{\bf S}}_{i,2}^{\top}\\
-{\bf \tilde{S}}_{i,2}^{\top}\tilde{{\bf S}}_{i,1} & {\bf I}-\tilde{{\bf S}}_{i,2}\tilde{{\bf S}}_{i,2}^{\top}
\end{array}\right)\left(\begin{array}{c}
\hat{{\bf S}}_{1}^{*}\\
\hat{{\bf S}}_{2}^{*}
\end{array}\right)\\
 & =\left(\tilde{{\bf A}}-\tilde{{\bf S}}_{i}^{*}\tilde{{\bf S}}_{i}^{*\top}\right)\hat{{\bf S}}^{*},
\end{align*}
and ${\bf A}{\bf H}_{i}^{*}{\bf S}^{*}=\left({\bf A}-{\bf S}_{i}^{*}{\bf S}_{i}^{*\top}\right){\bf S}^{*}$. Note that ${\bf A}{\bf H}_{i}^{*}{\bf S}^{*}{\bf S}^{*\top}{\bf H}_{i}^{*}{\bf A}{\bf s}_{i}^{*}=\left({\bf A}-{\bf S}_{i}^{*}{\bf S}_{i}^{*\top}\right){\bf S}^{*}{\bf S}^{*\top}\left({\bf A}-{\bf S}_{i}^{*}{\bf S}_{i}^{*\top}\right){\bf s}_{i}^{*}$.
Since ${\bf A}{\bf s}_{i}^{*}={\bf s}_{i}^{*}$ and ${\bf S}_{i}^{*}{\bf S}_{i}^{*\top}{\bf s}_{i}^{*}={\bf 0}$,
thus 
\[
{\bf A}{\bf H}_{i}^{*}{\bf S}^{*}{\bf S}^{*\top}{\bf H}_{i}^{*}{\bf A}{\bf s}_{i}^{*}=\left({\bf A}-{\bf S}_{i}^{*}{\bf S}_{i}^{*\top}\right){\bf S}^{*}{\bf S}^{*\top}{\bf s}_{i}^{*}=\left({\bf A}-{\bf S}_{i}^{*}{\bf S}_{i}^{*\top}\right){\bf s}_{i}^{*}={\bf s}_{i}^{*},
\]
which indicates that ${\bf s}_{i}^{*}$ is the leading eigenvector of ${\bf A}{\bf H}_{i}^{*}{\bf S}^{*}{\bf S}^{*\top}{\bf H}_{i}^{*}{\bf A}{\bf s}_{i}^{*}$.

By using Lemma \ref{lemma:boundprojection}, we have
\begin{align*}
\Vert\tilde{{\bf A}}\hat{{\bf S}}^{*}-{\bf A}{\bf S}^{*}\Vert_{2} & \leq\Vert\tilde{{\bf S}}_{i,1}\left(\tilde{{\bf S}}_{i,1}^{\top}\tilde{{\bf S}}_{i,1}\right)^{-1}\tilde{{\bf S}}_{i,1}^{\top}\hat{{\bf S}}_{1}^{*}-{\bf S}_{i,1}\left(\mathbf{S}_{i,1}^{\top}\mathbf{S}_{i,1}\right)^{-1}\mathbf{S}_{i,1}^{\top}{\bf S}_{1}^{*}\Vert_{2}+\Vert\hat{{\bf S}}_{2}^{*}-{\bf S}_{2}^{*}\Vert_{2}\\
 & =\Vert\tilde{{\bf S}}_{i,1}\left(\tilde{{\bf S}}_{i,1}^{\top}\tilde{{\bf S}}_{i,1}\right)^{-1}\tilde{{\bf S}}_{i,1}^{\top}\Vert_{2}\Vert\hat{{\bf S}}_{1}^{*}-{\bf S}_{1}^{*}\Vert_{2}+\\
 & \Vert\tilde{{\bf S}}_{i,1}\left(\tilde{{\bf S}}_{i,1}^{\top}\tilde{{\bf S}}_{i,1}\right)^{-1}\tilde{{\bf S}}_{i,1}^{\top}-{\bf S}_{i,1}\left(\mathbf{S}_{i,1}^{\top}\mathbf{S}_{i,1}\right)^{-1}\mathbf{S}_{i,1}^{\top}\Vert_{2}\Vert{\bf S}_{1}^{*}\Vert_{2}+\Vert\hat{{\bf S}}_{2}^{*}-{\bf S}_{2}^{*}\Vert_{2}\\
 & =O_{p}\left(\tau_{n,p,m}\right).
\end{align*}
And $
 \Vert\tilde{{\bf S}}_{i}^{*}\tilde{{\bf S}}_{i}^{*\top}\hat{{\bf S}}^{*}-{\bf S}_{i}^{*}{\bf S}_{i}^{*\top}{\bf S}^{*}\Vert_2
\leq  \Vert\tilde{{\bf S}}_{i}^{*}\tilde{{\bf S}}_{i}^{*\top}\Vert_{2}\Vert\hat{{\bf S}}^{*}-{\bf S}^{*}\Vert_{2}+\Vert\tilde{{\bf S}}_{i}^{*}\tilde{{\bf S}}_{i}^{*\top}-{\bf S}_{i}^{*}{\bf S}_{i}^{*\top}\Vert_{2}\Vert{\bf S}^{*}\Vert_{2}= O_{p}\left(\tau_{n,p,m}\right).
$
Since $\Vert{\bf A}{\bf H}_{i}^{*}{\bf S}^{*}\Vert_{2}\leq1$ and
$\Vert\tilde{{\bf A}}\hat{{\bf H}}_{i}^{*}\hat{{\bf S}}^{*}\Vert_2\leq1$,
thus 
\begin{align*}
 & \Vert\tilde{{\bf A}}\hat{{\bf H}}_{i}^{*}\hat{{\bf S}}^{*}\hat{{\bf S}}^{*\top}{\bf \hat{H}^{*}}_{i}\tilde{{\bf A}}-{\bf A}{\bf H}_{i}^{*}{\bf S}^{*}{\bf S}^{*\top}{\bf H}_{i}^{*}{\bf A}\Vert_{2}\\
\leq & 2\Vert\left(\tilde{{\bf A}}-\tilde{{\bf S}}_{i}^{*}\tilde{{\bf S}}_{i}^{*\top}\right)\hat{{\bf S}}^{*}-\left({\bf A}-{\bf S}_{i}^{*}{\bf S}_{i}^{*\top}\right){\bf S}^{*}\Vert_{2} =  O_{p}\left(\tau_{n,p,m}\right).
\end{align*}

Thus, from Lemma \ref{lemma_eigenvalue}, we have.
\[
\Vert\hat{{\bf s}}_{i}^{*}-{\bf s}_{i}^{*}\Vert_2=O_{p}\left(\tau_{n,p,m}\right).
\]

Next, we will consider the estimator for ${\bf q}_i^*$.

Since 
$
{\bf q}_{i}^{*}=\left({\bf I}-{\bf S}_{i,2}{\bf S}_{i,2}^{\top}\right)^{-1}{\bf s}_{i,2}={\bf s}_{i,2}+\mathbf{S}_{i2}\left({\bf S}_{i,1}^{\top}{\bf S}_{i,1}\right)^{-1}{\bf S}_{i,2}^{\top}{\bf s}_{i,2}
$
 and $\hat{{\bf q}}_{i}^{*}=\hat{{\bf s}}_{i,2}+\tilde{\mathbf{S}}_{i2}\left(\tilde{\bf {S}}_{i,1}^{\top}\tilde{\bf {S}}_{i,1}\right)^{-1}\tilde{{\bf S}}_{i,2}^{\top}{\bf \hat{s}}_{i,2}$,
thus 
$
\Vert\hat{{\bf q}}_{i}^{*}-{\bf q}_{i}^{*}\Vert_2\leq\Vert\hat{{\bf s}}_{i,2}-{\bf s}_{i,2}\Vert_{2}+\Vert\tilde{\mathbf{S}}_{i2}\left(\tilde{\bf {S}}_{i,1}^{\top}\tilde{\bf {S}}_{i,1}\right)^{-1}\tilde{{\bf S}}_{i,2}^{\top}{\bf \hat{s}}_{i,2}-\mathbf{S}_{i2}\left({\bf S}_{i,1}^{\top}{\bf S}_{i,1}\right)^{-1}{\bf S}_{i,2}^{\top}{\bf s}_{i,2}\Vert_{2}.
$

Using Lemma \ref{lemma:boundprojection}, we have
\begin{align*}
 & \Vert\tilde{\mathbf{S}}_{i2}\left(\tilde{\bf {S}}_{i,1}^{\top}\tilde{\bf {S}}_{i,1}\right)^{-1}\tilde{{\bf S}}_{i,2}^{\top}{\bf \hat{s}}_{i,2}-\mathbf{S}_{i2}\left({\bf S}_{i,1}^{\top}{\bf S}_{i,1}\right)^{-1}{\bf S}_{i,2}^{\top}{\bf s}_{i,2}\Vert_{2}\\
\leq & \Vert\tilde{\mathbf{S}}_{i2}\left(\tilde{\bf {S}}_{i,1}^{\top}\tilde{\bf {S}}_{i,1}\right)^{-1}\tilde{{\bf S}}_{i,2}^{\top}{\bf \hat{s}}_{i,2}\Vert_{2}\Vert\hat{{\bf s}}_{i,2}-{\bf s}_{i,2}\Vert_{2}+\\
 & \Vert\tilde{\mathbf{S}}_{i2}\left(\tilde{\bf {S}}_{i,1}^{\top}\tilde{\bf {S}}_{i,1}\right)^{-1}\tilde{{\bf S}}_{i,2}^{\top}-\mathbf{S}_{i2}\left({\bf S}_{i,1}^{\top}{\bf S}_{i,1}\right)^{-1}{\bf S}_{i,2}^{\top}\Vert_{2}\Vert{\bf s}_{i,2}\Vert_{2}\\
= & O_{p}\left(\tau_{n,p,m}\right).
\end{align*}

Thus, $\Vert\hat{{\bf q}}_{i}^{*}-{\bf q}_{i}^{*}\Vert_2=O_{p}\left(\tau_{n,p,m}\right).$

Recall that $\hat{{\bf q}}_{i[\mathcal{V}_i]}^{or} = \hat{{\bf q}}_i^*$ and $\hat{{\bf q}}_{i[-\mathcal{V}_i]}^{or} = {\bf 0}$. Thus,  $\Vert\hat{{\bf q}}_{i}^{or}-{\bf q}_{i}\Vert_2=O_{p}\left(\tau_{n,p,m}\right).$

{\bf Step 2}: Next step is to show that $\hat{{\bf q}}_{i}^{or}$ is a local minimizer
of $G\left({\bf q}_{i}\right)$ with the following form,
\begin{align*}
G\left({\bf q}_{i}\right)=\frac{1}{2}\Vert\widehat{{\bf S}}\widehat{{\bf S}}^{\top}-{\bf s}_{i}{\bf s}_{i}^{\top}\Vert_{F}^{2}+\sum_{j=1}^{p}\mathcal{P}_{\gamma}\left(\vert q_{ij}\vert;\lambda\right)\\
\text{subject to }{\bf s}_{i}=\left({\bf I}-\tilde{{\bf S}}_{i}\tilde{{\bf S}}_{i}^{\top}\right){\bf q}_{i}\text{ and }\Vert{\bf s}_{i}\Vert_2=1.
\end{align*}

Consider a neighbor of ${\bf q}_{i}$ such that $\Vert{\bf u}-{\bf q}_{i}\Vert_2=O_{p}\left(\tau_{n,p,m}\right)$, $\Vert{\bf u}-\hat{{\bf q}}_{i}^{or}\Vert_2 \leq \delta_n$, where $\delta_n = o(1)$,
and satisfies $\Vert\left({\bf I}-\tilde{{\bf S}}_{i}\tilde{{\bf S}}_{i}^{\top}\right){\bf u}\Vert_2=1$.
Define ${\bf u}^{*}[\mathcal{V}_{i}]={\bf u}_{[\mathcal{V}_{i}]}$
and ${\bf u}^{*}[-\mathcal{V}_{i}]=\bf{0}$, and $\alpha=\Vert\left({\bf I}-\tilde{{\bf S}}_{i}\tilde{{\bf S}}_{i}^{\top}\right){\bf u}^{*}\Vert_2$. Denote
$\tilde{{\bf u}}={\bf u}^{*}/\alpha$ , which indicates that $\tilde{{\bf u}}[-\mathcal{V}_{i}]={\bf 0}$ and $\Vert\left({\bf I}-\tilde{{\bf S}}_{i}\tilde{{\bf S}}_{i}^{\top}\right)\tilde{{\bf u}}\Vert_2=1$ based on the definition of $\tilde{\bu}$.

First we will compare $G\left(\hat{{\bf q}}_{i}^{or}\right)$ and
$G\left(\tilde{{\bf u}}\right)$.

From the definitions, we have $\Vert{\bf u}^{*}-{\bf q}_{i}\Vert_2=O_{p}\left(\tau_{n,p,m}\right)$,
$\Vert{\bf u}_{[-\mathcal{V}_{i}]}\Vert_2=\Vert{\bf u}-{\bf u}^{*}\Vert_2=O_{p}\left(\tau_{n,p,m}\right)$,
and 
$
\alpha=\Vert\left({\bf I}-\tilde{{\bf S}}_{i}\tilde{{\bf S}}_{i}^{\top}\right){\bf u}^{*}\Vert_2=\Vert\left({\bf I}-\tilde{{\bf S}}_{i}\tilde{{\bf S}}_{i}^{\top}\right){\bf u}+\left({\bf I}-\tilde{{\bf S}}_{i}\tilde{{\bf S}}_{i}^{\top}\right){\bf {\bf u}}_{[-\mathcal{V}_{i}]}\Vert_2 \geq 1-\Vert\left({\bf I}-\tilde{{\bf S}}_{i}\tilde{{\bf S}}_{i}^{\top}\right){\bf {\bf u}}_{[-\mathcal{V}_{i}]}\Vert_2
$ with $\alpha\leq1$.  We have 
\begin{align*}
\tilde{{\bf u}}-{\bf q}_{i} & =\frac{{\bf u}^{*}}{\alpha}-{\bf q}_{i}=\frac{{\bf u}-\left({\bf u}-{\bf u}^{*}\right)}{\alpha}-{\bf q}_{i}=\frac{{\bf u}-{\bf q}_{i}+\left({\bf u}-{\bf u}^{*}\right)}{\alpha}+\left(\frac{1}{\alpha}-1\right){\bf q}_{i}.
\end{align*}
Thus, 
\begin{align*}
\Vert\tilde{{\bf u}}-{\bf q}_{i}\Vert_2 & \leq\frac{1}{\alpha}\Vert{\bf u}-{\bf q}_{i}\Vert_2+\frac{1}{\alpha}\Vert{\bf u}-{\bf u}^{*}\Vert_2+\frac{1}{\alpha}-1\\
 & \leq\frac{\Vert{\bf u}-{\bf q}_{i}\Vert_2+\Vert{\bf u}-{\bf u}^{*}\Vert_2+\Vert\left({\bf I}-\tilde{{\bf S}}_{i}\tilde{{\bf S}}_{i}^{\top}\right){\bf {\bf u}}_{[-\mathcal{V}_{i}]}\Vert_2}{1-\Vert\left({\bf I}-\tilde{{\bf S}}_{i}\tilde{{\bf S}}_{i}^{\top}\right){\bf {\bf u}}_{[-\mathcal{V}_{i}]}\Vert_2} =O_{p}\left(\tau_{n,p,m}\right).
\end{align*}

Based on the assumption about the minimal signal and the assumption about $\lambda$, for $j\in\mathcal{V}_{i}$, we have
$\vert\hat{q}_{ij}^{or}\vert\geq\vert q_{ij}\vert- \vert q_{ij} -\hat{q}_{ij}^{or}\vert >\gamma\lambda$ since $\vert q_{ij} -\hat{q}_{ij}^{or}\vert = O_p(\tau_{n,p,m})$. Similarly $\vert\tilde{u}_{j}\vert>\gamma\lambda$ for $j\in \mathcal{V}_i$.
Then $\mathcal{P}_{\gamma}\left(\vert\hat{q}_{ij}\vert;\lambda\right)=\mathcal{P}_{\gamma}\left(\vert\tilde{u}_{j}\vert;\lambda\right)=\frac{1}{2}\gamma\lambda^{2}$.
So we have $\sum_{j=1}^{p}\mathcal{P}_{\gamma}\left(\vert\hat{q}_{ij}^{or}\vert;\lambda\right)=\sum_{j=1}^{p}\mathcal{P}_{\gamma}\left(\vert\tilde{u}_{j}\vert;\lambda\right)$.
Based on the definition of $\hat{{\bf q}}_{i}^{or}$, we have $\Vert\hat{{\bf S}}\hat{{\bf S}}^{\top}-\hat{{\bf s}}_{i}^{or}\left(\hat{{\bf s}}_{i}^{or}\right)^{\top}\Vert_{F}^{2}<\Vert\hat{{\bf S}}\hat{{\bf S}}^{\top}-\tilde{{\bf s}}\tilde{{\bf s}}^{\top}\Vert_{F}^{2}$
for $\tilde{{\bf u}}\neq\hat{{\bf q}}_{i}^{or}$, where $\hat{{\bf s}}_{i}^{or}=\left({\bf I}-\tilde{{\bf S}}_{i}\tilde{{\bf S}}_{i}^{\top}\right)\hat{{\bf q}}_{i}^{or}$
and $\tilde{{\bf s}}_{i}=\left({\bf I}-\tilde{{\bf S}}_{i}\tilde{{\bf S}}_{i}^{\top}\right)\tilde{{\bf u}}$.
This implies that $G\left(\hat{{\bf q}}_{i}^{or}\right)<G\left(\tilde{{\bf u}}\right)$.

Next, we will compare $G\left(\tilde{{\bf u}}\right)$ and $G\left({\bf u}\right)$.
We have that 
\begin{align}
\label{eq_qi_diff}
G\left(\tilde{{\bf u}}\right)-G\left({\bf u}\right) & =-\tilde{{\bf u}}^{\top}\hat{{\bf H}}\hat{{\bf {\bf S}}}\hat{{\bf S}}^{\top}\hat{{\bf H}}\tilde{{\bf u}}+{\bf u}^{\top}\hat{{\bf H}}{\bf S}{\bf S}^{\top}\hat{{\bf H}}{\bf u}+\sum_{j=1}^{p}\mathcal{P}_{\gamma}\left(\vert\tilde{u}_{j}\vert;\lambda\right)-\sum_{j=1}^{p}\mathcal{P}_{\gamma}\left(\vert u_{j}\vert;\lambda\right),
\end{align}
where $\hat{\bf H} = \hat{\bf H}_i$.
Since
$
-\tilde{{\bf u}}^{\top}\hat{{\bf H}}\hat{{\bf S}}\hat{\bS}^\top \hat{{\bf H}}\tilde{{\bf u}}=-\frac{{\bf u}^{*\top}\hat{{\bf H}}\hat{{\bf S}}\hat{\bS}^\top \hat{{\bf H}}{\bf u}^{*}}{\alpha^{2}}\leq-{\bf u}^{*\top}\hat{{\bf H}}\hat{{\bf S}}\hat{\bS}^\top \hat{{\bf H}}{\bf u}^{*},
$
thus, the first part in \eqref{eq_qi_diff} is equivalent to the following,
\begin{align}
\label{eq_qi_p1}
-\tilde{{\bf u}}^{\top}\hat{{\bf H}}\hat{{\bf {\bf S}}}\hat{{\bf S}}^{\top}\hat{{\bf H}}\tilde{{\bf u}}+{\bf u}^{\top}\hat{{\bf H}}\hat{{\bf {\bf S}}}\hat{{\bf S}}^{\top}\hat{{\bf H}}{\bf u} & \leq-{\bf u}^{*\top}\hat{{\bf H}}\hat{{\bf {\bf S}}}\hat{{\bf S}}^{\top}\hat{{\bf H}}{\bf u}^{*}+{\bf u}^{\top}\hat{{\bf H}}\hat{{\bf {\bf S}}}\hat{{\bf S}}^{\top}\hat{{\bf H}}{\bf u}\nonumber \\
 & ={\bf u}^{*\top}\hat{{\bf H}}\hat{{\bf {\bf S}}}\hat{{\bf S}}^{\top}\hat{{\bf H}}\left({\bf u}-{\bf u}^{*}\right)+\left({\bf u}-{\bf u}^{*}\right)^{\top}\hat{{\bf H}}\hat{{\bf {\bf S}}}\hat{{\bf S}}^{\top}\hat{{\bf H}}{\bf u}
\end{align}

Let ${\bf e}^{*}={\bf u}^{*}-{\bf q}_{i}$, ${\bf E}=\hat{{\bf S}}\hat{{\bf S}}^{\top}-{\bf S}{\bf S}^{\top}$
and ${\bf E}_{H}={\bf \hat{{\bf H}}-{\bf H}}$ with $\Vert{\bf E}_{H}\Vert_2=O_{p}\left(\tau_{n,p,m}\right)$.
For the first part in \eqref{eq_qi_p1}, 
\begin{align*}
 & \vert{\bf u}^{*\top}\hat{{\bf H}}\hat{{\bf {\bf S}}}\hat{{\bf S}}^{\top}\hat{{\bf H}}\left({\bf u}-{\bf u}^{*}\right)\vert
\leq  \vert{\bf u}^{*\top}\left({\bf H}+{\bf E}_{H}\right)\hat{{\bf {\bf S}}}\hat{{\bf S}}^{\top}\left({\bf H}+{\bf E}_{H}\right)\left({\bf u}-{\bf u}^{*}\right)\vert\\
\leq & \vert{\bf u}^{*\top}{\bf H}\hat{{\bf {\bf S}}}\hat{{\bf S}}^{\top}{\bf H}\left({\bf u}-{\bf u}^{*}\right)\vert+\vert{\bf u}^{*\top}{\bf E}_{H}\hat{{\bf {\bf S}}}\hat{{\bf S}}^{\top}\hat{{\bf H}}\left({\bf u}-{\bf u}^{*}\right)\vert\\
 & +\vert{\bf u}^{*\top}{\bf H}\hat{{\bf {\bf S}}}\hat{{\bf S}}^{\top}{\bf E}_{H}\left({\bf u}-{\bf u}^{*}\right)\vert+\vert{\bf u}^{*\top}{\bf E}_{H}\hat{{\bf {\bf S}}}\hat{{\bf S}}^{\top}{\bf E}_{H}\left({\bf u}-{\bf u}^{*}\right)\vert\\
\leq & \vert{\bf u}^{*\top}{\bf H}\hat{{\bf {\bf S}}}\hat{{\bf S}}^{\top}{\bf H}\left({\bf u}-{\bf u}^{*}\right)\vert+O_{p}\left(\tau_{n,p,m}\right)\Vert{\bf u}-{\bf u}^{*}\Vert_2.
\end{align*}
Furthermore, 
\begin{align*}
 & \vert{\bf u}^{*\top}{\bf H}\hat{{\bf {\bf S}}}\hat{{\bf S}}^{\top}{\bf H}\left({\bf u}-{\bf u}^{*}\right)\vert
\leq  \vert{\bf u}^{*\top}{\bf H}\left({\bf S}{\bf S}^{\top}+{\bf E}\right){\bf H}\left({\bf u}-{\bf u}^{*}\right)\vert\\
\leq & \vert{\bf u}^{*\top}{\bf H}{\bf S}{\bf S}^{\top}{\bf H}\left({\bf u}-{\bf u}^{*}\right)\vert+\vert{\bf u}^{*\top}{\bf H}{\bf E}{\bf H}\left({\bf u}-{\bf u}^{*}\right)\vert.
\end{align*}
These two parts can be bounded as follows,
\begin{align*}
 & \vert{\bf u}^{*\top}{\bf H}{\bf S}{\bf S}^{\top}{\bf H}\left({\bf u}-{\bf u}^{*}\right)\vert
\leq  \vert{\bf q}_{i}^{\top}{\bf H}{\bf S}{\bf S}^{\top}{\bf H}\left({\bf u}-{\bf u}^{*}\right)\vert+\vert{\bf e}^{*\top}{\bf H}{\bf S}{\bf S}^{\top}{\bf H}\left({\bf u}-{\bf u}^{*}\right)\vert
\leq  O_{p}\left(\tau_{n,p,m}\right)\Vert{\bf u}-{\bf u}^{*}\Vert,
\end{align*}
and 
\begin{align*}
 & \vert{\bf u}^{*\top}{\bf H}{\bf E}{\bf H}\left({\bf u}-{\bf u}^{*}\right)\vert
\leq  \vert{\bf q}_{i}^{\top}{\bf H}{\bf E}{\bf H}\left({\bf u}-{\bf u}^{*}\right)\vert+\vert{\bf e}^{*\top}{\bf H}{\bf E}{\bf H}\left({\bf u}-{\bf u}^{*}\right)\vert\\
\leq & \Vert{\bf s}_{i}^{\top}{\bf E}{\bf H}\Vert_{\max}\sum_{j\notin\mathcal{V}_{i}}\vert u_{j}\vert+O_{p}\left(\tau_{n,p,m}\right)\Vert{\bf u}-{\bf u}^{*}\Vert_2.
\end{align*}

If $m=o(p)$, then the bound of $\Vert{\bf s}_{i}^{\top}{\bf E}{\bf H}\Vert_{\max}$ is 
\begin{align*}
\Vert{\bf s}_{i}^{\top}{\bf E}{\bf H}\Vert_{\max} & \leq\Vert{\bf E}{\bf H}\Vert_{\max}\sum_{j=1}^{p}\vert s_{ij}\vert\le\sqrt{m}\Vert{\bf E}{\bf H}\Vert_{\max}\\
 & \leq\sqrt{m}\Vert{\bf E}\Vert_{2\rightarrow\infty}\Vert{\bf H}\Vert_{2}=\sqrt{m}\Vert{\bf E}\Vert_{2\rightarrow\infty}.
\end{align*}
Furthermore, we have
\begin{align*}
\Vert{\bf E}\Vert_{2\rightarrow\infty} & =\Vert\hat{{\bf S}}\hat{{\bf S}}^{\top}-{\bf S}{\bf S}^{\top}\Vert_{2\rightarrow\infty} \leq\Vert\hat{{\bf S}}\hat{{\bf S}}^{\top}-{\bf S}\hat{{\bf S}}^{\top}+{\bf S}\hat{{\bf S}}^{\top}-{\bf S}{\bf S}^{\top}\Vert_{2\rightarrow\infty}\\
 & \leq\Vert\hat{{\bf S}}\hat{{\bf S}}^{\top}-{\bf S}\hat{{\bf S}}^{\top}\Vert_{2\rightarrow\infty}+\Vert{\bf S}\hat{{\bf S}}^{\top}-{\bf S}{\bf S}^{\top}\Vert_{2\rightarrow\infty}\\
 & \leq\Vert\hat{{\bf S}}-{\bf S}\Vert_{2\rightarrow\infty}\Vert\hat{{\bf S}}^{\top}\Vert+\Vert{\bf S}\Vert_{\infty}\Vert\hat{{\bf S}}-{\bf S}\Vert_{2\rightarrow\infty}\\
 & \leq\Vert\hat{{\bf S}}-{\bf S}\Vert_{2\rightarrow\infty}+\sqrt{r}\Vert{\bf S}\Vert_{2\rightarrow\infty}\Vert\hat{{\bf S}}-{\bf S}\Vert_{2\rightarrow\infty}.
\end{align*}
Thus 
\[
\Vert{\bf s}_{i}^{\top}{\bf E}{\bf H}\Vert_{\max}\leq c\sqrt{m}\Vert\hat{{\bf S}}-{\bf S}\Vert_{2\rightarrow\infty}=O_{p}\left(\tau_{n,p,m}\right).
\]

If $m=O(p)$, then the bound of $\Vert{\bf s}_{i}^{\top}{\bf E}{\bf H}\Vert_{\max}$ is 
\begin{align*}
\Vert{\bf s}_{i}^{\top}{\bf E}{\bf H}\Vert_{\max} & \leq \Vert{\bf s}_{i}\Vert_2 \Vert {\bf E}\Vert_2 \Vert {\bf H}\Vert_2 \leq \Vert {\bf E}\Vert_2  = O_p(\tau_{n,p,m}). 
\end{align*}

Combine all together; we can bound the first part in \eqref{eq_qi_p1} by
\begin{align*}
 & \vert{\bf u}^{*\top}\hat{{\bf H}}\hat{{\bf {\bf S}}}\hat{{\bf S}}^{\top}\hat{{\bf H}}\left({\bf u}-{\bf u}^{*}\right)\vert
\leq  \vert{\bf u}^{*\top}{\bf H}\hat{{\bf {\bf S}}}\hat{{\bf S}}^{\top}{\bf H}\left({\bf u}-{\bf u}^{*}\right)\vert+O_{p}\left(\tau_{n,p,m}\right)\Vert{\bf u}-{\bf u}^{*}\Vert_2\\
\leq & O_{p}\left(\tau_{n,p,m}\right)\Vert{\bf u}-{\bf u}^{*}\Vert+O_{p}\left(\tau_{n,p,m}\right)\sum_{j\notin\mathcal{V}_{i}}\vert u_{j}\vert.
\end{align*}

For the second part in \eqref{eq_qi_p1}, we have
\begin{align*}
 & \vert\left({\bf u}-{\bf u}^{*}\right)^{\top}\hat{{\bf H}}\hat{{\bf {\bf S}}}\hat{{\bf S}}^{\top}\hat{{\bf H}}{\bf u}\vert
\leq  \vert\left({\bf u}-{\bf u}^{*}\right)^{\top}{\bf H}\hat{{\bf {\bf S}}}{\bf \hat{{\bf S}}^{\top}}{\bf H}{\bf u}\vert+O_{p}\left(\tau_{n,p,m}\right)\Vert{\bf u}-{\bf u}^{*}\Vert_2\\
\leq & \vert {\bf u}^{\top}{\bf H}{\bf S}{\bf S}^{\top}{\bf H}\left({\bf u}-{\bf u}^{*}\right)\vert+\vert{\bf u}^{\top}{\bf H}{\bf E}{\bf H}\left({\bf u}-{\bf u}^{*}\right)\vert+O_{p}\left(\tau_{n,p,m}\right)\Vert{\bf u}-{\bf u}^{*}\Vert_2.
\end{align*}
Let ${\bf e}={\bf u}-{\bf q}_{1}$, then 
\begin{align*}
 & \vert{\bf u}^{\top}{\bf H}{\bf S}{\bf S}^{\top}{\bf H}\left({\bf u}-{\bf u}^{*}\right)\vert+\vert{\bf u}^{\top}{\bf H}{\bf E}{\bf H}\left({\bf u}-{\bf u}^{*}\right)\vert\\
\leq & \vert{\bf q}_{i}^{\top}{\bf H}{\bf S}{\bf S}^{\top}{\bf H}\left({\bf u}-{\bf u}^{*}\right)\vert+\vert{\bf e}^{\top}{\bf H}{\bf S}{\bf S}^{\top}{\bf H}\left({\bf u}-{\bf u}^{*}\right)\vert\\
 & +\vert{\bf q}_{i}^{\top}{\bf H}{\bf E}{\bf H}\left({\bf u}-{\bf u}^{*}\right)\vert+\vert{\bf e}^{\top}{\bf H}{\bf E}{\bf H}\left({\bf u}-{\bf u}^{*}\right)\vert\\
\leq & \vert{\bf q}_{i}^{\top}{\bf H}{\bf E}{\bf H}\left({\bf u}-{\bf u}^{*}\right)\vert+O_{p}\left(\tau_{n,p,m}\right)\Vert{\bf u}-{\bf u}^{*}\Vert_2.
\end{align*}

Then, we can bound the second part in \eqref{eq_qi_p1}, 
\[
\vert\left({\bf u}-{\bf u}^{*}\right)^{\top}\hat{{\bf H}}\hat{{\bf {\bf S}}}\hat{{\bf S}}^{\top}\hat{{\bf H}}{\bf u}\vert\leq O_{p}\left(\tau_{n,p,m}\right)\Vert{\bf u}-{\bf u}^{*}\Vert_2+O_{p}\left(\tau_{n,p,m}\right)\sum_{j\notin\mathcal{V}_{i}}\vert u_{j}\vert.
\]

Thus, the first part in \eqref{eq_qi_diff} can be bounded by  
\[
-\tilde{{\bf u}}^{\top}\hat{{\bf H}}\hat{{\bf {\bf S}}}\hat{{\bf S}}^{\top}\hat{{\bf H}}\tilde{{\bf u}}+{\bf u}^{\top}\hat{{\bf H}}\hat{{\bf {\bf S}}}\hat{{\bf S}}^{\top}\hat{{\bf H}}{\bf u}\leq O_{p}\left(\tau_{n,p,m}\right)\sum_{j\notin\mathcal{V}_{i}}\vert u_{j}\vert.
\]

Same arguments above for proving $\hat{{\bf q}}_{1}$ in Part 1, we have $G\left(\tilde{{\bf u}}\right)-G\left({\bf u}\right)<0$.
Thus, we have shown that $G\left(\hat{{\bf q}}_{i}^{or}\right)<G\left(\tilde{{\bf u}}\right)<G\left({\bf u}\right)$,
which implies that $\hat{{\bf q}}_{i}^{or}$ is a local minimizer
of the objective function. This completes the proof.

\end{proof}

\subsection{Proof of Theorem 3}
\begin{proof}
    Based on the definition of ${\bf Q}$ and ${\bf S}$, we know that
\begin{align*}
{\bf s}_{1} & ={\bf q}_{1},\\
{\bf s}_{2} & ={\bf q}_{2}-{\bf s}_{1}{\bf s}_{1}^{\top}{\bf q}_{2},\\
 & \vdots\\
{\bf s}_{r} & ={\bf q}_{r}-{\bf s}_{1}{\bf s}_{1}^{\top}{\bf q}_{r}-\cdots-{\bf s}_{r-1}{\bf s}_{r-1}^{\top}{\bf q}_{r},
\end{align*}
and 
\[
{\bf Q=}\left({\bf q}_{1},{\bf q}_{2},\dots,{\bf q}_{r}\right)=\left({\bf s}_{1},{\bf s}_{2},\dots,{\bf s}_{r}\right)\left(\begin{array}{ccccc}
1 & {\bf s}_{1}^{\top}{\bf q}_{2} & \cdots & {\bf s}_{1}^{\top}{\bf q}_{r-1} & {\bf s}_{1}^{\top}{\bf q}_{r}\\
 & 1 & \cdots & {\bf s}_{2}^{\top}{\bf q}_{r-1} & {\bf s}_{2}^{\top}{\bf q}_{r}\\
 &  & \ddots & \vdots & \vdots\\
 &  &  & 1 & {\bf s}_{r-1}^{\top}{\bf q}_{r}\\
 &  &  &  & 1
\end{array}\right)={\bf S}{\bf R},
\]
where ${\bf R}$ is a full rank matrix since all the diagonal elements
are positive. Then, we can also write the model as ${\bf x}_{t}={\bf Q}_{t}{\bf z}_{t}+{\bf \epsilon}_{t}={\bf S}{\bf R}{\bf z_{t}}+\bve_{t}={\bf S}{\bf z}_{t}^{*}+\bve_{t}$. Similarly, we know that $\hat{{\bf Q}}=\tilde{{\bf S}}\hat{{\bf R}}$,
and $\hat{{\bf R}}$ is a full rank matrix. Thus, we have
\begin{align*}
\hat{{\bf Q}}\left(\hat{{\bf Q}}^{\top}\hat{{\bf Q}}\right)^{-1}\hat{{\bf Q}}^{\top} & =\tilde{{\bf S}}\hat{{\bf R}}\left(\hat{{\bf R}}^{\top}\tilde{{\bf S}}^{\top}\tilde{{\bf S}}\hat{{\bf R}}\right)^{-1}\hat{{\bf R}}^{\top}\tilde{{\bf S}}^{\top}\\
 & =\tilde{{\bf S}}\hat{{\bf R}}\hat{{\bf R}}^{-1}\left(\tilde{{\bf S}}^{\top}\tilde{{\bf S}}\right)^{-1}\left(\hat{{\bf R}}^{\top}\right)\hat{{\bf R}}^{-1\top}\tilde{{\bf S}}^{\top} =\tilde{{\bf S}}\tilde{{\bf S}}^{\top}.
\end{align*}

We know that $\hat{{\bf z}}_{t}=\left(\hat{{\bf Q}}^{\top}\hat{{\bf Q}}\right)^{-1}\hat{{\bf Q}}^{\top}{\bf x}_{t}$, then
\begin{align*}
\hat{{\bf Q}}\hat{{\bf z}}_{t}-{\bf A}^{s}{\bf f}_{t}^{s} & =\hat{{\bf Q}}\hat{{\bf z}}_{t}-{\bf {\bf Q}}{\bf z}_{t} =\hat{{\bf Q}}\left(\hat{{\bf Q}}^{\top}\hat{{\bf Q}}\right)^{-1}\hat{{\bf Q}}^{\top}{\bf x}_{t}-{\bf S}{\bf z}_{t}^{*}\\
 & =\tilde{{\bf S}}\tilde{{\bf S}}^{\top}{\bf {\bf S}}{\bf z}_{t}^{*}-{\bf S}{\bf z}_{t}^{*}+\tilde{{\bf S}}\tilde{{\bf S}}^{\top}\bve_{t} =\tilde{{\bf S}}\tilde{{\bf S}}^{\top}{\bf {\bf S}}{\bf z}_{t}^{*}-{\bf S}{\bf z}_{t}^{*}+\tilde{{\bf S}}\tilde{{\bf S}}^{\top}\bve_{t}\\
 & =\left(\tilde{{\bf S}}\tilde{{\bf S}}^{\top}-{\bf S}{\bf S}^{\top}\right){\bf S}{\bf z}_{t}^{*}+\tilde{{\bf S}}\left(\tilde{{\bf S}}^{\top}-{\bf S}^{\top}\right)\bve_{t}+\tilde{{\bf S}}{\bf S}^{\top}\bve_{t}.
\end{align*}

By the same arguments from \cite{lam2011}, we have that $\Vert\tilde{{\bf S}}{\bf S}^{\top}\bve_{t}\Vert_2 = O_p(1)$, 
$\tilde{{\bf S}}\left(\tilde{{\bf S}}^{\top}-{\bf S}^{\top}\right)\bve_{t}$ is dominated by $\Vert\tilde{{\bf S}}{\bf S}^{\top}\bve_{t}\Vert_2$ since $\Vert\tilde{{\bf S}}-{\bf S}\Vert_2 = o_p(1)$, and $\Vert \left(\tilde{{\bf S}}\tilde{{\bf S}}^{\top}-{\bf S}{\bf S}^{\top}\right){\bf S}{\bf z}_{t}^{*}\Vert_2 = O_p(\Vert\tilde{{\bf S}}\tilde{{\bf S}}^{\top}-{\bf S}{\bf S}^{\top}\Vert_2 \cdot \Vert \bz^*_t\Vert_2 )$. Furthermore, $\Vert \bz_t^*\Vert_2 \leq \Vert \bz_t\Vert_2 = O_p(m^{\frac{1-\delta}{2}}),$ which implies that $\Vert \left(\tilde{{\bf S}}\tilde{{\bf S}}^{\top}-{\bf S}{\bf S}^{\top}\right){\bf S}{\bf z}_{t}^{*}\Vert_2 = O_p(m^{\frac{1-\delta}{2}} \cdot \Vert \tilde{{\bf S}}\tilde{{\bf S}}^{\top}-{\bf S}{\bf S}^{\top}\Vert_2).$ Thus, we have
\[
p^{-1/2}\Vert\hat{{\bf Q}}\hat{{\bf z}}_{t}-{\bf A}^{s}{\bf f}_{t}^{s}\Vert_{2}=O_{p}\left(p^{-1/2}m^{1/2-\delta/2}\Vert\hat{{\bf Q}}-{\bf Q}\Vert_{2}+p^{-1/2}\right).
\]
\end{proof}

\section{Proof of Lemmas}
\label{appendix_lemma}

\subsection{Lemma \ref{lemma::Sigmax_norm}}


\begin{lemma}
    \label{lemma::Sigmax_norm}  
Under Conditions (C\ref{cond_alphamix})-(C\ref{cond_strength}),
\[
\Vert\bSigma_x\left(h\right)\Vert_{1}=\Vert\bSigma_x\left(h\right)\Vert_{\infty}=O\left(m^{1-\delta/2}\right).
\]
\end{lemma}

\begin{proof}
Let $a_{il}$ be the $(i,l)$th element of $\bA^s$ and $\sigma_{f,ll^\prime}$ be the $(l,l^\prime)$th element of $\bSigma_f^*(h)$.  We know that $\bSigma_x\left(h\right)=\bA^s\bSigma_{f}^s\left(h\right)\bA^{s\top}$. Then, the $(i,j)$th element of $\bSigma_x\left(h\right)$ is $\bSigma_{x,ij}\left(h\right)=\sum_{l=1}^{r}\sum_{l^{\prime}=1}^{r}a_{il}^s\sigma_{f,ll^{\prime}}a_{jl^{\prime}}$.
Thus 
\begin{align*}
\Vert\bSigma_x\left(h\right)\Vert_{1}=\max_{j}\sum_{i=1}^{p}\vert\bSigma_{x,ij}\left(h\right)\vert & =\max_{j}\sum_{i=1}^{p}\vert\sum_{l=1}^{r}\sum_{l^{\prime}=1}^{r}a_{il}\sigma_{f,ll^{\prime}}a_{jl^{\prime}}\vert\\
 & \leq\max_{j}\sum_{l=1}^{r}\sum_{l^{\prime}=1}^{r}\vert\sigma_{f,ll^{\prime}}a_{jl^{\prime}}\vert\sum_{i=1}^{p}\vert a_{il}\vert.
\end{align*}

Based Cauchy-Schwartz inequality and the sparsity of $\bA^s$ in (C\ref{cond_strength}), 
$\sum_{i=1}^{p}\vert a_{il}\vert\leq\sqrt{m}\sqrt{\Vert \ba_{l}^s\Vert_2^{2}}\asymp\sqrt{m\times m^{1-\delta}}=m^{1-\delta/2}.$
Thus, $\Vert\bSigma_x\left(h\right)\Vert_{1}\asymp r^{2}m^{1-\delta/2}\asymp m^{1-\delta/2}$.

 Since $\bSigma_x(h)$ is symmetric, we have $\Vert\bSigma_x\left(h\right)\Vert_{\infty} = \Vert\bSigma_x\left(h\right)\Vert_{1}$.
\end{proof}

\subsection{Lemma \ref{lemma::maxSigmax}}

\begin{lemma}
    \label{lemma::maxSigmax}
Under Conditions (C\ref{cond_alphamix})-(C\ref{cond_strength}), (C\ref{cond_subgaussian}), and $\log p = o(n)$, it holds that 
\[
\Vert\hat{{\bf \Sigma}}_{x}\left(h\right)-{\bf \Sigma}_{x}\left(h\right)\Vert_{\max}=\max_{1\leq i,j\leq p}\vert\hat{{\bf \Sigma}}_{x,ij}\left(h\right)-{\bf \Sigma}_{x,ij}\left(h\right)\vert=O_{p}\left(\sqrt{\frac{\log p}{n}}\right),
\]
where $\hat{{\bf \Sigma}}_{x,ij}(h)$ and ${\bf \Sigma}_{x,ij}\left(h\right)$ is the $(i,j)$th element of $\hat{\bSigma}_x(h)$ and $\bSigma_x(h)$, respectively.
\end{lemma}

Lemma \ref{lemma::maxSigmax} gives an element-wise bound for the estimate of $\hat{\bSigma}_x(h)$, which plays an important role in the proof of our analysis when $p$ goes to infinity.

\begin{proof}
Define
\[
{\bSigma}_f^s(h)=\frac{1}{n-h}\sum_{t=1}^{n-h}{\rm E}(\bff_t^s \bff_t^{s\top}), \quad \hat{\bSigma}_f^s(h)=\frac{1}{n-h}\sum_{t=1}^{n-h}(\bff_t^s \bff_t^{s\top}).
\]
Based on the definition, we have 
\begin{eqnarray*}
\lefteqn{\hat{\bSigma}_x(h)- \bSigma_x(h)}\\
&=& \bA^s[\hat{\bSigma}_f^s(h)- \bSigma_f^s(h)]\bA^{s\top} +\frac{1}{n}\sum_{t=1}^{n-h}\bA^s \bff_t^s \bve_{t+h}^\top +\frac{1}{n}\sum_{t=1}^{n-h}\bve_{t} \bff_{t+h}^{s\top}\bA^{s\top} +\frac{1}{n}\sum_{t=1}^{n-h}\bve_t \bve_{t+h}^\top\\
&=& I_1 +I_2 +I_3 +I_4.
\end{eqnarray*}

We will bound $I_{1},I_{2},I_{3}$ and $I_{4}$, respectively for each element. 

\paragraph{bound for $I_1$.}
Let $a_{il}$ be the $(i,l)$th element of $\bA^s$ for simplicity, $\hat{\sigma}_{f,ll^\prime}$ and $\sigma_{f,ll^\prime}$ be the $(l,l^{s\prime})$th element of $\hat{\bSigma}_f^s(h)$ and $\bSigma_f^s(h)$, respectively.  Then the $ij$th element of $I_{1}$ is $\sum_{l^{\prime}=1}^{r}\sum_{l=1}^{r}a_{il}\left(\hat{\sigma}_{f,ll^{\prime}}-\sigma_{f,ll^{\prime}}\right)a_{jl^{\prime}}$.

With Condition (C\ref{cond_strength}), we know that 
\[
\vert\sum_{l^{\prime}=1}^{r}\sum_{l=1}^{r}b_{il}\left(\hat{\sigma}_{f,ll^{\prime}}-\sigma_{f,ll^{\prime}}\right)b_{jl^{\prime}}\vert\leq C_{1}^{2}\sum_{l^{\prime}=1}^{r}\sum_{l=1}^{r}\vert\hat{\sigma}_{f,ll^{\prime}}-\sigma_{f,ll^{\prime}}\vert,
\]
thus from Lemma \ref{lemma::sig_f},
\begin{align*}
\Vert I_1\Vert_{\max} & =\max_{1\leq i,j\leq p}\vert\sum_{l^{\prime}=1}^{r}\sum_{l=1}^{r}a_{il}\left(\hat{\sigma}_{f,ll^{\prime}}-\sigma_{f,ll^{\prime}}\right)a_{jl^{\prime}}\vert \leq C_{1}^{2}\sum_{l^{\prime}=1}^{r}\sum_{l=1}^{r}\vert\hat{\sigma}_{f,ll^{\prime}}-\sigma_{f,ll^{\prime}}\vert =O_{p}\left(n^{-1/2}\right).
\end{align*}

\paragraph{bound for $I_2$.} Let $f_{t,l}$ be the $l$-th element in $\bff_t^s$ for simplicity. The $(i,j)$the element of $I_{2}$ is $\frac{1}{n}\sum_{t=1}^{n-h}\sum_{l=1}^{r}a_{il}f_{t,l}\epsilon_{t+h,j}=\sum_{l=1}^{r}a_{il}\frac{1}{n}\sum_{t=1}^{n-h}f_{t,l}\epsilon_{t+h,j}$. Thus,
\[
\Vert I_2\Vert_{\max}=\max_{i,j}\vert\sum_{l=1}^{r}a_{il}\frac{1}{n}\sum_{t=1}^{n-h}f_{t,l}\epsilon_{t+h,j}\vert\leq C_{1}\sum_{l=1}^{r} \max_{j}\vert\frac{1}{n}\sum_{t=1}^{n-h}f_{t,l}\epsilon_{t+h,j}\vert.
\]

Next, we will study the order of $\vert\frac{1}{n}\sum_{t=1}^{n-h}f_{t,l}\epsilon_{t+h,j}\vert$. Let $\sigma^2_{ff} = E(f_{t,l}^2)$ and $\sigma^2_e = E(\epsilon_{t,j}^2)$. 

It is known that $4f_{t,l}\epsilon_{t+h,j}=\left[\left(f_{t,l}+\epsilon_{t+h,j}\right)^{2}-\left(\sigma_{ff}^{2}+\sigma_{e}^{2}\right)\right]-\left[\left(f_{t,l}-\epsilon_{t+h,j}\right)^{2}-\left(\sigma_{ff}^{2}+\sigma_{e}^{2}\right)\right]$. Let $\tilde{x}_{t}=f_{t,l}+\epsilon_{t+h,j}$, and $\tilde{\bf{x}}=\left(\tilde{x}_{1},\dots,\tilde{x}_{n-h}\right)^{\top},$
we have $V\left(\tilde{\bf{x}}\right)=\bf{V}$. Then, \\ $\frac{1}{n}\sum_{t=1}^{n-h}\left[\left(f_{t,l}+\epsilon_{t+h,j}\right)^{2}-\left(\sigma_{ff}^{2}+\sigma_{e}^{2}\right)\right]$ can be written as $\frac{1}{n}\bz^{\top}\bV\bz$, where $\bz$ has independent random variables with mean 0 and variance 1. Next, we will consider the upper
bound of $\Vert\bV\Vert_2$. Let 
\[
\sigma_{tt^{\prime}}=Cov(\tilde{x}_{t},\tilde{x}_{t^{\prime}})=Cov\left(f_{t,l}+\epsilon_{t+h,j},f_{t^{\prime},l}+\epsilon_{t^{\prime}+h,j}\right)=\begin{cases}
\sigma_{ff}^{2}+\sigma_{e}^{2} & t=t^{\prime},\\
\sigma_{f,tt^{\prime}} & t\neq t^{\prime},
\end{cases}
\]
where $\sigma_{f,tt^{\prime}}=Cov(f_{t,l},f_{t^{\prime}l})=E\left(f_{t,l}f_{t^{\prime},l}\right)-Ef_{t,l}Ef_{t^{\prime}l}$.
From \cite{roussas1987moment}, we have 
\[
\vert\sigma_{f,tt^{\prime}}\vert\leq10\alpha\left(\vert t-t^{\prime}\vert\right)^{1/2}\left[E\left(f_{t,l}^{4}\right)\right]^{1/4}\left[E\left(f_{t^{\prime},l}^{4}\right)\right]^{1/4}\leq10\alpha\left(\vert t-t^{\prime}\vert\right)^{1/2}\sigma_{f}^{2},
\]
since $E\left(f_{t,l}^{4}\right)\leq\sigma_{f}^{4}$.

For $\Vert\bw\Vert=1$, consider 
\begin{align*}
\bf{w}^{\top}\bf{V}\bf{w} & =\sum_{t=1}^{n-h}\sum_{t^{\prime}=1}^{n-h}w_{t}\sigma_{tt^{\prime}}w_{t^{\prime}}=\sum_{t=1}^{n-h}w_{t}^{2}\left(\sigma_{ff}^{2}+\sigma_{e}^{2}\right)+2\sum_{1\leq t<t^{\prime}\leq n-h}w_{t}w_{t^{\prime}}\sigma_{f,tt^{\prime}}w\\
 & =\sigma_{ff}^{2}+\sigma_{e}^{2}+2\sum_{s=1}\sum_{t^{\prime}=t+s}w_{t}w_{t^{\prime}}\sigma_{f,s}=\sigma_{ff}^{2}+\sigma_{e}^{2}+2\sum_{s=1}\sigma_{f,s}\sum_{t=1}w_{t}w_{t+s}\\
 & =\sigma_{ff}^{2}+\sigma_{e}^{2}+2\sum_{s=1}\sigma_{f,s}\sqrt{\sum_{t}w_{t}^{2}}\sqrt{\sum_{t=1}w_{t+s}^{2}}\\
 & \leq\sigma_{ff}^{2}+\sigma_{e}^{2}+2\sum_{s=1}10\alpha\left(s\right)^{1/2}\sigma_{f}^{2} =\sigma_{ff}^{2}+\sigma_{e}^{2}+20\sigma_{f}^{2}\sum_{s=1}\alpha\left(s\right)^{1/2},
\end{align*}
where $\sigma_{f,s}=\sigma_{f,tt^{\prime}}$. Based on the assumption
of $\alpha$-mixing, we have $\Vert\bV\Vert_2$ is bounded. Thus,

\begin{equation}
\label{eq_I2_p1}
    P\left(\vert\frac{1}{n}\sum_{t=1}^{n-h}\left[\left(f_{t,l}+\epsilon_{t+h,j}\right)^{2}-\left(\sigma_{ff}^{2}+\sigma_{e}^{2}\right)\right]\vert>c_{0}\rho\sqrt{\frac{\log p}{n}}\right)\leq\frac{2}{p^{\tilde{c}}},
\end{equation}
based on Lemma \ref{lemma_inequal}. 

Similarly, we can show that 
\begin{equation}
\label{eq_I2_p2}
 P\left(\vert\frac{1}{n}\sum_{t=1}^{n-h}\left[\left(f_{t,l}-\epsilon_{t+h,j}\right)^{2}-\left(\sigma_{ff}^{2}+\sigma_{e}^{2}\right)\right]\vert>c_{0}\rho\sqrt{\frac{\log p}{n}}\right)\leq\frac{2}{p^{\tilde{c}}}.   
\end{equation}

Combine \eqref{eq_I2_p1} and \eqref{eq_I2_p2}, we have
\begin{align*}
 & P\left(4\vert\frac{1}{n}\sum_{t=1}^{n-h}f_{t,l}\epsilon_{t+h,j}\vert>2c_{0}\rho\sqrt{\frac{\log p}{n}}\right)\\
\leq & P\left(\vert\frac{1}{n}\sum_{t=1}^{n-h}\left[\left(f_{t,l}+\epsilon_{t+h,j}\right)^{2}-\left(\sigma_{f}^{2}+\sigma_{e}^{2}\right)\right]\vert >c_{0}\rho\sqrt{\frac{\log p}{n}}\right)\\
 & +P\left(\vert\frac{1}{n}\sum_{t=1}^{n-h}\left[\left(f_{t,l}-\epsilon_{t+h,j}\right)^{2}-\left(\sigma_{f}^{2}+\sigma_{e}^{2}\right)\right]\vert>c_{0}\rho\sqrt{\frac{\log p}{n}}\right)
\leq  \frac{4}{p^{\tilde{c}}}.
\end{align*}

Thus 
\begin{align*}
 & P\left(\max_{j}\sum_{l=1}^{r}\vert\frac{4}{n}\sum_{t=1}^{n-h}f_{t,l}\epsilon_{t+h,j}\vert>2c_{0}\rho\sqrt{\frac{\log p}{n}}\right)\\
\leq & \sum_{j=1}^{p}\sum_{l=1}^{r}P\left(\vert\frac{4}{n}\sum_{t=1}^{n-h}f_{t,l}\epsilon_{t+h,j}\vert>2c_{0}\rho\sqrt{\frac{\log p}{n}}\right)
\leq  pr\frac{4}{p^{\tilde{c}}}=\frac{4r}{p^{\tilde{c}-1}}.
\end{align*}

This implies that $\Vert I_2\Vert_{\max} = O_p \left( \sqrt{\frac{\log p}{n}}\right)$.

Similarly it can be showed for $\Vert I_3\Vert_{\max} =O_p \left( \sqrt{\frac{\log p}{n}}\right)$.

\paragraph{bound for $I_4$.} The $(i,j)$the element of $I_4$ is $\frac{1}{n}\sum_{t=1}^{n-h}\epsilon_{t,i}\epsilon_{t+h,j}$.

If $i=j$, then we the element is $\frac{1}{n}\sum_{t=1}^{n-h}\epsilon_{t,i}\epsilon_{t+h,i}=\frac{1}{n}\bve_{i}^{\top}{\bV}\bve_{i}$, where $\bve_i=\left(\epsilon_{1,i},\epsilon_{2,i},\dots,\epsilon_{n,i}\right)^{\top}$. Based on the assumption, we know that $E\left(\frac{1}{n}\sum_{t=1}^{n-h}\epsilon_{t,i}\epsilon_{t+h,i}\right)=0$ and  $\Vert\bV\Vert_2=1$. Thus, based on Lemma \ref{lemma_inequal}, we have
\begin{align*}
P\left(\vert\frac{1}{n}\sum_{t=1}^{n-h}\epsilon_{t,i}\epsilon_{t+h,i}\vert>c_{0}\rho\sqrt{\frac{\log p}{n}}\right) & \leq\frac{2}{p^{\tilde{c}}}.
\end{align*}

If $i\neq j$, $\frac{1}{n}\sum_{t=1}^{n-h}\epsilon_{t,i}\epsilon_{t+h,j}=\frac{1}{4}\left[\frac{1}{n}\sum_{t=1}^{n-h}\left(\epsilon_{t,i}+\epsilon_{t+h,j}\right)^{2}-2\sigma_{e}^{2}+\left(\frac{1}{n}\sum_{t=1}^{n-h}\left(\epsilon_{t,i}-\epsilon_{t+h,j}\right)^{2}\right)-2\sigma_{e}^{2}\right]$.

Let $\tilde{\epsilon}_{t}=\epsilon_{t,i}+\epsilon_{t+h,j},$ and $\frac{1}{n}\sum_{t=1}^{n-h}\left(\epsilon_{t,i}+\epsilon_{t+h,j}\right)^{2}=\frac{1}{n}\sum_{t=1}^{n-h}\tilde{\epsilon}_{t}^{2}$.
And $V\left(\tilde{\bf{\epsilon}}_{t}\right)$ is a $\left(n-h\right)\times\left(n-h\right)$ matrix,
with diagonal elements $2\sigma_{e}^{2}$ and off $\sigma_{e,i,j}$
($2\times(n-2h)$) elements. It can be written as $\frac{1}{n}\bz^{\top}\bV\bz$, where
$\bf{z}$ has independent random variables with mean 0 and variance 1. Now consider the upper bound
of $\bf{V}$. For $\Vert\bw\Vert=1$,
\[
\bw^{\top}\bV\bw\leq2\sigma_{e}^{2}\sum_{t=1}w_{t}^{2}+2\sigma_{e,ij}\sum_{t=1}^{n-h}w_{t}w_{t+h}\leq2\sigma_{e}^{2}+2\vert\sigma_{e,ij}\vert\sqrt{\sum w_{t}^{2}}\sqrt{\sum w_{t+h}^{2}}\leq2\sigma_{e}^{2}+2\vert\sigma_{e,ij}\vert,
\]
which indicates that $\Vert\bV\Vert_2$ is bounded. Thus,
\[
P\left(\vert\frac{1}{n}\sum_{t=1}^{n-h}\left(\epsilon_{t,i}+\epsilon_{t+h,j}\right)^{2}-2\sigma_{e}^{2}\vert>c_{0}\rho\sqrt{\frac{\log p}{n}}\right)\leq\frac{2}{p^{\tilde{c}}}.
\]
Similarly, 
\[
P\left(\vert\frac{1}{n}\sum_{t=1}^{n-h}\left(\epsilon_{t,i}-\epsilon_{t+h,j}\right)^{2}-2\sigma_{e}^{2}\vert>c_{0}\rho\sqrt{\frac{\log p}{n}}\right)\leq\frac{2}{p^{\tilde{c}}}.
\]
These two imply that 
\begin{align*}
 & P\left(\vert\frac{4}{n}\sum_{t=1}^{n-h}\epsilon_{t,i}\epsilon_{t+h,j}\vert>2c_{0}\rho\sqrt{\frac{\log p}{n}}\right)
\leq  \frac{4}{p^{\tilde{c}}}.
\end{align*}

All implies that 
\begin{align*}
 & P\left(\Vert I_4\Vert _{\max}>\frac{1}{2}c_{0}\rho\sqrt{\frac{\log p}{n}}\right)
\leq  \sum_{i=1}^{p}\sum_{j=1}^{p}P\left(\vert I_{4,i,j}\vert> \frac{1}{2}c_{0}\rho\sqrt{\frac{\log p}{n}}\right)\\
= & \sum_{i=1}^{p}\sum_{j=1}^{p}P\left(\vert\frac{4}{n}\sum_{t=1}^{n-h}\epsilon_{t,i}\epsilon_{t+h,j}\vert>2c_{0}\rho\sqrt{\frac{\log p}{n}}\right)
\leq  p^{2}\times\frac{4}{p^{\tilde{c}}}=\frac{4}{p^{\tilde{c}-2}}.
\end{align*}

Combined the results for $I_1$, $I_2$, $I_3$ and $I_4$, we have 
\[
\Vert \hat{{\bf \Sigma}}_{y,ij}\left(h\right)-{\bf \Sigma}_{y,ij}\left(h\right)\Vert_{\max} = \max_{1\leq i,j\leq p}\vert\hat{{\bf \Sigma}}_{y,ij}\left(h\right)-{\bf \Sigma}_{y,ij}\left(h\right)\vert=O_{p}\left(\sqrt{\frac{\log p}{n}}\right).
\]
\end{proof}

\subsection{Lemma \ref{lemma_Sstar}}

\begin{lemma}
    \label{lemma_Sstar}
    Consider an index $\mathcal{A}$ with size $\vert\mathcal{A}\vert \leq m$,  $\hat{\bS}^{*}$ and $\bS^{*}$ are subrows of $\hat{\bS}$ and $\bS$ for
index $\mathcal{A}$, respectively. Under Conditions (C\ref{cond_alphamix}) - (C\ref{cond_coherence}),

\[
\Vert\hat{{\bf S}}^{*}\hat{{\bf S}}^{*\top}-{\bf S}^{*}{\bf S}^{*\top}\Vert_{2}=\begin{cases}
O_{p}\left(\max\left(m^{2\delta-2}p^{2}n^{-1/2},m^{\delta}\right)\sqrt{\frac{\log p}{n}}\right) & \text{if } m = o(p), \\
O_p \left(m^{\delta-1}pn^{-1/2}\right)  = O_p \left(p^{\delta}n^{-1/2}\right)& \text{if } m=O(p).\\
\end{cases}
\]

\end{lemma}

\begin{proof}
If $m = o(p)$, from the results in \cite{cape2019two} about the relationship between different norms,  we have
\begin{align*}
\Vert\hat{\bS}^{*}\hat{\bS}^{*\top}-\bS^{*}\bS^{*\top}\Vert_{2} & =  \Vert\hat{\bS}^{*}\hat{\bS}^{*\top}-\bS^{*}\hat{\bS}^{*\top}+\bS^{*}\hat{\bS}^{*\top}-\bS^{*}\bS^{*}\Vert_2\\
 & \leq\Vert\hat{\bS}^{*}-{\bS}^{*}\Vert_{2}\Vert\hat{\bS}\Vert_{2}+\Vert{\bS}^{*}\Vert_{2}\Vert\hat{\bS}^{*}-{\bS}^{*}\Vert_{2} \leq2\Vert\hat{\bS}^{*}-\bS^{*}\Vert_{2}\\
 & \leq 2\sqrt{\vert\mathcal{A}\vert}\Vert\hat{\bS}^{*}-{\bS}^{*}\Vert_{2\rightarrow\infty} \leq 2 \sqrt{\vert\mathcal{A}\vert}\Vert\hat{\bS}-{\bS}\Vert_{2\rightarrow\infty}.
\end{align*}

From Theorem 4.2 in \cite{cape2019two}, we also know that,
\begin{align*}
\Vert\hat{{\bf S}}^{*}\hat{{\bf S}}^{*\top}-{\bf S}^{*}{\bf S}^{*\top}\Vert_{2} & \leq2\sqrt{m}\Vert\hat{{\bf S}}-{\bf S}\Vert_{2\rightarrow\infty} \leq28\sqrt{m}\left(\frac{\Vert\hat{{\bf M}}-{\bf M}\Vert_{\infty}}{\lambda_{r}}\right)\Vert{\bf S}\Vert_{2\rightarrow\infty}.
\end{align*}

From Lemma \ref{lemma_Minfty} and Lemma \ref{lemma::M}, we know that $\Vert\hat{{\bf M}}-{\bf M}\Vert_{\infty}=O_{p}\left(\max\left(p^{2}n^{-1/2},m^{2-\delta}\right)\sqrt{\frac{\log p}{n}}\right)$ and $\lambda_r = O_p(m^{2-2\delta})$ 
Then
\begin{align*}
\Vert\hat{{\bf S}}^{*}\hat{{\bf S}}^{*\top}-{\bf S}^{*}{\bf S}^{*\top}\Vert_{2} & =O_{p}\left(\frac{\max\left(p^{2}n^{-1/2},m^{2-\delta}\right)\sqrt{\frac{\log p}{n}}}{m^{2-2\delta}}\right)\\
 & =O_{p}\left(\max\left(m^{2\delta-2}p^{2}n^{-1/2},m^{\delta}\right)\sqrt{\frac{\log p}{n}}\right) 
\end{align*}

If $m=O(p)$,  we have 
\begin{align*}
\Vert\hat{\bS}^{*}\hat{\bS}^{*\top}-\bS^{*}\bS^{*\top}\Vert_{2} & =\sqrt{\lambda_{\max}\left[\left(\hat{\bS}^{*}\hat{\bS}^{*\top}-\bS^{*}\bS^{*\top}\right)^{\top}\left(\hat{\bS}^{*}\hat{\bS}^{*\top}-\bS^{*}\bS^{*\top}\right)\right]}\\
 & \leq\sqrt{\text{tr}\left[\left(\hat{\bS}^{*}\hat{\bS}^{*\top}-\bS^{*}\bS^{*\top}\right)^{\top}\left(\hat{\bS}^{*}\hat{\bS}^{*\top}-\bS^{*}\bS^{*\top}\right)\right]}\\
 & =\Vert\hat{\bS}^{*}\hat{\bS}^{*\top}-\bS^{*}\bS^{*\top}\Vert_{F}\leq\Vert\hat{\bS}\hat{\bS}^{\top}-\bS^ {}\bS^{\top}\Vert_{F} \\
 & = O_p(m^{\delta-1}pn^{-1/2}),
\end{align*}
from Theorem 1.
\end{proof}

\section{Technical Lemmas and their proof}
\label{appendix_lemma_tech}

\setcounter{lemma}{0}
\renewcommand{\thelemma}{\Alph{section}.\arabic{lemma}}
\begin{lemma}
    Suppose $\bH$ is a $p\times p$ symmetric matrix, maximizing  $\by^\top \bH \by$ with respect to $\by$ is equivalent to following problem 
    \begin{equation}
        \min_{\by: \Vert \by\Vert_2  = 1}\Vert \bH - \by \by^\top\Vert_F^2,
    \end{equation}
    where $\by$ is an $p\times 1$ vector. 
    \label{lemma_obj}
\end{lemma}

\begin{proof}
    It is known that $\Vert \bH - \by \by^\top \Vert_F^2 = \tr\left( (\bH - \by \by^\top)^\top (\bH - \by \by^\top)\right)$. And we have 
    \begin{align*}
        \tr\left( (\bH - \by \by^\top)^\top (\bH - \by \by^\top)\right) &= \tr(\bH\bH)-2\tr(\by^T\bH \by) + \tr(\by\by^\top) \\
        & = \tr(\bH\bH)-2\tr(\by^T\bH \by) + \tr(\by^\top\by) \\
        & = \tr(\bH\bH)-2\tr(\by^\top\bH \by) + 1.
    \end{align*}
    Thus minimizing $\Vert \bH - \by \by^\top \Vert_F^2$ is equivalent to maximizing $\by^\top \bH \by$ with respect to $\Vert \by \Vert_2 =1$.
\end{proof}

\begin{lemma}
Let $\bS_1$ and $\bS_2$ be two orthonormal matrix, then maximizing $\tr(\bS_1 \bS_1^\top \bS_2\bS_2^\top )$ is equivalent to minimizing $\sum_{i=1}^r\Vert \bS_1\bS_1^\top - \bs_{2i}\bs_{2i}^\top\Vert_F^2$, where $\bs_{2i}$ is the $i$th column of $\bS_2$.  
    \label{lemma_obj0}
\end{lemma}

\begin{proof}
    From the definition, we know that 
    \begin{align*}
\tr(\bS_1 \bS_1^\top \bS_2 \bS_2^\top)= \|\bS_1^\top \bS_2\|_F^2= \sum_{i=1}^r \|\bS_1^\top \bs_{2i}\|_2^2  = \sum_{i=1}^r \bs_{2i}^\top \bS_1\bS_1^\top \bs_{2i}.
\end{align*}

From proof in Lemma \ref{lemma_obj}, $\bs_{2i}^\top \bS_1\bS_1^\top \bs_{2i} = \frac{1}{2}\left( r+1 - \Vert \bS_1\bS_1^\top - \bs_{2i}\bs_{2i}^\top \Vert_F^2\right)$. Thus, we have 
\begin{equation*}
    \tr(\bS_1 \bS_1^\top \bS_2 \bS_2^\top) = \frac{1}{2}r(r+1) -\frac{1}{2}\sum_{i=1}^r \Vert\bS_1 \bS_1^\top - \bs_{2i}\bs_{2i}^\top \Vert_F^2.
\end{equation*}
Thus, maximizing $\tr(\bS_1 \bS_1^\top \bS_2\bS_2^\top )$ is equivalent to minimizing $\sum_{i=1}^r\Vert \bS_1\bS_1^\top - \bs_{2i}\bs_{2i}^\top\Vert_F^2$.
\end{proof}

\begin{lemma}
Let ${\bf L}$ be a $k\times k$ matrix with a fixed rank $r^{*}$, which has distinct eigenvalues with $1=\lambda_{1}>\lambda_{2}>\cdots>\lambda_{r^{*}}>0$.
\label{lemma_eigenvalue}
Let ${\bf u}_1$ be a $k\times1$ vector such that ${\bf L}{\bf u}_1={\bf u}_1$.
And ${\bf L}+{\bf E}$ is a $k\times k$ symmetric matrix with $\Vert \bE\Vert_2 = o_p(1)$, $\hat{{\bf u}}_1$
is the maximal eigenvector of ${\bf L+{\bf E}}$ that is ${\bf \hat{u}}_1=\arg\max_{\Vert{\bf u}\Vert_2=1}\Vert({\bf L} + {\bf E}){\bf u}\Vert_2$,
then, we have 
\begin{equation*}
 \Vert{\bf u}_1-\hat{{\bf u}}_1\Vert_2=O_{p}\left(\Vert{\bf E}\Vert_2\right).   
\end{equation*}  
\end{lemma}

\begin{proof}

We will prove the result in two cases. 

\paragraph{Case 1: $r^* = k$\\}

Let$\{\bu_{1},\ldots,\bu_{k}\}$ as the orthogonal basis of $\mathbb{R}^{m}$,
and ${\bf u}\in\mathbb{R}^{k}$ with $\Vert{\bf u}\Vert=1$ can be
written as $\bu=\sum_{i=1}^{k}a_{j}\bu_{j},$ where $\sum_{j=1}^{k}a_{j}^{2}=1$. Let $G(\bu) = \Vert({\bf L} + {\bf E}){\bf u}\Vert_2$. 
Since $\bu_{j}$'s are eigenvectors of ${\bf L}$, then we have
\begin{align*}
G(\bu)&= \| \sum_{i=1}^k a_j \lambda_j \bu_j + \sum_{j=1}^k a_j \bE\bu_j\|_2 \geq \| \sum_{i=1}^k a_j \lambda_j \bu_j\|_2 -  \sum_{j=1}^k |a_j| \cdot \|\bE\|_2\\
&\geq \left( \sum_{i=1}^k a_j^2 \lambda_j^2\right)^{1/2} -\sqrt{k}\cdot \|\bE\|_2.
\end{align*}
On the other hand,
\begin{align*}
G(\bu)&= \| \sum_{i=1}^k a_j \lambda_j \bu_j + \sum_{j=1}^k a_j \bE\bu_j\|_2 \leq  \| \sum_{i=1}^k a_j \lambda_j \bu_j\|_2 +  \sum_{j=1}^k |a_j| \cdot \|\bE\|_2\\
&\leq  \left(\sum_{i=1}^k a_j^2 \lambda_j ^2\right)^{1/2}+\sqrt{k} \cdot \|\bE\|_2.
\end{align*}
Then, 
\[
G(\bu)= \left(\sum_{i=1}^k a_j^2 \lambda_j^2\right)^{1/2}+O_p(\Vert \bE\Vert_2 ), \mbox{ for } \|\bu\|_2=1.
\]
It follows
\[
G(\hat{\bu}_1)= \max_{\|\bu\|_2=1} G(\bu ) = \lambda_1 + O_p(\Vert \bE\Vert_2),
\]
where $\hat{\bu}_1$ is the maximal eigenvector of $\bL+\bE$.

Since $G(\bu_1)=\lambda_1$, we have
\[
\hat{\bu}_1= \bu_1 +O_p(\Vert \bE\Vert_2).
\]

\paragraph{Case 2: $r^* < k$\\}

Under this case, we can write $\bL$ in the following form
\[
\left(\begin{array}{c}
{\bf u}_{1}^{\top}\\
{\bf B}^{\top}
\end{array}\right){\bf L}\left(\begin{array}{cc}
{\bf u}_{1} & {\bf B}\end{array}\right)=\left(\begin{array}{cccc}
1\\
 & \lambda_{2}\\
 &  & \ddots\\
 &  &  & \lambda_{r^{*}}
\end{array}\right),
\]
where $({\bf u}_{1} \quad {\bf B})$ is a orthonormal matrix, ${\bf B}$ is a $k\times\left(r_{*}-1\right)$
matrix. 

We can write ${\bf B}=\left({\bf B}_{1},{\bf B}_{2}\right)$, where
the $\lambda_{\min}\left({\bf B}_{1}\right)=O_{p}\left(1\right)$
and $\lambda_{\max}\left({\bf B}_{2}\right)=o_{p}\left(1\right)$.
A special case is that ${\bf B}_{1}={\bf B}$. Let's define ${\bf C}$, an $k\times\left(k-r^{*}\right)$
matrix, as an orthogonal complement of $\left({\bf u}_{1},{\bf B}\right)$,
such that $\bL{\bf C}=\bzero$. Then, we have 
\begin{align*}
\left(\begin{array}{c}
{\bf X}_{1}^{\top}\\
{\bf X}_{2}^{\top}
\end{array}\right){\bf L}\left(\begin{array}{cc}
{\bf X}_{1} & {\bf X}_{2}\end{array}\right) & =\left(\begin{array}{cc}
{\bf D}_{1} & 0\\
0 & {\bf D_{2}}
\end{array}\right)
\end{align*}
where ${\bf X}_{1}=\left({\bf u}_{1},{\bf B}_{1}\right)$ and
${\bf X}_{2}=\left({\bf B}_{2},{\bf C}\right)$, with $\lambda_{\min}\left({\bf D}_{1}\right)=O_{p}\left(1\right)$
and $\lambda_{\max}\left({\bf D}_{2}\right)=o_{p}\left(1\right)$.
Then $\text{sep}\left({\bf D}_{1},{\bf D}_{2}\right)=O_{p}\left(1\right)$.

Since $\Vert \bE\Vert_2 =o_p(1)$, thus, $\Vert{\bf E}\Vert_2\leq\text{sep}\left({\bf D}_{1},{\bf D}_{2}\right)/5$. Based on the result in Lemma 3 in \cite{lam2011}, there exists a matrix ${\bf P}$ such that 
\[
\Vert{\bf P}\Vert_2\leq\frac{4}{\text{sep}\left({\bf D}_{1},{\bf D}_{2}\right)}\Vert{\bf E}\Vert_2,
\]
and $\hat{\bX}_1 = (\bX_1 + \bC\bP)(\bI + \bP^\top \bP)^{-1/2}$ is an estimator for $\bX_1$. Then, we have
\begin{align*}
\Vert\hat{{\bf X}}_{1}-{\bf X}_{1}\Vert_2 & \leq\Vert[{\bf X}_{1}({\bf I}-({\bf I}+{\bf P}^{\top}{\bf P})^{1/2})+{\bf B}{\bf P}]({\bf I}+{\bf P}^{\top}{\bf P})^{-1/2}\Vert_2\\
 & \leq\Vert{\bf I}-({\bf I}+{\bf P}^{\top}{\bf P})^{1/2}\Vert_2+\Vert{\bf P}\Vert_2\leq2\Vert{\bf P}\Vert_2 = O_p(\Vert \bE\Vert_2).
\end{align*}
Thus, this holds for each component, that is $\Vert \hat{\bu}  - \bu_1\Vert_2 = O_p(\Vert \bE\Vert_2)$. 

These two cases complete the proof. 

\end{proof}

\begin{lemma} Under Conditions (C\ref{cond_alphamix})-(C\ref{cond_fbound}), it holds that \label{lemma::sig_f}
\begin{align*}
{\rm E}\| \hat{\bSigma}_f^s(h) -\bSigma_f^s(h)\|_2^2 \asymp {\rm E} \| \hat{\bSigma}_f^s(h) -\bSigma_f^s(h)\|_F^2 =O(n^{-1}), \quad
\|\bSigma_f^s(h)\|_2 =O(1).
\end{align*}
\end{lemma}

\begin{proof}
By Condition (C\ref{cond_fbound}) and Jensen's inequality we have that ${\rm E}(f_{t,i}^2) < \sigma_f^2$ and ${\rm E}(f_{t,i}^4) < \sigma_f^4$, for $i=1,\ldots, r$ and $t=1, \ldots, n$. Let $\sigma_{f,ij}(h)$ and $\hat{\sigma}_{f,ij}(h)$ be the $(i,j)$-th elements in $\bSigma_f^s(h)$ and $\bSigma_f^s(h)$, respectively. By Cauchy-Schwarts inequality,
\begin{align*}
|\sigma_{f,ij}(h)|^2 =\Big| \frac{1}{n-h} \sum_{t=1}^{n-h} {\rm E}(f_{t,i}f_{t+h,j})\Big|^2 \leq \Bigg| \frac{1}{n} \sum_{t=1}^{n-h} \sqrt{{\rm E} (f_{t,i}^2) {\rm E}( f_{t+h, j}^2)} \Bigg|^2  < \sigma_f^4.
\end{align*}
It follows that $\|\bSigma_f(h)\|_2^2 \leq \|\bSigma_f(h)\|_F^2 < r^2 \sigma_f^4= O(1)$.

With Condition (C\ref{cond_alphamix}) and Proposition 2.5 in \cite{fan2003}, we can get
\begin{eqnarray*}
\lefteqn{{\rm E}(\hat{\sigma}_{f,ij} -\sigma_{f,ij})^2=\frac{1}{(n-h)^2} {\rm E} \left( \sum_{t=1}^{n-h} f_{t,i} f_{t+h, j}- {\rm E}(f_{t,i}f_{t+h,j})\right)^2}\\
&=& \frac{1}{(n-h)^2} \sum_{|t_1-t_2|\leq h} {\rm E} [f_{t_1,i} f_{t_1+h,j} -{\rm E}(f_{t_1,i} f_{t_1+h, j})][f_{t_2,i} f_{t_2+h,j} -{\rm E}(f_{t_2,i} f_{t_2+h, j})]\\
&&+ \frac{1}{(n-h)^2} \sum_{|t_1-t_2|> h} {\rm E} [f_{t_1,i} f_{t_1+h,j} -{\rm E}(f_{t_1,i} f_{t_1+h, j})][f_{t_2,i} f_{t_2+h,j} -{\rm E}(f_{t_2,i} f_{t_2+h, j})]\\
&\leq &\frac{[(2h+1)n-h^2-h]\sigma_f^4}{(n-h)^2}+ \frac{\sigma_f^4}{n-h}\sum_{u=1}^{n-2h-1} \alpha^{1-2/\gamma}=O(n^{-1}).
\end{eqnarray*}
Hence, ${\rm E}\| \hat{\bSigma}_f(h) -\bSigma_f(h)\|_2^2 \asymp {\rm E} \| \hat{\bSigma}_f(h) -\bSigma_f(h)\|_F^2 =O(n^{-1})$.
\end{proof}

\begin{lemma} Under Conditions (C1)-(C4), it holds that \label{lemma::sig_x}
\begin{align*}
\| \hat{\bSigma}_x(h)- \bSigma_x(h)\|_2=O_p(pn^{-1/2}), \quad \|\bSigma_x(h)\|_2=O(m^{1-\delta}).
\end{align*}
\end{lemma}

\begin{proof} 
Based on the definition, we have 
\begin{eqnarray*}
\lefteqn{\hat{\bSigma}_x(h)- \bSigma_x(h)}\\
&=& \bA^s[\hat{\bSigma}_f^s(h)- \bSigma_f^s(h)]\bA^{s\top} +\frac{1}{n-h}\sum_{t=1}^{n-h}\bA^s \bff_t^s \bve_{t+h}^\top +\frac{1}{n-h}\sum_{t=1}^{n-h}\bve_{t} \bff_{t+h}^{s\top}\bA^{s\top} +\frac{1}{n-h}\sum_{t=1}^{n-h}\bve_t \bve_{t+h}^\top\\
&=& I_1 +I_2 +I_3 +I_4.
\end{eqnarray*}
By Lemma \ref{lemma::sig_f} and Condition (C\ref{cond_strength}), we have
\begin{align}
{\rm E}\| I_1\|_2^2 \leq \|\bA^s\|_2^4 \cdot {\rm E} \|\hat{\bSigma}_f^s(h) -\bSigma_f^s(h)\|_2^2
=O(m^{2-\delta}n^{-1}).
\label{eq::I1}
\end{align}
Condition (C\ref{cond_cov}) implies that 
\begin{eqnarray}
\lefteqn{{\rm E}\|I_2\|_2^2 \nonumber}\\
 &\leq& \|\bA^s \| _2^2 \cdot {\rm E} \Big\| \frac{1}{n} \sum_{t=1}^{n-h} \bff_t \bve_{t+h}\Big\|_F^2
\leq  \|\bA^s \| _2^2 \cdot \sum_{i=1}^r \sum_{j=1}^p{\rm E}\left( \sum_{t=1}^{n-h} f_{t,i} \epsilon_{t+h,j}\right)^2 \nonumber\\
& \leq&    \|\bA^s \| _2^2 \cdot \sum_{i=1}^r \sum_{j=1}^p {\rm E} \left( \sum_{t=1}^{n-h}f_{t,i}^2 \epsilon_{t+h, j}^2\right)^2
=O(m^{2-\delta}n^{-1}),
\label{eq::I2}
\end{eqnarray}
where $\epsilon_{t,j}$ is the $j$-th element in $\bve_t$ and  $\sigma_e^2$ is the marginal variance of $\epsilon_{t,j}$. We can show that ${\rm E}\|I_2\|_2^2 =O(m^{2-\delta}n^{-1})$ in a similar way. About $I_4$,
\begin{eqnarray}
\lefteqn{{\rm E}\|I_4\|_2^2 \nonumber}\\
 &\leq&  {\rm E} \Big\| \frac{1}{n-h} \sum_{t=1}^{n-h} \bve_t \bve_{t+h}\Big\|_F^2
\leq  \frac{1}{(n-h)^2} \sum_{i=1}^p \sum_{j=1}^p{\rm E}\left( \sum_{t=1}^{n-h} \epsilon_{t,i} \epsilon_{t+h,j}\right)^2 \nonumber\\
 &\leq&    \frac{1}{(n-h)^2}\sum_{i=1}^p \sum_{j=1}^p {\rm E} \left( \sum_{t=1}^{n-h}\epsilon_{t,i}^2 \epsilon_{t+h, j}^2\right)^2
=O(p^2n^{-1}).
\label{eq::I4}
\end{eqnarray}
Together with (\ref{eq::I1}), (\ref{eq::I2}), and (\ref{eq::I4}), we have
\[
\| \hat{\bSigma}_x(h)- \bSigma_x(h)\|_2=O_p(pn^{-1/2}).
\]
Condition (C\ref{cond_strength}) and Lemma \ref{lemma::sig_f} show the following,
\begin{align*}
\|\bSigma_x(h)\|_2 \leq \|\bA^s\|_2^2\cdot \|\bSigma_f^s(h)\|_2 = O(m^{1-\delta}).
\end{align*}
\end{proof}

\begin{lemma}
Under Conditions (C1)-(C4), it holds that
    \label{lemma::sig_x_F}
    \begin{align*}
\| \hat{\bSigma}_x(h)- \bSigma_x(h)\|_F=O_p(pn^{-1/2}).
\end{align*}
\end{lemma}
\begin{proof} 
This result can be proved using the same techniques used in Lemma \ref{lemma::sig_x}. Since $\Vert \bA^s\Vert_F\leq \sqrt{r}\Vert \bA^s\Vert$, with Lemma \ref{lemma::sig_f} about F norm, we can have the conclusion. 
\end{proof}

\begin{lemma} Under Conditions (C1)-(C5) and $m^{\delta-1}p n^{-1/2}=o(1)$, it holds that
\begin{align*}
\| \hat{\bM}- \bM\|_2=O_p(m^{1-\delta}p n^{-1/2}), \quad \|\bM\|_{\min}=O(m^{2-2\delta}).
\end{align*}\label{lemma::M}
\end{lemma}

\begin{proof}
  With Lemma \ref{lemma::sig_x}, we can show that
\begin{eqnarray*}
\lefteqn{ \| \hat{\bM}- \bM\|_2^2}\\
&\leq & h_0\sum_{h=1}^{h_0} \| \hat{\bSigma}_x(h)\hat{\bSigma}_x(h)^\top -\bSigma_x(h)\bSigma_x(h)^\top\|_2^2\\
&\leq & 2h_0 \sum_{h=1}^{h_0} \left[  \|\hat{\bSigma}_x(h) -\bSigma_x(h)\|_2^4 + \|\bSigma_x(h)\|_2^2 \cdot  \|\hat{\bSigma}_x(h)-\bSigma_x(h)\|_2^2    \right]\\
&\leq & O_p(p^4 n^{-2}) +O_p(m^{2-2\delta}p^2 n^{-1})=O_p(m^{2-2\delta}p^2 n^{-1}) ,
\end{eqnarray*}
and
\begin{eqnarray*}
\|\bM\|_{\min} \geq  \|\bSigma_x(h)\|_{\min}^2 \geq \|\bA^s\|_{\min}^2  \cdot \| \bSigma_{f}^s(h)\|_{\min}^2=O(m^{2-2\delta}).
\end{eqnarray*}
\end{proof}

\begin{lemma}
Let $\bZ$ be a $n\times 1$ random vector with independent components $Z_{i}$, which satisfy $EZ_{i}=0$ and $V\left(Z_{i}\right)=1$ and $\Vert Z_{i}\Vert_{\psi_{2}}\leq K<\infty$.
$\bV$ is a $n\times n$ semi-positive matrix such that $0<\Vert\bV\Vert_2<c_{0}$, where $c_0 < \infty$ is a constant. Then, 
There exists positive constants $\eta$ and $c$ such that 
\[
P\left(\vert\frac{1}{n}{\bf{Z}^{\top}\bf{V}\bf{Z}}-\frac{1}{n}E{\bf{Z}^{\top}\bf{V}\bf{Z}}\vert>\eta\Vert\bV\Vert_2\right)\leq2\exp\left[-nc\min\left(\eta^{2},\eta\right)\right].
\]

Furthermore, for any $p>0$, there exists constants $0<\rho<\infty$ and $2<\tilde{c}<\infty$
such that
\[
P\left(\vert\frac{1}{n}\bZ^{\top}\bV\bZ-\frac{1}{n}E\bZ^{\top}\bV\bZ\vert>c_{0}\rho\sqrt{\frac{\log p}{n}}\right)\leq\frac{2}{p^{\tilde{c}}}.
\]
\label{lemma_inequal}
\end{lemma}

\begin{proof}
The proof is similar to the one in in \cite{guo2023consistency} and uses the similar technique.
We know that $\Vert\bV\Vert_{F}^{2}\leq n\Vert\bV\Vert_2^{2}$, then by the Hanson-Wright inequality in \cite{Rudelson2013},
\[
P\left(\vert\frac{1}{n}\bZ^{\top}\bV\bZ-\frac{1}{n}E\bZ^{\top}\bV\bZ\vert>t\right)\leq2\exp\left[-c\min\left(\frac{n^{2}t^{2}}{K^{4}\Vert\bV\Vert_{F}^{2}},\frac{nt}{K^{2}\Vert\bV\Vert_2}\right)\right].
\]
Let $t=K^{2}\eta\Vert\bV\Vert_2,$ then 
\[
P\left(\vert\frac{1}{n}{\bf{Z}^{\top}\bf{V}\bf{Z}}-\frac{1}{n}E{\bf{Z}^{\top}\bf{V}\bf{Z}}\vert>\eta\Vert\bV\Vert_2\right)\leq2\exp\left[-nc\min\left(\eta^{2},\eta\right)\right].
\]

Let $\eta=\rho\sqrt{\frac{\log p}{n}}\leq1$ with $\log p=o\left(n\right)$,
and $\tilde{c}=c\rho^{2}>2$, then 
\begin{align*}
 P\left(\vert\frac{1}{n}{\bf{Z}^{\top}\bf{V}\bf{Z}}-\frac{1}{n}E{\bf{Z}^{\top}\bV\bf{Z}}\vert >\eta c_{0}\right)  &\leq P\left(\vert\frac{1}{n}{\bf{Z}^{\top}\bf{V}\bf{Z}}-\frac{1}{n}E{\bf{Z}^{\top}\bV\bf{Z}}\vert>\eta\Vert\bV\Vert_2\right)\\
 & \leq 2\exp\left(-nc\rho^{2}\frac{\log p}{n}\right)=2\exp\left(-c\rho^{2}\log p\right) =  \frac{2}{p^{\tilde{c}}}.
\end{align*}
\end{proof}

\begin{lemma}
Under Conditions (C\ref{cond_alphamix})-(C\ref{cond_subgaussian}), we have  \label{lemma_Minfty}
\[
\Vert\hat{{\bf M}}-{\bf M}\Vert_{\infty}=O_{p}\left(\max\left(p^2n^{-1/2},m^{2-\delta}\right)\sqrt{\frac{\log p}{n}}\right)
\]
\end{lemma}

\begin{proof}
We know that 
\begin{align*}
\hat{{\bf M}}-{\bf M} & =\sum_{h=1}^{h_{0}}\left[\hat{{\bf \Sigma}}_{x}\left(h\right)-{\bf \Sigma}_{x}\left(h\right)\right]\left[\hat{{\bf \Sigma}}_{x}\left(h\right)-{\bf \Sigma}_{x}\left(h\right)\right]^{\top}\\
 & +\sum_{h=1}^{h_{0}}\left(\hat{{\bf \Sigma}}_{x}\left(h\right)-{\bf \Sigma}_{x}\left(h\right)\right){\bf \Sigma}_{x}\left(h\right)^{\top}+\sum_{h=1}^{h_{0}}{\bf \Sigma}_{x}\left(h\right)\left(\hat{{\bf \Sigma}}_{x}\left(h\right)-{\bf \Sigma}_{x}\left(h\right)\right)^{\top}.
\end{align*}

It is known that,
\[
\left(\hat{{\bf \Sigma}}_{x}\left(h\right)-{\bf \Sigma}_{x}\left(h\right)\right){\bf \Sigma}_{x}\left(h\right)^{\top}=\left(\hat{{\bf \Sigma}}_{x}\left(h\right)-{\bf \Sigma}_{x}\left(h\right)\right)\bA^s \bSigma_{f}^s\left(h\right)\bA^{s\top}.
\]
Let $\Delta_{ih}$ be the $(i,h)$th element of $\hat{{\bf \Sigma}}_{x}\left(h\right)-{\bf \Sigma}_{x}\left(h\right)$,
then the $(i,j)$th element of $\left(\hat{{\bf \Sigma}}_{x}\left(h\right)-{\bf \Sigma}_{x}\left(h\right)\right){\bf \Sigma}_{x}\left(h\right)^{\top}$
is $\sum_{l^{\prime}=1}^{r}\sum_{l=1}^{r}\sum_{h=1}^{p}\Delta_{ih}a_{hl}\sigma_{f,ll^{\prime}}a_{jl^{\prime}}$. Then
\[
\Vert\left(\hat{{\bf \Sigma}}_{x}\left(h\right)-{\bf \Sigma}_{x}\left(h\right)\right){\bf \Sigma}_{x}\left(h\right)^{\top}\Vert_{\infty}=\max_{1\leq i\leq p}\sum_{j=1}^{p}\vert\sum_{l^{\prime}=1}^{r}\sum_{l=1}^{r}\sum_{h=1}^{p}\Delta_{ih}a_{hl}\sigma_{f,ll^{\prime}}a_{jl^{\prime}}\vert.
\]
Since $\Delta_{ih}=O_{p}\left(\sqrt{\frac{\log p}{n}}\right)$ from Lemma \ref{lemma::maxSigmax}, then
\[
\Vert\left(\hat{{\bf \Sigma}}_{x}\left(h\right)-{\bf \Sigma}_{x}\left(h\right)\right){\bf \Sigma}_{x}\left(h\right)^{\top}\Vert_{\infty}=O_{p}\left(\sqrt{\frac{\log p}{n}}\right)\sum_{l=1}^{r}\sum_{l^{\prime}=1}^{r}\sum_{j=1}^{p}\sum_{h=1}^{p}\vert a_{hl}\vert\vert a_{jl^{\prime}}\vert.
\]
Furthermore,
\[
\sum_{j=1}^{p}\sum_{h=1}^{p}\vert a_{hl}\vert \vert a_{jl^{\prime}}\vert=\sum_{j=1}^{p}\vert a_{jl^{\prime}}\vert\sum_{h=1}^{p}\vert a_{hl}\vert,
\]
and 
$\sum_{j=1}^{p}\vert a_{jl^{\prime}}\vert\leq\sqrt{m}\Vert\ba^s_{l^{\prime}}\Vert\asymp m^{1-\delta/2}$ from Condition (C\ref{cond_strength}).
Thus $
\sum_{j=1}^{p}\sum_{h=1}^{p}\vert a_{hl}\vert \vert a_{jl^{\prime}}\vert\asymp m^{2-\delta}.$
This implies that  
\begin{equation}
\label{eq_siginf_p1}
 \Vert\left(\hat{{\bf \Sigma}}_{x}\left(h\right)-{\bf \Sigma}_{x}\left(h\right)\right){\bf \Sigma}_{x}\left(h\right)^{\top}\Vert_{\infty}=O_{p}\left(m^{2-\delta}\sqrt{\frac{\log p}{n}}\right). 
\end{equation}

For the first term, with Lemma \ref{lemma::maxSigmax} and Lemma \ref{lemma::sig_x_F}, we have
\begin{align}
\label{eq_siginf_p2}
 & \Vert\left[\hat{{\bf \Sigma}}_{x}\left(h\right)-{\bf \Sigma}_{x}\left(h\right)\right]\left[\hat{{\bf \Sigma}}_{x}\left(h\right)-{\bf \Sigma}_{x}\left(h\right)\right]^{\top}\Vert_{\infty} \nonumber\\
 =&\max_{1\leq i \leq p}\sum_{j=1}^p\vert \sum_{h=1}^p\Delta_{ih}\Delta_{jh}\vert \leq \Vert \hat{{\bf \Sigma}}_{x}\left(h\right)-{\bf \Sigma}_{x}\Vert_{\max}  \vert \sum_{j=1}^p\sum_{h=1}^p\vert \Delta_{jh}\vert \nonumber \\
 = & \Vert \hat{{\bf \Sigma}}_{x}\left(h\right)-{\bf \Sigma}_{x}\Vert_{\max} \sqrt{p^2\Vert \hat{{\bf \Sigma}}_{x}\left(h\right)-{\bf \Sigma}_{x}\Vert_F^2} =  O_p\left(p^2n^{-1/2}\sqrt{\frac{\log p}{n}}\right).
\end{align}
Combine \eqref{eq_siginf_p1} and \eqref{eq_siginf_p2}, we have
\[
\Vert\hat{{\bf M}}-{\bf M}\Vert_{\infty}=O_{p}\left(\max\left(p^{2}n^{-1/2},m^{2-\delta}\right)\sqrt{\frac{\log p}{n}}\right).
\]
\end{proof}

\begin{lemma}
Let ${\bf u}$ be a $p\times 1$ vector such that $\Vert{\bf u}\Vert_2=1$ 
and $\mathcal{A}$ be an index, ${\bf u}_{[\mathcal{A}]}^{*}={\bf u}_{[\mathcal{A}]}$, ${\bf u}_{[-\mathcal{A}]}^{*}={\bf 0}$ and $\vert \mathcal{A}\vert \asymp m$. Then, (the max element) \label{lemma_uE}
\[
\Vert{\bf u}^{*\top}\left(\hat{\bS}\hat{\bS}^{\top}-\bS\bS^{\top}\right)\Vert_{\max} =\begin{cases}
 O_p\left(\max\left(m^{2\delta-2}p^{2}n^{-1/2},m^{\delta}\right)\sqrt{\frac{\log p}{n}}\right), & \text{if } m= o(p)  \\
O_p\left( m^{\delta-1} p n^{-1/2}\right)  = O_p\left(p^\delta n^{-1/2}\right), & \text{if } m= O(p).  \\
\end{cases}
\]
\end{lemma}


\begin{proof}

If $m=o(p)$, we know that 
\begin{align*}
 & \Vert{\bf u}^{*\top}\left(\hat{\bS}\hat{\bS}^{\top}-\bS\bS^{\top}\right)\Vert_{\max} \leq  \Vert\hat{\bS}\hat{\bS}^{\top}-\bS\bS^{\top}\Vert_{\max}\sum_{i=1}^{p}\vert u_{i}^{*}\vert \leq \sqrt{m}\Vert\hat{\bS}\hat{\bS}^{\top}-\bS\bS^{\top}\Vert_{\max}.
\end{align*}

Furthermore, we have 
\begin{align*}
\Vert\hat{\bS}\hat{\bS}^{\top}-\bS\bS^{\top}\Vert_{\max} & \leq \sum_{j=1}^{r}\Vert\hat{\bs}_{j}\hat{\bs}_{j}^{\top}-\bs_{j}\bs_{j}^{\top}\Vert_{\max}\leq2\sum_{j=1}^{r}\max_{i}\vert\hat{s}_{ij}-s_{ij}\vert\\
 & =2\max_{i}\sum_{j=1}^{r}\vert\hat{s}_{ij}-s_{ij}\vert=2\Vert \hat{\bS}-\bS\Vert_{\infty} \leq2\sqrt{r}\Vert \hat{\bS}-\bS\Vert_{2\rightarrow\infty}.
\end{align*}

Using the same result about $\Vert \hat{\bS}-\bS\Vert_{2\rightarrow\infty}$ as used in the proof of Lemma \ref{lemma_Sstar}, we know that 
\begin{align*}
& \Vert{\bf u}^{*\top}\left(\hat{\bS}\hat{\bS}^{\top}-\bS\bS^{\top}\right)\Vert_{\max}\\
\leq & 2\sqrt{r} \sqrt{m} \Vert \hat{\bS}-\bS\Vert_{2\rightarrow\infty} 
\leq  28 \sqrt{r} \sqrt{m} \left(\frac{\Vert\hat{{\bf M}}-{\bf M}\Vert_{\infty}}{\lambda_{r}}\right)\Vert{\bf S}\Vert_{2\rightarrow\infty}\\
= & O_p\left(\max\left(m^{2\delta-2}p^{2}n^{-1/2},m^{\delta}\right)\sqrt{\frac{\log p}{n}}\right).
\end{align*}

If $m=O(p)$, we have 
\begin{align*}
 \Vert{\bf u}^{*\top}\left(\hat{\bS}\hat{\bS}^{\top}-\bS\bS^{\top}\right)\Vert_{\max} & \leq  \Vert{\bf u}^{*\top}\left(\hat{\bS}\hat{\bS}^{\top}-\bS\bS^{\top}\right)\Vert_2\\
 & \leq \Vert {\bf u} \Vert_2 \Vert\hat{\bS}\hat{\bS}^{\top}-\bS\bS^{\top} \Vert_2 \leq \Vert\hat{\bS}\hat{\bS}^{\top}-\bS\bS^{\top} \Vert_2 \\
 &   = O_p\left( m^{\delta-1} p n^{-1/2}\right)  = O_p\left(p^\delta n^{-1/2}\right),
\end{align*}
where the last result is Theorem 1. 
\end{proof}

Let $\mathcal{V}_{i}$ be the nonzero index of
${\bf q}_{i}$ and $\mathcal{N}_{i}$ be the zero index of ${\bf q}_{i}$
and denote ${\bf q}_{i}^{*}={\bf q}_{i[\mathcal{V}_{i}]}$, a subvector of $\bq_i$ with nonzero elements. Let $\mathcal{V}_{s_{i}}$
be the nonzero index of ${\bf s}_{i}$. Based on the relationship between
${\bf s}_{i}$ and ${\bf q}_{i}$, we know that $\mathcal{V}_{s_{i}}\subseteq\mathcal{V}_{s_{1}}\cup\mathcal{V}_{s_{2}}\dots\cup\mathcal{V}_{s_{i-1}}\cup\mathcal{V}_{i}$.
Let $\mathcal{V}_{i}^{*}=\mathcal{V}_{s_{1}}\cup\mathcal{V}_{s_{2}}\dots\cup\mathcal{V}_{s_{i-1}}\cup\mathcal{V}_{i},$
and $\mathcal{N}_{i}^{*}=\mathcal{V}_{i}^{*}\backslash\mathcal{V}_{i}$. From the definition, we know that $\mathcal{V}_i^*$ contains the nonzero index of $\bs_i$ and  $\bq_i$, and the elements out of $\mathcal{V}_i^*$ in $\bs_i$ and $\bq_i$ are zeros. $\mathcal{N}_i^*$ is the index that $\bq_i$ has zeros while $\bs_i$ may not. Denote ${\bf S}_{i,1}={\bf S}_{i[\mathcal{N}_{i}^{*}]}$ , ${\bf S}_{i,2}={\bf S}_{i[\mathcal{V}_{i}]}$,
${\bf s}_{i,1}={\bf s}_{i[\mathcal{N}_{i}^{*}]}$ and ${\bf s}_{i,2}={\bf s}_{i[\mathcal{V}_{i}]}$.
With loss of generality, we can write ${\bf s}_{i}$ as below
\begin{align}
\label{eq:s_constraint}
{\bf s}_{i}= \left(\begin{array}{c}
{\bf 0}\\
{\bf s}_{i,1}\\
{\bf s}_{i,2}
\end{array}\right)= (\bI - \bS_i\bS_i^\top) \bq_i =  & \left(\begin{array}{ccc}
{\bf I} & {\bf 0} & {\bf 0}\\
{\bf 0} & {\bf I} & -{\bf S}_{i,1}{\bf S}_{i,2}^{\top}\\
{\bf 0} & -{\bf S}_{i,2}{\bf S}_{i,1}^{\top} & {\bf I}-{\bf S}_{i,2}{\bf S}_{i,2}^{\top}
\end{array}\right)\left(\begin{array}{c}
{\bf 0}\\
{\bf 0}\\
{\bf q}_{i}^{*}
\end{array}\right) \nonumber\\
= & \left(\begin{array}{c}
{\bf 0}\\
-{\bf S}_{i,1}{\bf S}_{i,2}^{\top}{\bf q}_{i}^{*}\\
\left({\bf I}-{\bf S}_{i,2}{\bf S}_{i,2}^{\top}\right){\bf q}_{i}^{*}
\end{array}\right).
\end{align}

\begin{lemma}
The matrix defined in \eqref{eq:s_constraint} ${\bf I}-{\bf S}_{i,2}{\bf S}_{i,2}^{\top}$ and the matrix ${\bf I}-{\bf S}^\top_{i,2}{\bf S}_{i,2}$ are invertible. 
    \label{lemma:inv_s2}
\end{lemma}

\begin{proof}
    Suppose ${\bf I}-{\bf S}_{i,2}{\bf S}_{i,2}^{\top}$ is not is invertible,
then there exists a nonzero vector ${\bf x}$ such that $\left({\bf I}-{\bf S}_{i,2}{\bf S}_{i,2}^{\top}\right){\bf x}={\bf 0}$,
which indicates that ${\bf x}={\bf S}_{i,2}{\bf S}_{i,2}^{\top}{\bf x}$.
Then, we have 
\[
\Vert{\bf x}\Vert_{2}^{2}={\bf x}^{\top}{\bf x}={\bf x}^{\top}{\bf S}_{i,2}{\bf S}_{i,2}^{\top}{\bf x}\leq\lambda_{\max}\left({\bf S}_{i,2}{\bf S}_{i,2}^{\top}\right){\bf x}^{\top}{\bf x}.
\]

However, we know that $\lambda_{\max}\left({\bf S}_{i,2}{\bf S}_{i,2}^{\top}\right)=\Vert{\bf S}_{i,2}{\bf S}_{i,2}^{\top}\Vert_{2}<\Vert{\bf S}_{i}{\bf S}_{i}^{\top}\Vert_{2}=1$,
since the nonzero index of each column of ${\bf S}_{i}$ cannot be
a subset of $\mathcal{V}_{i}$, otherwise extra 0 elements in ${\bf q}_{i}$
can be constructed. Since $\Vert{\bf x}\Vert_{2}\neq0$, thus we have
a contradiction ${\bf x}^{\top}{\bf x}<{\bf x}^{\top}{\bf x}$. 

This completes the proof that ${\bf I}-{\bf S}_{i,2}{\bf S}_{i,2}^{\top}$ is invertible. 

By the similar arguments, we can show that ${\bf I}-{\bf S}^\top_{i,2}{\bf S}_{i,2}$ is also invertible.
\end{proof}

\begin{lemma}
For ${\bf s}_{i}$ defined in \eqref{eq:s_constraint}, we have ${\bf s}_{i,1}={\bf S}_{i,1}\left(\mathbf{S}_{i,1}^{\top}\mathbf{S}_{i,1}\right)^{-1}\mathbf{S}_{i,1}^{\top}{\bf s}_{i,1}$. 
\label{lemma:si1}
\end{lemma}
\begin{proof}
    Based on the relationship, we now know that ${\bf s}_{i,2}=\left({\bf I}-{\bf S}_{i,2}{\bf S}_{i,2}^{\top}\right){\bf q}_{i}^{*}$
and ${\bf s}_{i,1}=-{\bf S}_{i,1}{\bf S}_{i,2}^{\top}{\bf q}_{i}^{*}$.
This implies that ${\bf q}_{i}^{*}=\left({\bf I}-{\bf S}_{i,2}{\bf S}_{i,2}^{\top}\right)^{-1}{\bf s}_{i,2}$
and ${\bf s}_{i,1}=-{\bf S}_{i,1}{\bf S}_{i,2}^{\top}\left({\bf I}-{\bf S}_{i,2}{\bf S}_{i,2}^{\top}\right)^{-1}{\bf s}_{i,2}$.

We know that $\left({\bf I}-{\bf S}_{i,2}{\bf S}_{i,2}^{\top}\right)^{-1}=\mathbf{I}+\mathbf{S}_{i2}\left(\mathbf{I}-\mathbf{S}_{i,2}^{\top}\mathbf{S}_{i,2}\right)^{-1}\mathbf{S}_{i,2}^{\top}$,
and $\mathbf{S}_{i,2}^{\top}\mathbf{S}_{i,2}=\mathbf{I}-\mathbf{S}_{i,1}^{\top}\mathbf{S}_{i,1}$,
thus 
\begin{align*}
 & -{\bf S}_{i,1}{\bf S}_{i,2}^{\top}\left({\bf I}-{\bf S}_{i,2}{\bf S}_{i,2}^{\top}\right)^{-1}\\
= & -{\bf S}_{i,1}{\bf S}_{i,2}^{\top}\left(\mathbf{I}+\mathbf{S}_{i2}\left(\mathbf{I}-\mathbf{S}_{i,2}^{\top}\mathbf{S}_{i,2}\right)^{-1}\mathbf{S}_{i,2}^{\top}\right)\\
= & -{\bf S}_{i,1}{\bf S}_{i,2}^{\top}-{\bf S}_{i,1}{\bf S}_{i,2}^{\top}\mathbf{S}_{i2}\left(\mathbf{I}-\mathbf{S}_{i,2}^{\top}\mathbf{S}_{i,2}\right)^{-1}\mathbf{S}_{i,2}^{\top}\\
= & -{\bf S}_{i,1}{\bf S}_{i,2}^{\top}-{\bf S}_{i,1}\left(\mathbf{I}-\mathbf{S}_{i,1}^{\top}\mathbf{S}_{i,1}\right)\left(\mathbf{S}_{i,1}^{\top}\mathbf{S}_{i,1}\right)^{-1}\mathbf{S}_{i,2}^{\top}\\
= & -{\bf S}_{i,1}\left(\mathbf{S}_{i,1}^{\top}\mathbf{S}_{i,1}\right)^{-1}\mathbf{S}_{i,2}^{\top}
\end{align*}
Then, ${\bf s}_{i,1}=-{\bf S}_{i,1}\left(\mathbf{S}_{i,1}^{\top}\mathbf{S}_{i,1}\right)^{-1}\mathbf{S}_{i,2}^{\top}{\bf s}_{i,2}.$
In additional, we know that $\mathbf{S}_{i,2}^{\top}{\bf s}_{i,2}=-\mathbf{S}_{i,1}^{\top}{\bf s}_{i,1}$,
thus ${\bf s}_{i,1}={\bf S}_{i,1}\left(\mathbf{S}_{i,1}^{\top}\mathbf{S}_{i,1}\right)^{-1}\mathbf{S}_{i,1}^{\top}{\bf s}_{i,1}$.
\end{proof}

\begin{lemma}
   Under Conditions (C\ref{cond_alphamix})-(C\ref{cond_si1_eigen}), let $\tau_{n,p,m} = \max\left(m^{2\delta-2}p^{2}n^{-1/2},m^{\delta}\right)\sqrt{\frac{\log p}{n}}$ for $m=o(p)$ and $\tau_{n,p,m} = p^\delta n^{-1/2}$ for $m=O(p)$. If  $\Vert \tilde{\bf{S}}_i - {{\bf S}_i}\Vert_2 = O_p(\tau_{n,p,m})$, for different cases of $m$ and $p$, then, we have 
  $$
  \Vert\left(\tilde{{\bf S}}_{i,1}^{\top}\tilde{{\bf S}}_{i,1}\right)^{-1}-\left(\mathbf{S}_{i,1}^{\top}\mathbf{S}_{i,1}\right)^{-1}\Vert_{2} =  \begin{cases}
      O_{p}\left(\max\left(m^{2\delta-2}p^{2}n^{-1/2},m^{\delta}\right)\sqrt{\frac{\log p}{n}}\right) & \text{if } m=o(p)\\
      O_{p}\left(p^{\delta}n^{-1/2}\right) & \text{if }  m=O(p). 
  \end{cases}
   $$
   $$
  \Vert\tilde{{\bf S}}_{i,1}\left(\tilde{{\bf S}}_{i,1}^{\top}\tilde{{\bf S}}_{i,1}\right)^{-1}\tilde{{\bf S}}_{i,1}^{\top}-\mathbf{S}_{i,1}^{\top}\left(\mathbf{S}_{i,1}^{\top}\mathbf{S}_{i,1}\right)^{-1}\mathbf{S}_{i,1}\Vert_2=\begin{cases}
      O_{p}\left(\max\left(m^{2\delta-2}p^{2}n^{-1/2},m^{\delta}\right)\sqrt{\frac{\log p}{n}}\right) & \text{if } m=o(p)\\
      O_{p}\left(p^{\delta}n^{-1/2}\right) & \text{if }  m=O(p). 
  \end{cases}
   $$
   and 
  \[
\Vert\tilde{{\bf S}}_{i,2}\left(\tilde{{\bf S}}_{i,1}^{\top}\tilde{{\bf S}}_{i,1}\right)^{-1}\tilde{{\bf S}}_{i,2}^{\top}-\mathbf{S}_{i,2}^{\top}\left(\mathbf{S}_{i,1}^{\top}\mathbf{S}_{i,1}\right)^{-1}\mathbf{S}_{i,2}\Vert_{2}=\begin{cases}
      O_{p}\left(\max\left(m^{2\delta-2}p^{2}n^{-1/2},m^{\delta}\right)\sqrt{\frac{\log p}{n}}\right) & \text{if } m=o(p)\\
      O_{p}\left(p^{\delta}n^{-1/2}\right) & \text{if }  m=O(p). 
  \end{cases}
\]
   \label{lemma:boundprojection}
\end{lemma}


\begin{proof}
 Based on Weyl's inequality, we have 
\[
\lambda_{j}\left(\mathbf{S}_{i,1}^{\top}\mathbf{S}_{i,1}\right)+\lambda_{\min}\left(\tilde{\bf {S}}_{i,1}^{\top}\tilde{{\bf {S}}}_{i,1}-\mathbf{S}_{i,1}^{\top}\mathbf{S}_{i,1}\right)\leq\lambda_{j}\left(\tilde{\bf {S}}_{i,1}^{\top}\tilde{\bf {S}}_{i,1}\right)\leq\lambda_{j}\left(\mathbf{S}_{i,1}^{\top}\mathbf{S}_{i,1}\right)+\lambda_{\max}\left(\tilde{\bf {S}}_{i,1}^{\top}\tilde{\bf {S}}_{i,1}-\mathbf{S}_{i,1}^{\top}\mathbf{S}_{i,1}\right),
\]
where $\lambda_j(\cdot)$ is the $jth$ largest eigenvalue. 

We know that $\Vert \tilde{\bf {S}}_{i,1}^{\top}\tilde{\bf {S}}_{i,1}-\mathbf{S}_{i,1}^{\top}\mathbf{S}_{i,1}\Vert_{2}=O_{p}\left(\tau_{n,p,m}\right)$ since $\Vert \tilde{\bf{S}}_i - {{\bf S}_i}\Vert_2 = O_p(\tau_{n,p,m})$. Thus,
thus $\lambda_{j}\left(\tilde{\bf {S}}_{i,1}^{\top}\tilde{\bf {S}}_{i,1}\right)=\lambda_{j}\left(\mathbf{S}_{i,1}^{\top}\mathbf{S}_{i,1}\right)+O_{p}\left(\tau_{n,p,m}\right)$
for $j=1,\dots,i-1$. 

Denote $\tilde{\lambda}_{j}=\lambda_{j}\left(\tilde{\bf {S}}_{i,1}^{\top}\tilde{\bf {S}}_{i,1}\right)$
and $\lambda_{j}=\lambda_{j}\left(\mathbf{S}_{i,1}^{\top}\mathbf{S}_{i,1}\right)$, then we have 
\begin{equation}
\label{eq_boundlambda}
\vert\tilde{\lambda}_{j}^{-1}-\lambda_{j}^{-1}\vert\leq\frac{\vert\lambda_{j}-\tilde{\lambda}_{j}\vert}{\lambda_{j}\tilde{\lambda}_{j}}=O_{p}\left(\tau_{n,p,m}\right).
\end{equation}

 We also know that $\tilde{{\bf S}}_{i,1}^{\top}\tilde{{\bf S}}_{i,1}=\hat{{\bf V}}\tilde{\Lambda}\hat{{\bf V}}^{\top}$
and ${\bf S}_{i,1}^{\top}{\bf S}_{i,1}={\bf V}\Lambda{\bf V}^{\top},$ where
$\tilde{\Lambda}=\text{diag}\left(\tilde{\lambda}_{1},\dots,\tilde{\lambda}_{i-1}\right)$
and $\Lambda=\text{diag}\left(\lambda_{1},\dots,\lambda_{i-1}\right)$.
And $\lambda_{i-1}=\Vert{\bf S}_{i-1}\Vert_{\min}^{2}$. Thus 
\begin{equation}
\label{eq_boundv}
 \Vert\hat{{\bf V}}-{\bf V}\Vert_{2}=O_{p}\left(\frac{\Vert\tilde{{\bf S}}_{i,1}^{\top}\tilde{{\bf S}}_{i,1}-{\bf S}_{i,1}^{\top}{\bf S}_{i,1}\Vert_{2}}{\lambda_{i-1}}\right)=O_{p}\left(\tau_{n,p,m}\right).  
\end{equation}

We know that $\left(\tilde{{\bf S}}_{i,1}^{\top}\tilde{{\bf S}}_{i,1}\right)^{-1}=\hat{{\bf V}}\tilde{\Lambda}^{-1}\hat{{\bf V}}^{\top}=\sum_{j=1}^{i-1}\tilde{\lambda}_{j}^{-1}\hat{{\bf v}}_{j}\hat{{\bf v}}_{j}^{\top}$
and $\left({\bf S}_{i,1}^{\top}{\bf S}_{i,1}\right)^{-1}={\bf V}\Lambda^{-1}{\bf V}^{\top}=\sum_{j=1}^{i-1}\lambda_{j}^{-1}{\bf v}_{j}{\bf v}_{j}^{\top}$.
Thus we have 
\begin{align*}
\Vert\left(\tilde{{\bf S}}_{i,1}^{\top}\tilde{{\bf S}}_{i,1}\right)^{-1}-\left(\mathbf{S}_{i,1}^{\top}\mathbf{S}_{i,1}\right)^{-1}\Vert_{2} & \leq\sum_{j=1}^{i-1}\Vert\tilde{\lambda}_{j}^{-1}\hat{{\bf v}}_{j}\hat{{\bf v}}_{j}^{\top}-\lambda_{j}^{-1}{\bf v}_{j}{\bf v}_{j}^{\top}\Vert_{2}\\
 & \leq\sum_{j=1}^{i-1}\Vert\tilde{\lambda}_{j}^{-1}-\lambda_{j}^{-1}\Vert_{2}\Vert\hat{{\bf v}}_{j}\hat{{\bf v}}_{j}^{\top}\Vert_{2}+\vert\lambda_{j}^{-1}\vert\Vert\hat{{\bf v}}_{j}\hat{{\bf v}}_{j}^{\top}-{\bf v}_{j}{\bf v}_{j}^{\top}\Vert_{2}\\
 & =O_{p}\left(\tau_{n,p,m}\right). 
\end{align*}
based on \eqref{eq_boundlambda} and \eqref{eq_boundv}.

We know that, 
\begin{align*}
 & \Vert\tilde{{\bf S}}_{i,1}\left(\tilde{{\bf S}}_{i,1}^{\top}\tilde{{\bf S}}_{i,1}\right)^{-1}-\mathbf{S}_{i,1}^{\top}\left(\mathbf{S}_{i,1}^{\top}\mathbf{S}_{i,1}\right)^{-1}\Vert_{2}\\
\leq & \Vert\tilde{{\bf S}}_{i,1}\Vert_{2}\Vert\left(\tilde{{\bf S}}_{i,1}^{\top}\tilde{{\bf S}}_{i,1}\right)^{-1}-\left(\mathbf{S}_{i,1}^{\top}\mathbf{S}_{i,1}\right)^{-1}\Vert_{2}+\Vert\tilde{{\bf S}}_{i,1}-\mathbf{S}_{i,1}^{\top}\Vert_{2}\Vert\left(\mathbf{S}_{i,1}^{\top}\mathbf{S}_{i,1}\right)^{-1}\Vert_{2}\\
= & O_{p}\left(\tau_{n,p,m}\right).
\end{align*}
Thus,
\begin{align*}
 & \Vert\tilde{{\bf S}}_{i,1}\left(\tilde{{\bf S}}_{i,1}^{\top}\tilde{{\bf S}}_{i,1}\right)^{-1}\tilde{{\bf S}}_{i,1}^{\top}-\mathbf{S}_{i,1}^{\top}\left(\mathbf{S}_{i,1}^{\top}\mathbf{S}_{i,1}\right)^{-1}\mathbf{S}_{i,1}\Vert_{2}\\
\leq & \Vert\tilde{{\bf S}}_{i,1}\left(\tilde{{\bf S}}_{i,1}^{\top}\tilde{{\bf S}}_{i,1}\right)^{-1}\Vert_{2}\Vert\tilde{{\bf S}}_{i,1}-{\bf S}_{i,1}\Vert_{2}+\Vert\tilde{{\bf S}}_{i,1}\left(\tilde{{\bf S}}_{i,1}^{\top}\tilde{{\bf S}}_{i,1}\right)^{-1}-\mathbf{S}_{i,1}^{\top}\left(\mathbf{S}_{i,1}^{\top}\mathbf{S}_{i,1}\right)^{-1}\Vert_{2}\Vert{\bf S}_{i,1}\Vert_{2}\\
= & O_{p}\left(\tau_{n,p,m}\right).
\end{align*}

By using the similar arguments, we can show the result for $
\Vert\tilde{{\bf S}}_{i,2}\left(\tilde{{\bf S}}_{i,1}^{\top}\tilde{{\bf S}}_{i,1}\right)^{-1}\tilde{{\bf S}}_{i,2}^{\top}-\mathbf{S}_{i,2}^{\top}\left(\mathbf{S}_{i,1}^{\top}\mathbf{S}_{i,1}\right)^{-1}\mathbf{S}_{i,2}\Vert_{2}=O_{p}\left(\tau_{n,p,m}\right).
$

\end{proof}


\section{Existing Definitions and Results}
\label{appendix_existing}

We provide the following definitions as provided in \cite{vershynin2018high}.

\begin{definition}
    \label{def_subgaussian}
    A random variable $X$ is called a sub-Gaussian random variable if there exists a $K>0$ such that 
    \begin{equation}
           E(X^2/K^2) \leq 2.\label{eq_subgaussian}
    \end{equation}
    And the sub-Gaussian norm of $X$, denoted $\Vert X\Vert_{\psi_2}$ is defined as the the smallest $K$ in \eqref{eq_subgaussian}. That is $\Vert X\Vert_{\psi_2} = \text{inf }\{k>0,\, E(X^2/k^2)\leq 2\}$.
\end{definition}

\begin{definition}
    \label{def_subvector}
A random vector $\bX$ in $\mathcal{R}^p$ is called sub-Gaussian if the one-dimensional marginals $\bx^\top \bX$ are sub-Gaussian random variables for $\bx\in \mathcal{R}^p$.
\end{definition}

We provide the existing results from \cite{cape2019two}.
\subsection*{Results 1}
From \cite{cape2019two}
Suppose ${\bf A}$ is a $p_{1}\times p_{2}$ matrix, then
\[
\frac{1}{\sqrt{p_{2}}}\Vert{\bf A}\Vert_{2\rightarrow\infty}\leq\Vert{\bf A}\Vert_{\max}\leq\Vert{\bf A}\Vert_{2\rightarrow\infty}\leq\Vert{\bf A}\Vert_{\infty}\leq\sqrt{p_{2}}\Vert{\bf A}\Vert_{2\rightarrow\infty}
\]
and 
\[
\Vert{\bf A}\Vert_{2\rightarrow\infty}\leq\Vert{\bf A}\Vert_{2}\leq\sqrt{p_{1}}\Vert{\bf A}\Vert_{2\rightarrow\infty}.
\]

{\bf Proposition 6.5} For $A\in \mathbb{R}^{p_1\times p_2}$, $B\in \mathbb{R}^{p_2\times p_3}$ and $C\in \mathbb{R}^{p_4\times p_1}$, then
\begin{align*}
    \Vert AB\Vert _{2\rightarrow\infty} \leq \Vert A\Vert_{2\rightarrow\infty}\Vert B\Vert_2;\\
   \Vert CA\Vert _{2\rightarrow\infty} \leq \Vert C\Vert_{\infty}\Vert A\Vert_{2\rightarrow\infty}.
\end{align*}

\subsection*{Result 2}

\textbf{Theorem 4.2 from \cite{cape2019two}}. Let $X$ and $E$ be $p\times p$ symmetric
matrixs where $X$ with rank $(X)=r$ has spectral decomposition $X=U\Lambda U^{\top}$and
leading eigenvalues $\vert\lambda_{1}\vert\geq\vert\lambda_{2}\vert\geq\cdots\geq\vert\lambda_{r}\vert>0$.
Suppose $\hat{X}=X+E$. Suppose $\vert\lambda_{r}\vert\geq4\Vert E\Vert_{\infty}$.
Then there exisits an orthogonal matrix $W$ ($r\times r$) such that
\[
\Vert\hat{U}-UW\Vert_{2\rightarrow\infty}\leq14\left(\frac{\Vert E\Vert_{\infty}}{\vert\lambda_{r}\vert}\right)\Vert U\Vert_{2\rightarrow\infty}.
\]

We also know that when all eigenvalues are distinct, then $U$ is identical, with $W$ being the identity matrix.

\bibliographystyle{apalike}

\bibliography{reference_JTSA}
\end{document}